\documentclass[11pt,a4paper]{emulateapj}

\usepackage{natbib}
\usepackage{color}
\usepackage{amsmath}
\usepackage{graphicx}

\newcommand{\chandra}{\emph{Chandra}}

\newcommand{\jvla}{\emph{JVLA}}

\definecolor{green}{RGB}{102,153,102}
\definecolor{darkred}{RGB}{153,0,51}
\definecolor{blue}{RGB}{51,102,153}

\begin{document}
\slugcomment{LLNL-JRNL-676771}

\title{The discovery of lensed radio and X-ray sources behind the Frontier Fields cluster MACS~J0717.5+3745 with the JVLA and Chandra}


\author{R.~J.~van~Weeren\altaffilmark{1,$\star$}, G.~A. Ogrean\altaffilmark{1,$\dagger$}, C.~Jones\altaffilmark{1}, W.~R.~Forman\altaffilmark{1},  F.~Andrade-Santos\altaffilmark{1}, 
A.~Bonafede\altaffilmark{2}, M.~Br\"uggen\altaffilmark{2}, E.~Bulbul\altaffilmark{1}, T.~E.~Clarke\altaffilmark{3}, E.~Churazov\altaffilmark{4,5}, L.~David\altaffilmark{1}, W.~A.~Dawson\altaffilmark{6}, M.~Donahue\altaffilmark{7}, A.~Goulding\altaffilmark{8},  R.~P.~Kraft\altaffilmark{1}, B.~Mason\altaffilmark{9}, J.~Merten\altaffilmark{10}, T.~Mroczkowski\altaffilmark{3,$\dagger\dagger$}, S.~S.~Murray\altaffilmark{1,11}, P.~E.~J.~Nulsen\altaffilmark{1,12}, P.~Rosati\altaffilmark{13}, E.~Roediger\altaffilmark{14,$\dagger\dagger\dagger$},   S.~W.~Randall\altaffilmark{1}, J.~Sayers\altaffilmark{15},  K.~Umetsu\altaffilmark{16}, A.~Vikhlinin\altaffilmark{1}, A.~Zitrin\altaffilmark{15,$\dagger$}}


\affil{\altaffilmark{1}Harvard-Smithsonian Center for Astrophysics, 60 Garden Street, Cambridge, MA 02138, USA}
\affil{\altaffilmark{2}Hamburger Sternwarte, Universit\"at Hamburg, Gojenbergsweg 112, D-21029 Hamburg, Germany}
\affil{\altaffilmark{3}U.S. Naval Research Laboratory, 4555 Overlook Ave SW, Washington, D.C. 20375, USA}
\affil{\altaffilmark{4}Max Planck Institute for Astrophysics, Karl-Schwarzschild-Str. 1, 85741, Garching, Germany}
\affil{\altaffilmark{5}Space Research Institute, Profsoyuznaya 84/32, Moscow, 117997, Russia}
\affil{\altaffilmark{6}Lawrence Livermore National Lab, 7000 East Avenue, Livermore, CA 94550, USA}
\affil{\altaffilmark{7}Department of Physics and Astronomy, Michigan State University, East Lansing, MI 48824, USA}
\affil{\altaffilmark{8}Department of Astrophysical Sciences, Princeton University, Princeton, NJ 08544, USA}
\affil{\altaffilmark{9}National Radio Astronomy Observatory, 520 Edgemont Road, Charlottesville, VA 22903, USA}
\affil{\altaffilmark{10}Department of Physics, University of Oxford, Keble Road, Oxford OX1 3RH, UK}
\affil{\altaffilmark{11}Department of Physics and Astronomy, Johns Hopkins University, 3400 North Charles Street, Baltimore, MD 21218, USA}
\affil{\altaffilmark{12}ICRAR, University of Western Australia, 35 Stirling Hwy, Crawley WA 6009, Australia}
\affil{\altaffilmark{13}Dipartimento di Fisica e Scienze della Terra, Universit\`a di Ferrara, Via Saragat 1, I-44122 Ferrara, Italy}
\affil{\altaffilmark{14}E.A. Milne Centre for Astrophysics, Department of Physics \& Mathematics, University of Hull, Cottinton Road, Hull, HU6 7RX, UK}
\affil{\altaffilmark{15}Cahill Center for Astronomy and Astrophysics, California Institute of Technology, MC 249-17, Pasadena, CA 91125, USA}
\affil{\altaffilmark{16}Institute of Astronomy and Astrophysics, Academia Sinica, PO Box 23-141, Taipei 10617, Taiwan}

\email{rvanweeren@cfa.harvard.edu}

\altaffiltext{$\star$}{Einstein Fellow}
\altaffiltext{$\dagger$}{Hubble Fellow}
\altaffiltext{$\dagger\dagger$}{National Research Council Fellow}
\altaffiltext{$\dagger\dagger\dagger$}{Visiting Scientist}

\shorttitle{Lensed radio and X-ray sources behind MACS~J0717.5+3745}
\shortauthors{van Weeren et al.}

\vspace{0.5cm}
\begin{abstract}
\noindent We report on high-resolution \jvla\ and \chandra\ observations of the HST Frontier Cluster MACS~J0717.5+3745. 
MACS~J0717.5+3745 offers the largest contiguous magnified area of any known cluster, making it a promising target to search for lensed radio and X-ray sources. With the high-resolution 1.0--6.5~GHz \jvla\ imaging in A and B configuration, we detect a total of 51 compact radio sources within the area covered by the HST imaging. Within this sample we find 7~lensed sources with amplification factors larger than $2$. None of these sources are identified as multiply-lensed. Based on the radio luminosities, the majority of these sources are likely star forming galaxies with star formation rates of 10--50 M$_\odot$~yr$^{-1}$ located at $1 \lesssim z \lesssim 2$. Two of the lensed radio sources are also detected in the \chandra\ image of the cluster. These two sources are likely AGN, given their $2-10$~keV X-ray luminosities of  $\sim 10^{43-44}$ erg~s$^{-1}$. From the derived radio luminosity function, we find evidence for an increase in the number density of radio sources at $0.6<z<2.0$, compared to a $z < 0.3$ sample. Our observations indicate that deep radio imaging of  lensing clusters can be used to study star forming galaxies, with star formation rates as low as $\sim10$~M$_{\odot}$~yr$^{-1}$, at the peak of cosmic star formation history.

% can be used to study typical star forming galaxies, with star formation rates of \sim 10$~M$_{\odot}$~yr$^{-1}$ ( with around the peak of the cosmic star formation history.
 %radio source population which would otherwise only be possible with a Square Kilometre Array class of telescope. 
 %The presence of thermal Bremsstrahlung from the intracluster medium that covers most areas with higher magnification/amplification makes searches for lensed X-ray sources comparatively less efficient.

\vspace{4mm}
\end{abstract}



\section{Introduction}

Strong lensing clusters are excellent targets to study high-redshift galaxies due to their large magnification. This offers the advantage of studying these sources at improved spatial resolution and allows the detection of very faint sources due to the amplification of their integrated fluxes \citep[for a review see][]{2011A&ARv..19...47K}. 

Most blind searches for lensed galaxies behind galaxy clusters have traditionally been carried out at optical, near-infrared (near-IR), and submillimeter wavelengths, with the submillimeter observations mostly targeting rare but very luminous dusty galaxies with high star formation rates. Very few blind and deep studies have been carried out at radio or X-ray wavelengths. One likely reason for this is that the number of lensed AGN is much lower than that of typical lensed galaxies that are observed at optical and near-IR wavelengths. 

Non-thermal radio emission is also emitted by the more numerous star forming galaxies, with the radio luminosity correlating with the star formation rate \citep[SFR, e.g.,][]{1992ARA&A..30..575C,2003ApJ...586..794B,2009MNRAS.397.1101G}. Such studies have the advantage of being free from dust extinction. The cosmic SFR increases with redshift, peaking at z $\sim 2$ \citep[for a recent review see][]{2014ARA&A..52..415M}. However at $z\sim2$, only the ``tip of the iceberg'' can be currently observed, with extreme SFR of $\gtrsim 10^{2}$~M$_{\odot}$~yr$^{-1}$. A more typical star forming galaxy, with SFR $\sim 10$~M$_{\odot}$~yr$^{-1}$ at $z\sim 1.5$, has an integrated flux density of about one $\mu$Jy at 3~GHz, below the detection limit of current generation radio telescopes. However, with the power of lensing these galaxies should come within reach of deep  {\emph Jansky Very Large Array} (\jvla) observations. Combining deep \jvla\ imaging with lensing will therefore be the only way to study these more typical SF galaxies in the radio at the peak of cosmic SF,  before the advent of the Square Kilometre Array.

MACS~J0717.5+3745 was discovered by \cite{2003MNRAS.339..913E} as part of the MAssive Cluster Survey \citep[MACS;][]{2001ApJ...553..668E}. It is an extremely massive, hot  merging galaxy cluster located at $z=0.5458$, with a global temperature of $11.6 \pm 0.5$~keV \citep{2007ApJ...661L..33E}. A large-scale galaxy filament that is connected to the cluster, with a projected  length of $\sim 4.5$~Mpc, has also been reported \citep{2004ApJ...609L..49E,2012MNRAS.426.3369J}. The cluster is one of the most complex and dynamically disturbed clusters known, with the merger involving four separate substructures and shock heated $\sim20$~keV gas \citep{2008ApJ...684..160M,2009ApJ...693L..56M,2012A&A...544A..71L}.

The large total mass of $M_{\rm{vir}} = \left(3.5 \pm 0.6\right) \times 10^{15} M_\odot$ \citep{2014ApJ...795..163U} and relatively shallow mass profile of the cluster boosts the gravitational lens magnification and results in a total lensed area that is about 3.5 arcmin$^{2}$ for a galaxy located at $z \sim 8$. This area is higher than any other known massive cluster \citep{2009ApJ...707L.102Z}. For this reason, the cluster is also part of the Cluster Lensing And Supernova survey with Hubble \citep[CLASH,][]{2012ApJS..199...25P,2013ApJ...777...43M} and the HST Frontier Fields program\footnote{http://www.stsci.edu/hst/campaigns/frontier-fields/}.

Since MACS~J0717.5+3745 is the largest known cosmic lens, it is a prime target to search for radio and X-ray emission associated with lensed background galaxies. In this work, we present deep high-resolution \jvla\ observations which can be used for this purpose. In addition we carry out a search for lensed X-ray sources with newly acquired deep \chandra\ data. We adopt a $\Lambda$CDM cosmology with $H_{\rm 0} = 70$~km~s$^{-1}$~Mpc$^{-1}$, $\Omega_{\rm m} = 0.3$, and $\Omega_{\Lambda} = 0.7$. With this cosmology, 1\arcsec~corresponds to a physical scale of 6.387~kpc at $z=0.5458$.

\section{Observations \& data reduction}

\begin{figure*}[h]
\begin{center}
\includegraphics[angle =180, trim =0cm 0cm 0cm 0cm,width=0.95\textwidth]{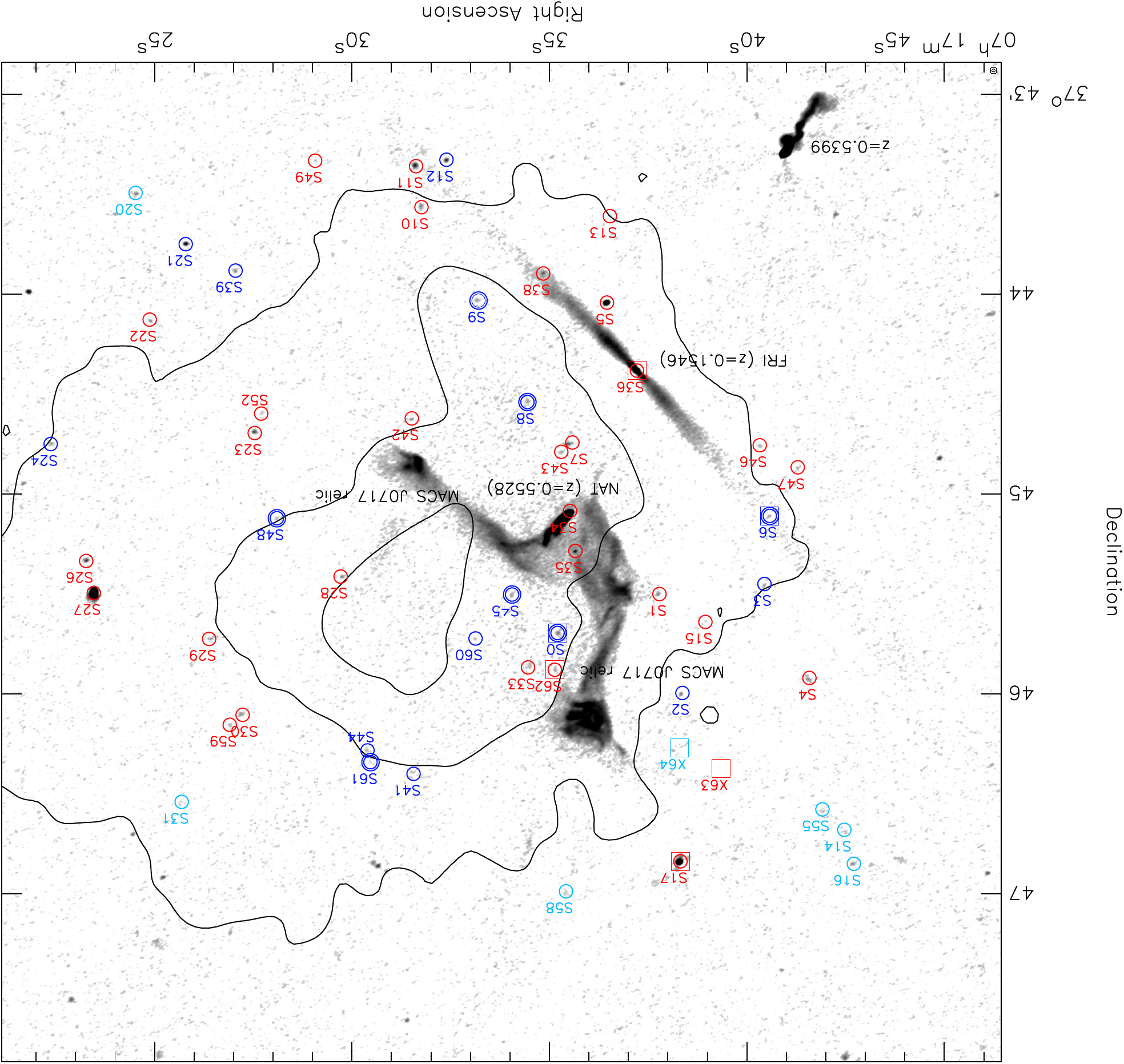}
\end{center}
\caption{S-band B+A-array image made with robust=0.75 weighting \citep{briggs_phd}. The image has a resolution of $1.0\arcsec\times 0.8\arcsec$ and a noise level of 1.8~$\mu$Jy~beam$^{-1}$. Compact radio and X-ray sources that fall within the HST coverage are indicated. Blue circles represent objects for which we could confirm that they are located behind the cluster. Blue double circled sources have amplification factors $>2$ and are individually discussed in Sect.~\ref{sec:lensedsources}. {Red circles represent cluster and foreground radio sources. Light blue circled sources have too uncertain redshifts to determine if they are cluster members, foreground objects, or background objects. Boxes indicate X-ray detected sources, with the color coding being identical to the radio sources.} Black contours show the X-ray emission from \chandra\, smoothed with a Gaussian with a FWHM of 10\arcsec. X-ray contours are drawn at levels of $[5,20,50] \times$ the background level (determined by measuring the background level around $2.5$~Mpc, i.e., $\simeq R_{200}$) from the cluster center. Both sky and instrumental background were included here.}
\label{fig:Sband}
\end{figure*}

\subsection{JVLA observations}
\jvla\ observations of MACS~J07175+3745 were obtained in the L-band (1--2~GHz) in the A-array configuration, in the S-band (2--4~GHz) in A- and B-array configurations, and in the C-band (4.5--6.5~GHz) in B-array configuration. All observations were done using single 6~hr runs, resulting in a typical on-source time of $\sim 5$~hr. The total recorded bandwidth was 1~GHz for the L-band, and 2~GHz for the S and C-bands. The primary calibrators were 3C138 and 3C147. The secondary calibrator (J0713+4349) was observed for a couple of minutes at 30--40~min intervals. All four circular polarization products were recorded. An overview of the observations is given in Table~\ref{tab:jvlaobs}. 

The data were reduced with {\tt CASA} version 4.2.1 \citep{2007ASPC..376..127M}. The data from the different observing runs were all reduced in the same way. For the two primary calibrators, we used the clean-component models provided by CASA. We also took the overall spectral index of the primary calibrator sources into account, scaling the flux-density for each channel.

As a first step, the data were Hanning smoothed and pre-determined elevation dependent gain tables and antenna position offsets  were applied.
This was followed by automatic flagging of radio frequency interference (RFI) using the {\it tfcrop} mode of the {\tt CASA} task {\tt flagdata}. We then determined an initial bandpass correction using 3C147. This bandpass was applied and additional RFI was identified and flagged with the {\tt AOFlagger} \citep{2010MNRAS.405..155O}. The reason for applying the bandpass is to avoid flagging good data due to the bandpass roll off at the edges of the spectral windows.

Next, we determined complex gain solutions on 3C147 for the central 10 channels of each spectral window to remove possible time variations of the gains during the calibrator observations. We pre-applied these solutions to find the delay terms ({\tt gaintype=`K'}) and bandpass calibration. Applying the bandpass and delay solutions, we re-determined the complex gain solutions for both primary calibrators using the full bandwidth. 
We then determined the global cross-hand delay solutions ({\tt gaintype=`KCROSS'}) from the polarized calibrator 3C138. For 3C138 we assumed a Rotation Measure (RM) of 0~rad~m$^{-1}$, and for the RL-phase difference we took $-15\degr$. 
All relevant solutions tables were applied on the fly to determine the complex gain solution for the secondary calibrator J0713+4349 and to establish its flux density scale accordingly. The absolute flux scale uncertainty due to bootstrapping from the calibrators is assumed to be a few percent \citep{2013ApJS..204...19P}. As a next step we used J0713+4349 to find the channel-dependent polarization leakage terms. 3C138 was used to perform the polarization angle calibration for each channel\footnote{The polarization observations will be discussed in a forthcoming paper.}. Finally, all solutions were applied to the target field. The corrected data were then averaged by a factor of 3 in time and a factor of 6 in frequency.

To refine the calibration for the target field, we performed three rounds of phase-only self-calibration and a final round of amplitude and phase self-calibration. For the imaging we employed w-projection \citep{2008ISTSP...2..647C,2005ASPC..347...86C} to take the non-coplanar nature of the array into account. Image sizes of up to $12288^2$ pixels were needed (A-array configuration) to deconvolve a few bright sources outside the main lobe of the primary beam. For each frequency band, the full bandwidth was used to make a single deep Stokes I continuum image. We used \cite{briggs_phd} weighting with a robust factor of 0. The spectral index was taken into account during the deconvolution of each observing run \citep[{\tt nterms=3};][]{2011A&A...532A..71R}. We manually flagged some additional data during the self-calibration process by visually inspecting the self-calibration solutions. Clean masks were employed during the deconvolution. The clean masks were made with the {\tt PyBDSM} source detection package \citep{2015ascl.soft02007M}. The S-band A-array and B-array configuration  data were combined after the self-calibration to make a single deep 2--4~GHz image. 
The final images were made with Briggs weighting and a robust factor of $0.75$, except for the C-band image for which we employed natural weighting. Images were corrected for the primary beam attenuation, with the frequency dependence of the beam taken into account. An overview of the resulting image properties, such as rms noise and resolution, is given in Table~\ref{tab:jvlaimages}.

\label{sec:obs}

\begin{table*}[]
\begin{center}
\caption{JVLA Observations}
\begin{tabular}{lllll}
&L-band A-array & S-band A-array & S-band B-array  & C-band B-array\\
\hline
\hline
Observation dates &    Mar 28, 2013 &  Feb 22, 2013   & Nov 5, 2013 &  Sep 30, 2013 \\
Frequencies coverage (GHz)           & 1--2 & 2--4 & 2--4 & 4.5--6.5\\
On source time   (hr)   & $\sim5$&$\sim5$ &$\sim5$ & $\sim5$\\
Correlations   & full stokes & full stokes & full stokes & full stokes\\
Channel width (MHz) &1  & 2 & 2&2\\
Integration time (s) & 1 & 1 & 3 & 3\\
LAS (arcsec)  & 36&18 & 58 & 29 \\
\hline
\hline
\end{tabular}
\label{tab:jvlaobs}
\end{center}
%$^{a}$ bal sflasdkasfdfad \\
\end{table*}

\begin{table*}[th!]
\begin{center}
\caption{Image properties }
\begin{tabular}{lllll}
&L-band A-array & S-band A+B-array &  C-band B-array\\
\hline
\hline
resolution (arcsec$\times$arcsec) 			& $1.5 \times 1.3$   & $1.04 \times 0.79$   &  $1.80 \times  1.38$   \\
noise ($\mu$Jy~beam$^{-1}$)      & 5.2   &  1.8 &  1.9      \\
\hline
\hline
\end{tabular}
\label{tab:jvlaimages}
\end{center}
\end{table*}

\subsection{Chandra Observations}

\begin{table*}
\begin{center}
\caption{Summary of the  \emph{Chandra} observations.}
\begin{tabular}{lccccc}
ObsID & Instrument & Mode & Start date & Exposure time (ksec) & Filtered exposure time (ksec)\\
\hline
  1655 & ACIS-I &    FAINT & 2001-01-29 & 19.87 & 17.06  \\
  4200 & ACIS-I & VFAINT & 2004-01-10 & 59.04 & 58.02  \\
16235 & ACIS-I &    FAINT & 2013-12-16 & 70.16 & 68.37  \\
16305 & ACIS-I & VFAINT & 2013-12-11 & 94.34 & 90.42  \\
\hline
\end{tabular}
\label{tab:chandraobs}
\end{center}
\end{table*}

\begin{figure*}[h]
\centering
\includegraphics[angle =180, trim =0cm 0cm 0cm 0cm,width=0.24\textwidth]{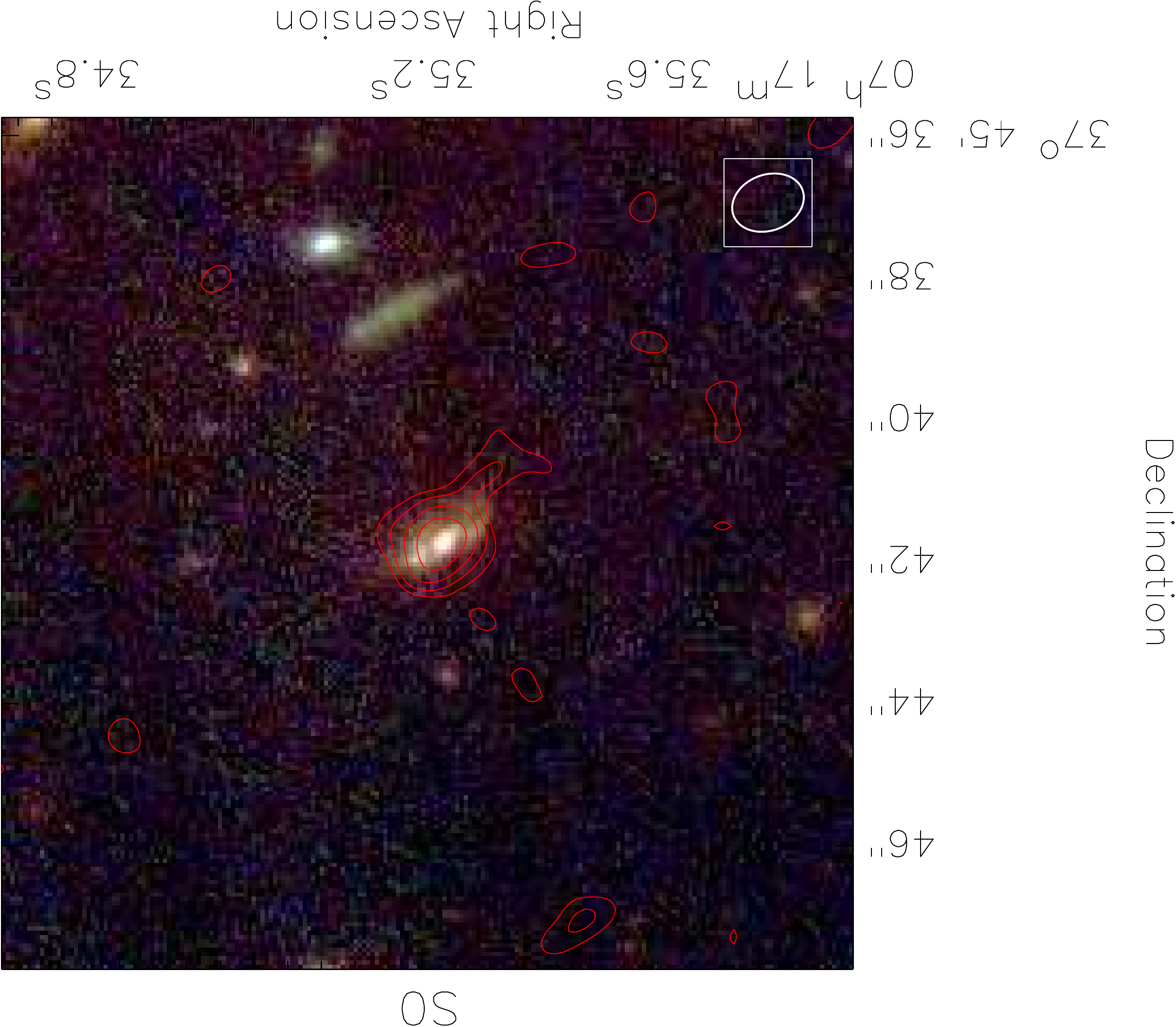}
\includegraphics[angle =180, trim =0cm 0cm 0cm 0cm,width=0.24\textwidth]{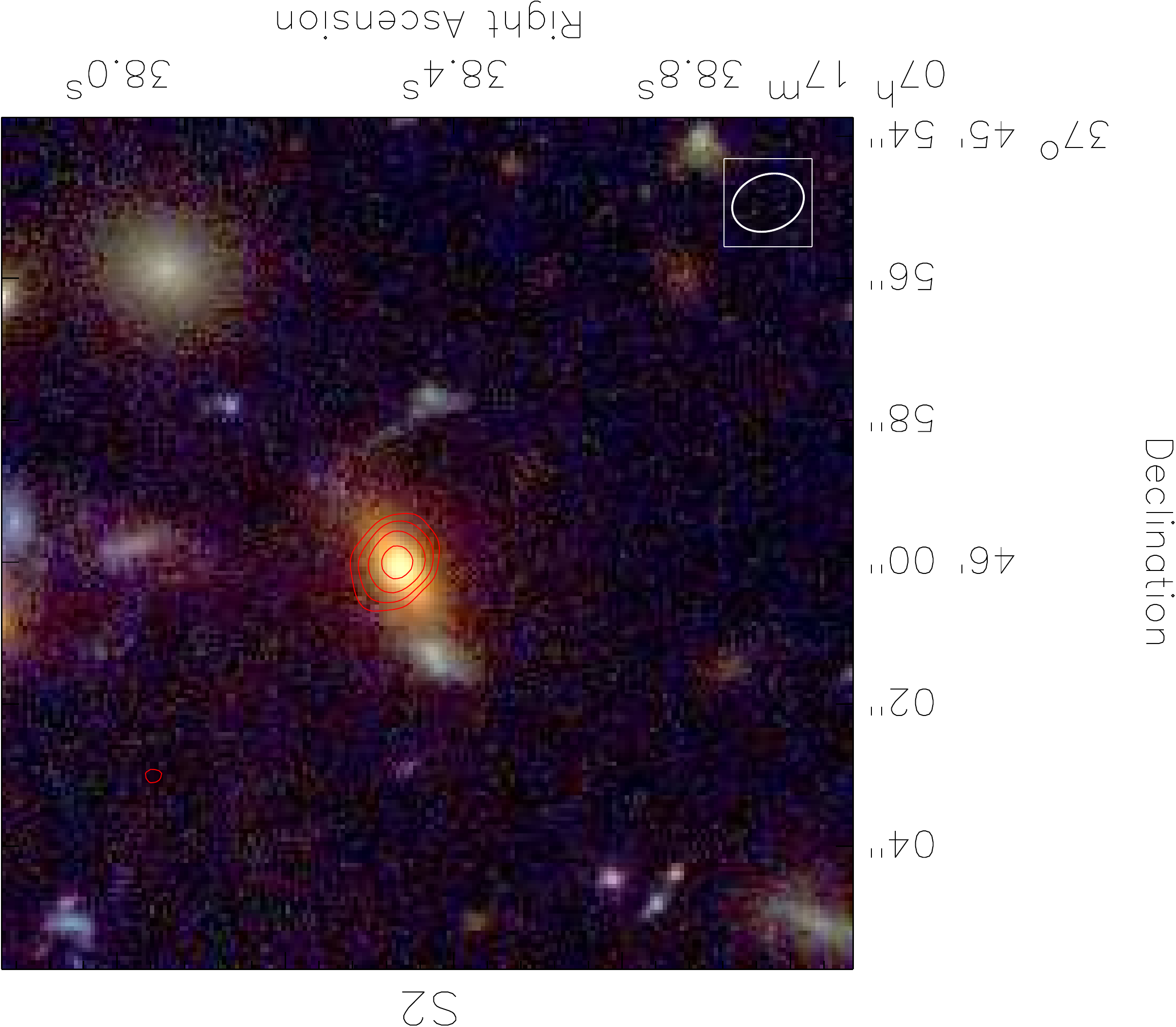}
\includegraphics[angle =180, trim =0cm 0cm 0cm 0cm,width=0.24\textwidth]{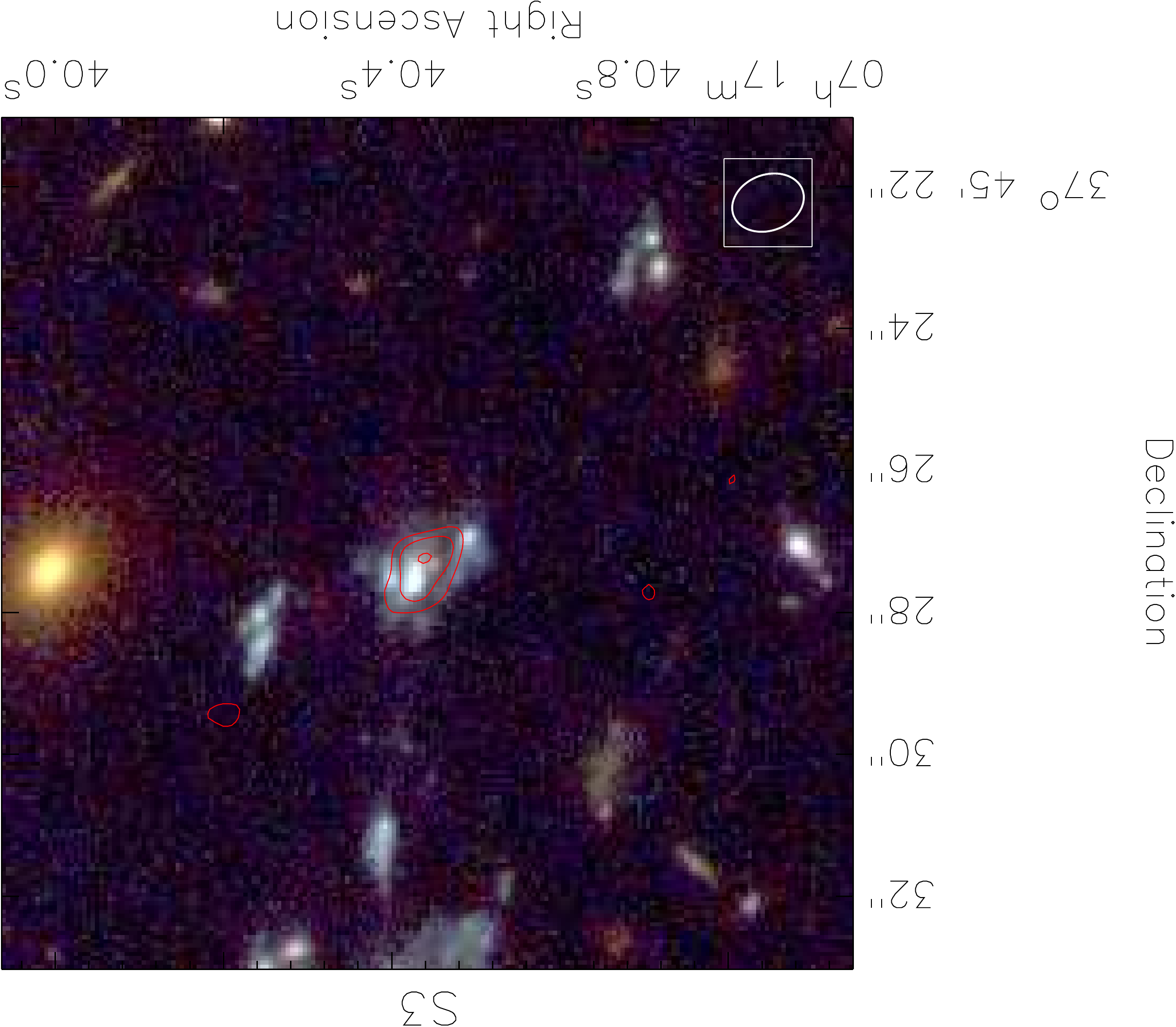}
\includegraphics[angle =180, trim =0cm 0cm 0cm 0cm,width=0.24\textwidth]{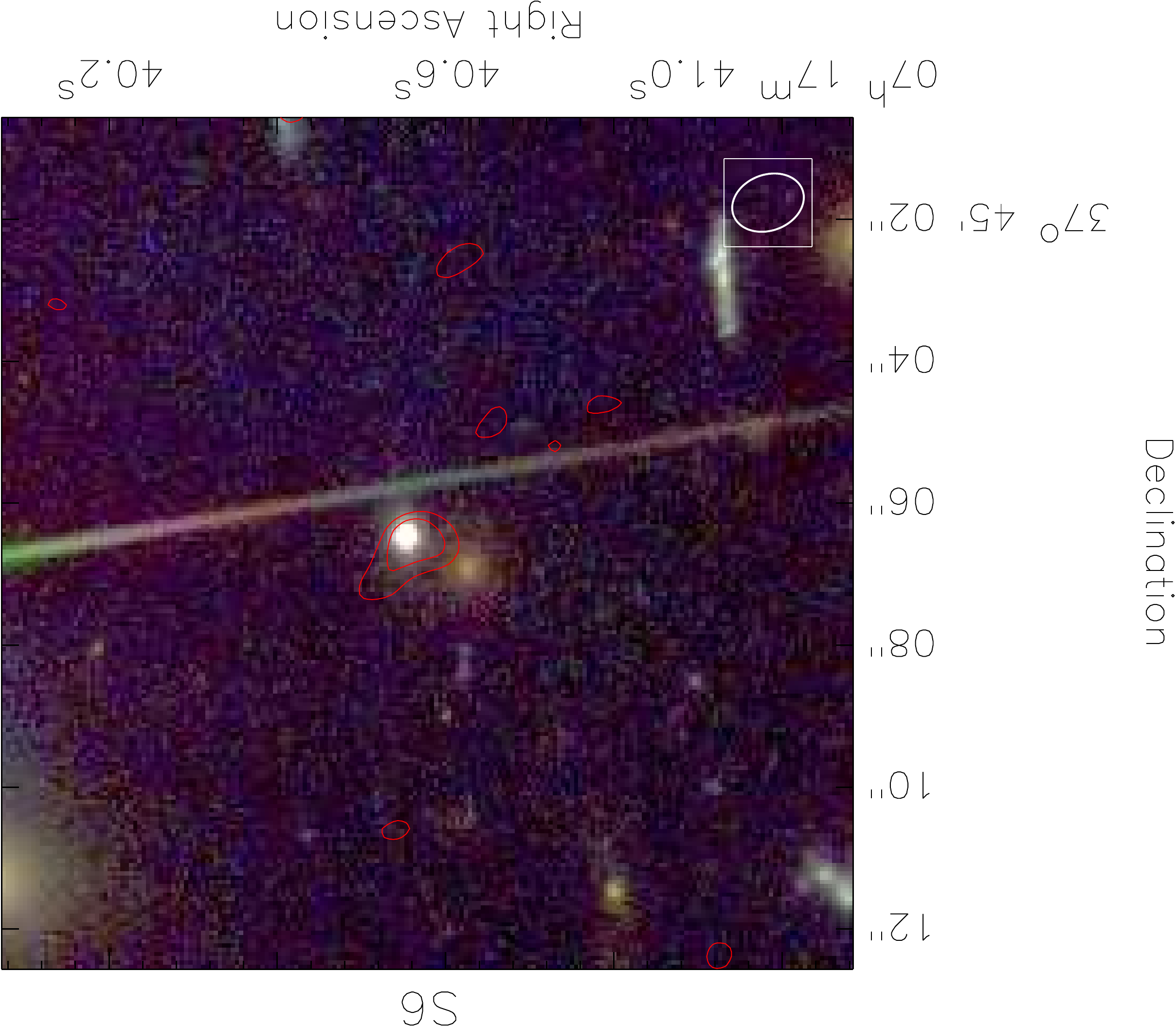}
\includegraphics[angle =180, trim =0cm 0cm 0cm 0cm,width=0.24\textwidth]{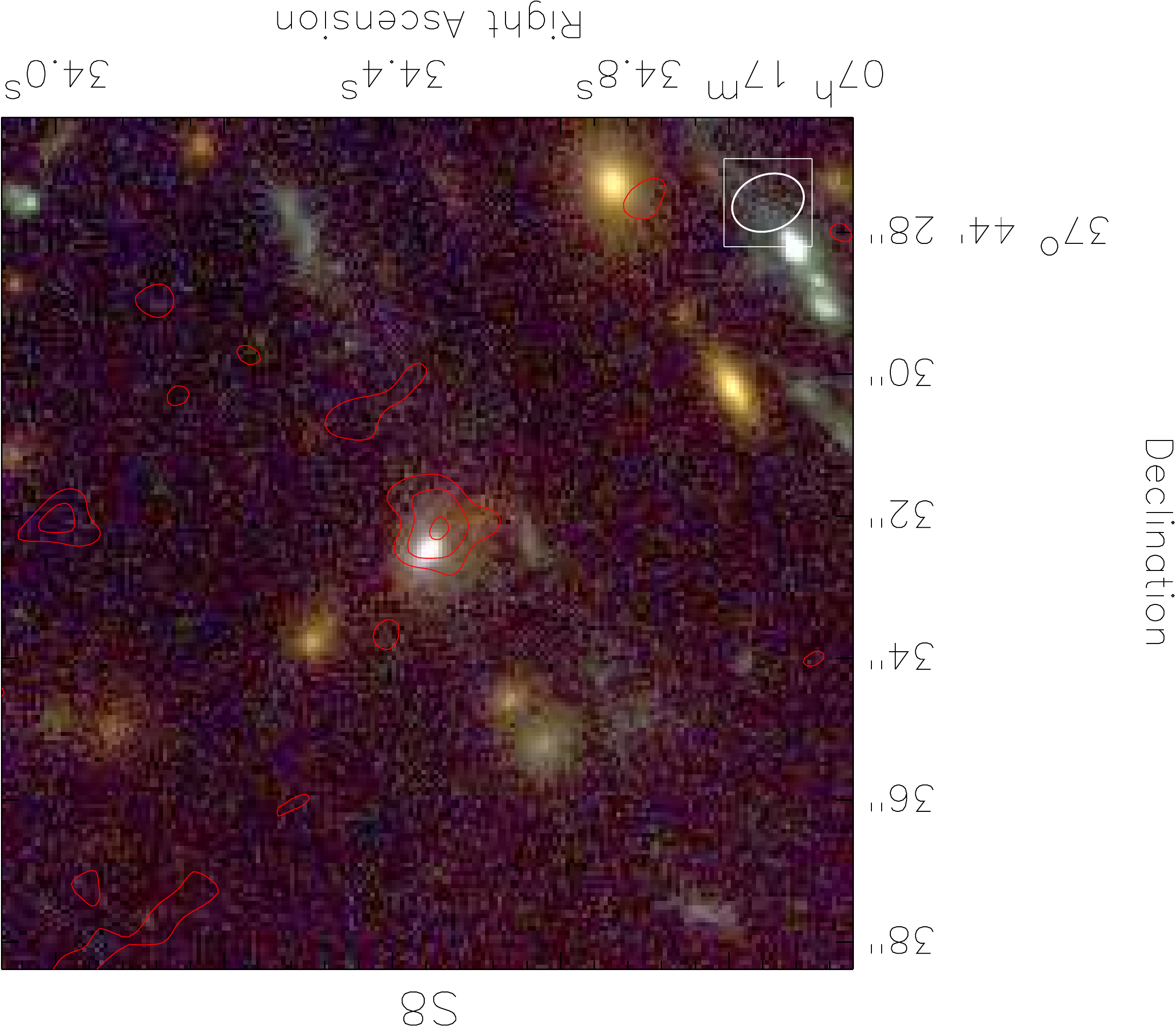}
\includegraphics[angle =180, trim =0cm 0cm 0cm 0cm,width=0.24\textwidth]{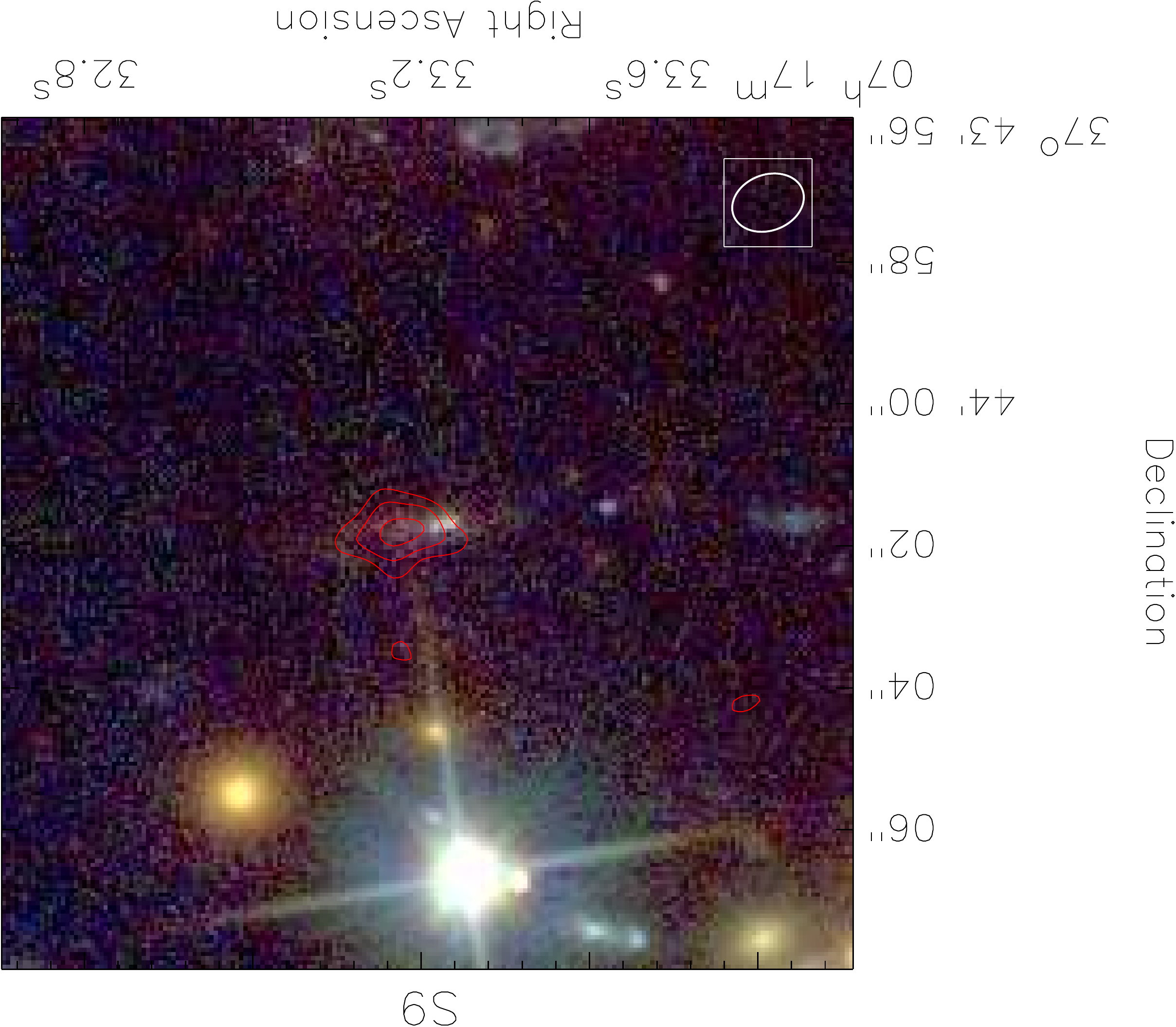} % F
\includegraphics[angle =180, trim =0cm 0cm 0cm 0cm,width=0.24\textwidth]{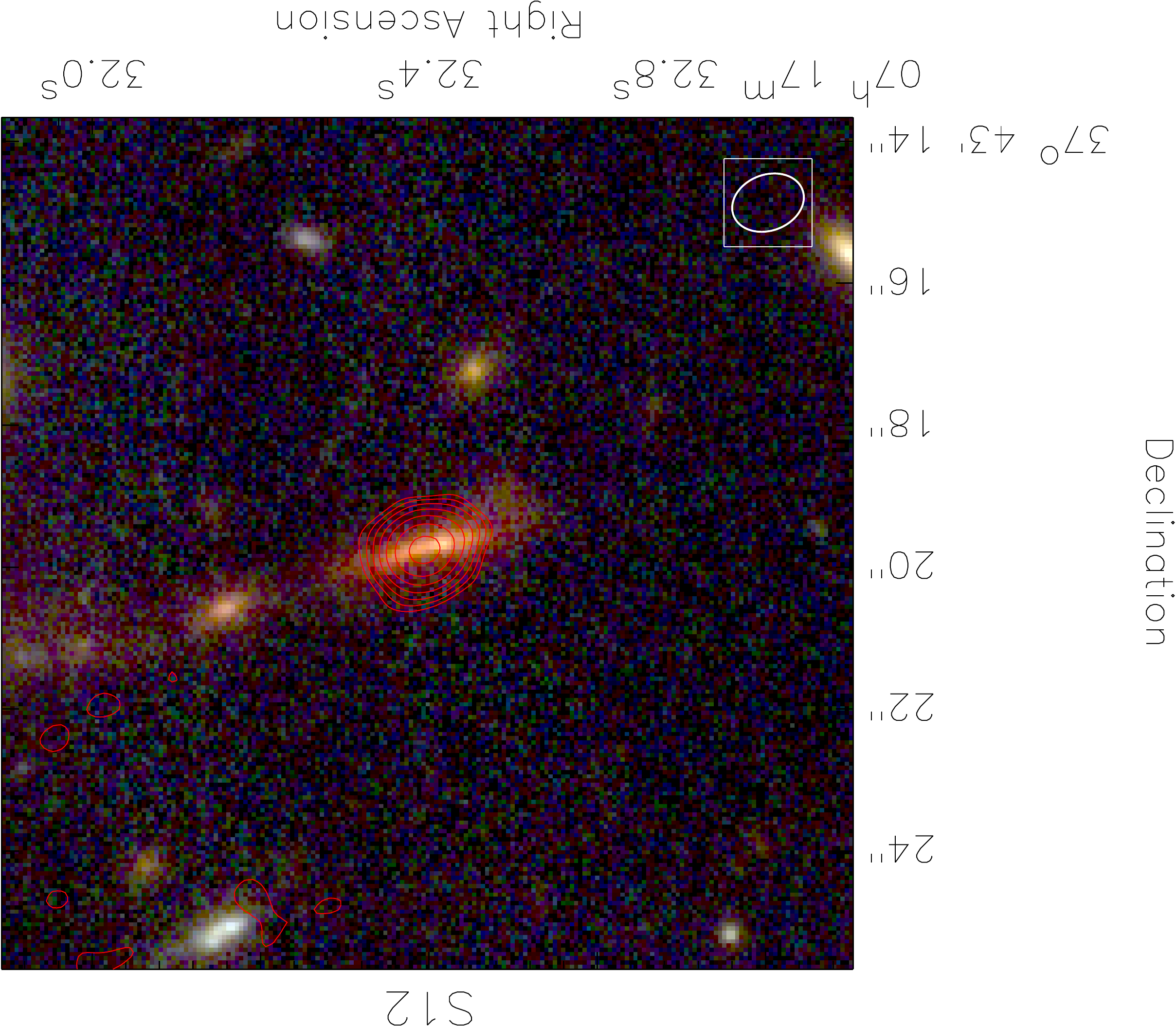}
\includegraphics[angle =180, trim =0cm 0cm 0cm 0cm,width=0.24\textwidth]{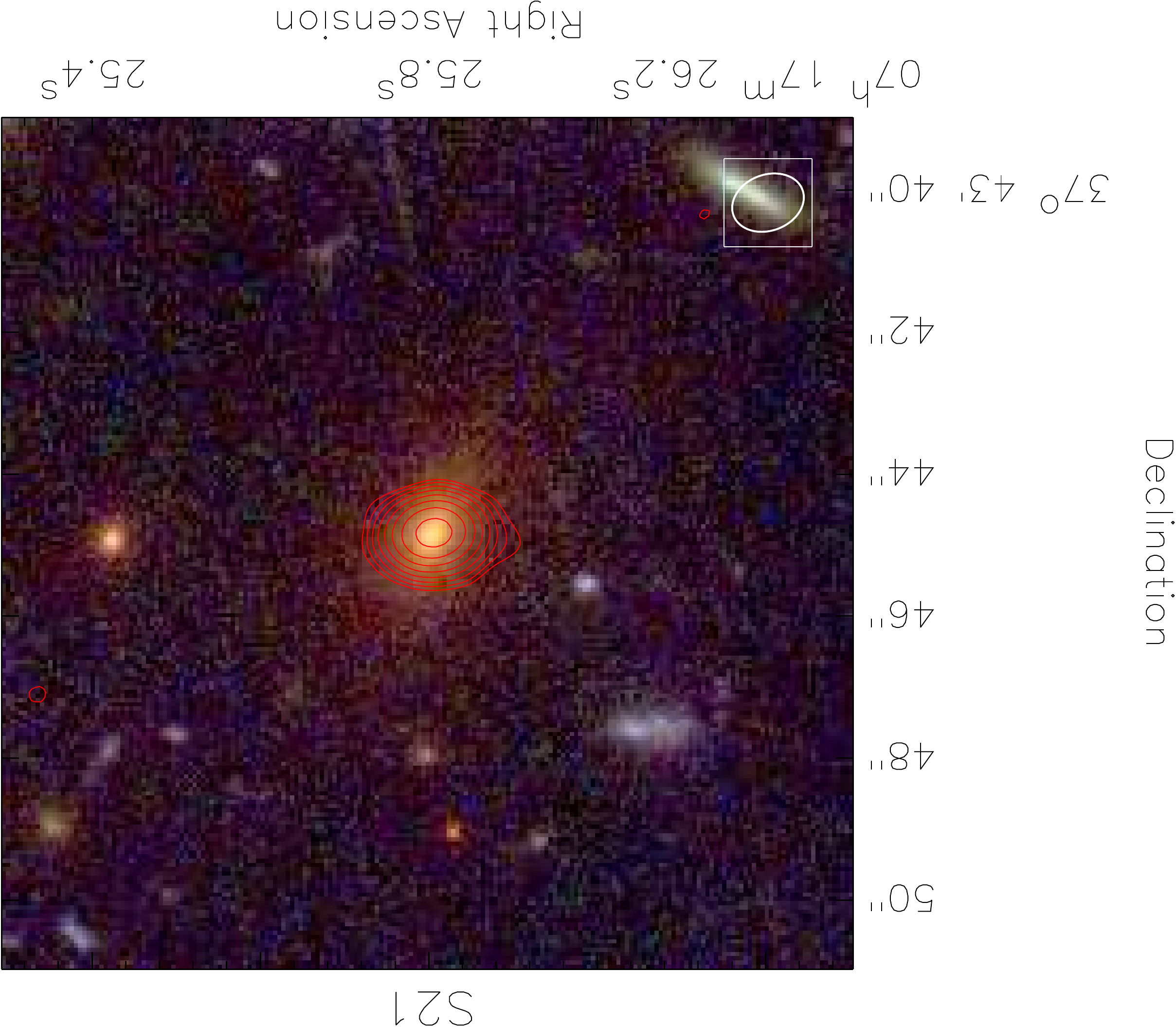} % H
\includegraphics[angle =180, trim =0cm 0cm 0cm 0cm,width=0.24\textwidth]{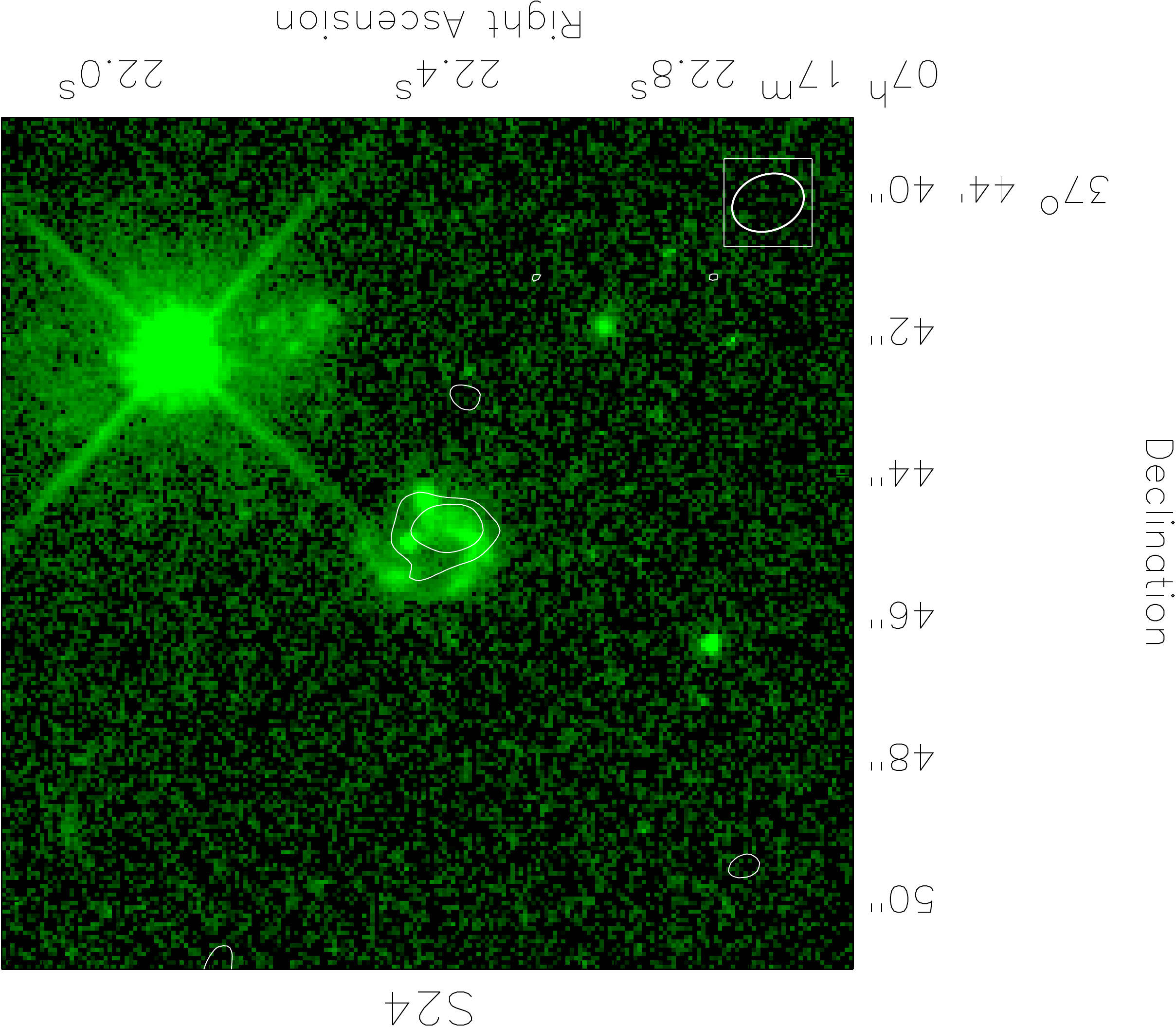} % I
\includegraphics[angle =180, trim =0cm 0cm 0cm 0cm,width=0.24\textwidth]{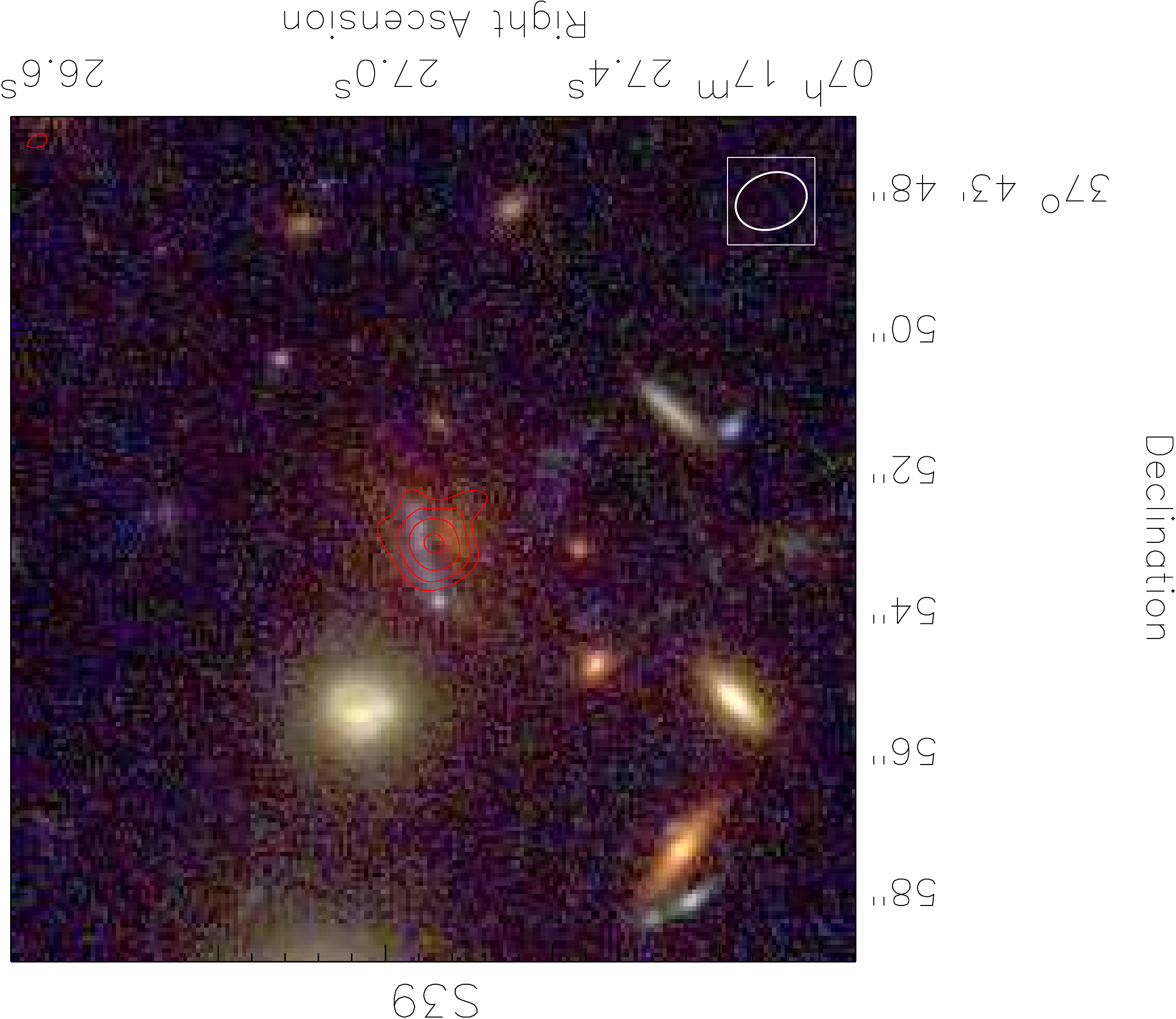} % J
\includegraphics[angle =180, trim =0cm 0cm 0cm 0cm,width=0.24\textwidth]{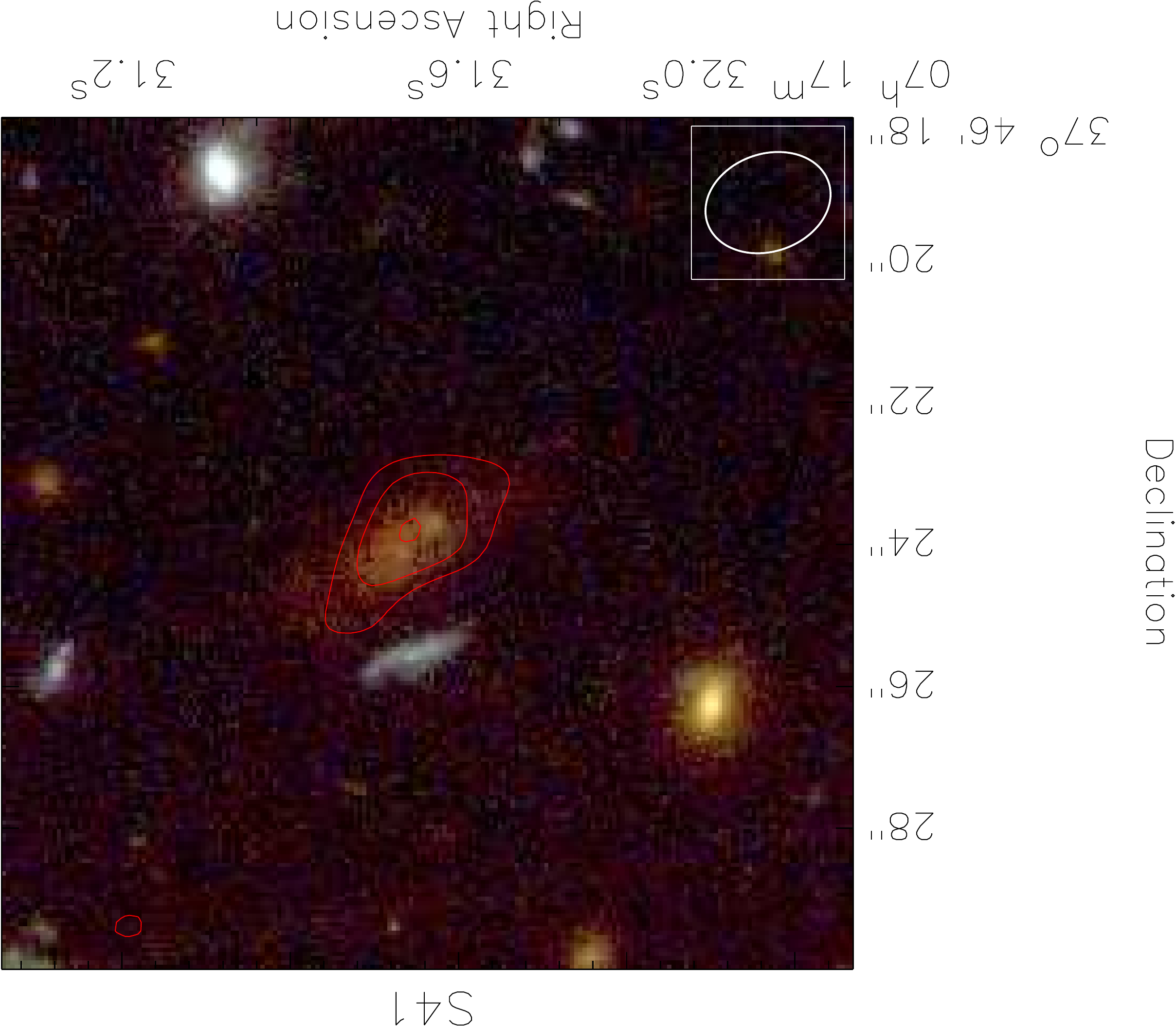} % K
\includegraphics[angle =180, trim =0cm 0cm 0cm 0cm,width=0.24\textwidth]{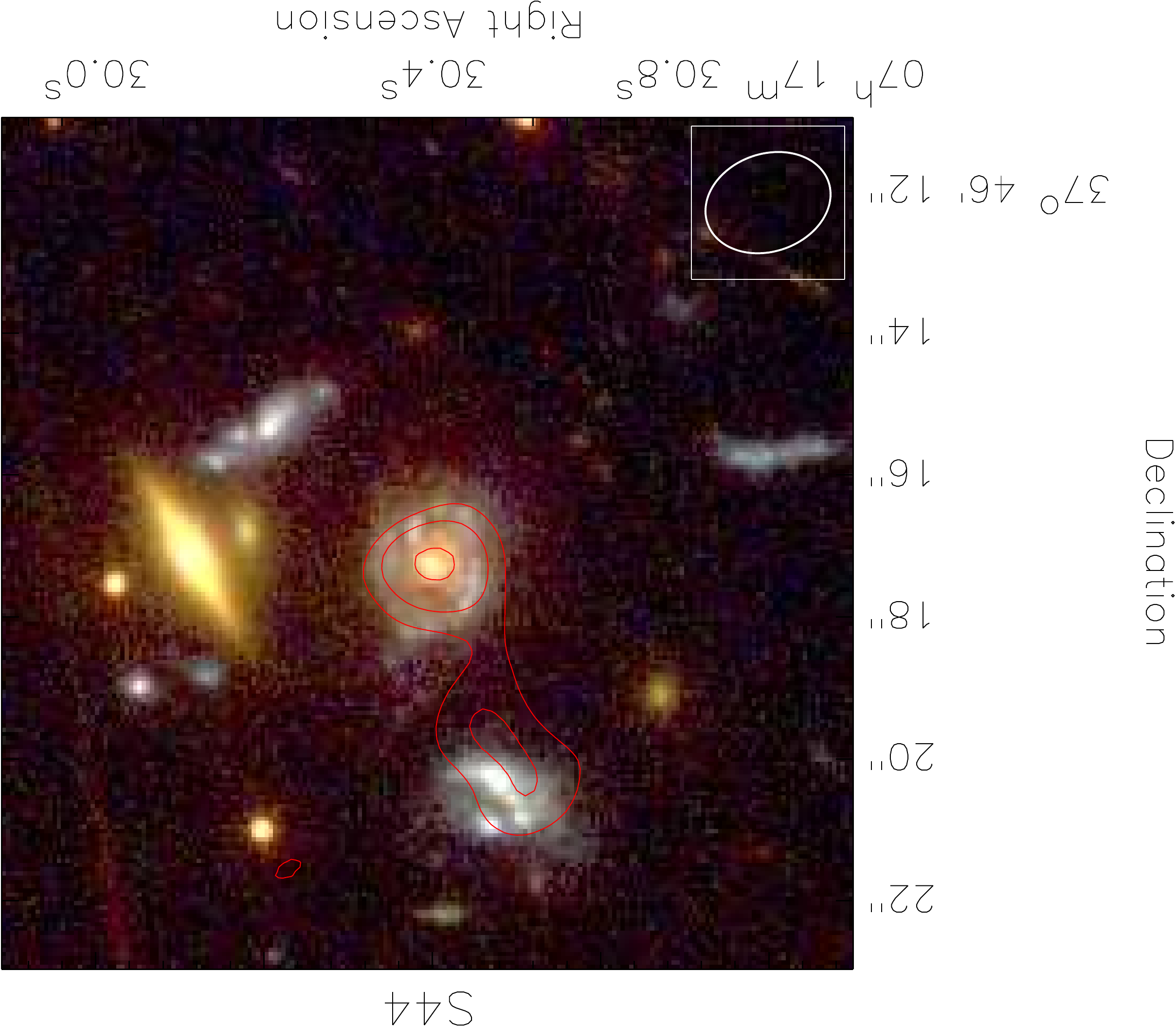} % L
\includegraphics[angle =180, trim =0cm 0cm 0cm 0cm,width=0.24\textwidth]{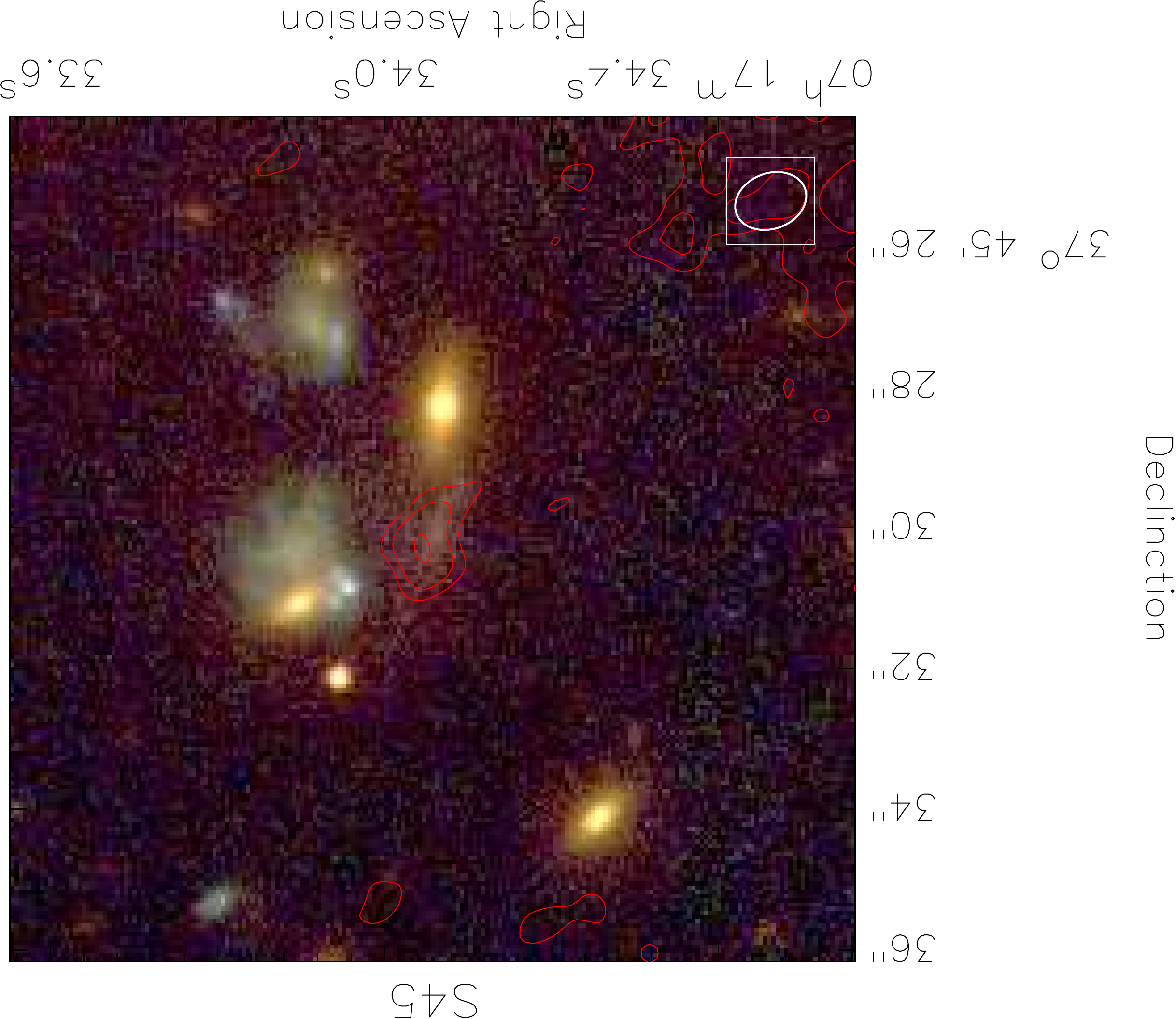} % M
\includegraphics[angle =180, trim =0cm 0cm 0cm 0cm,width=0.24\textwidth]{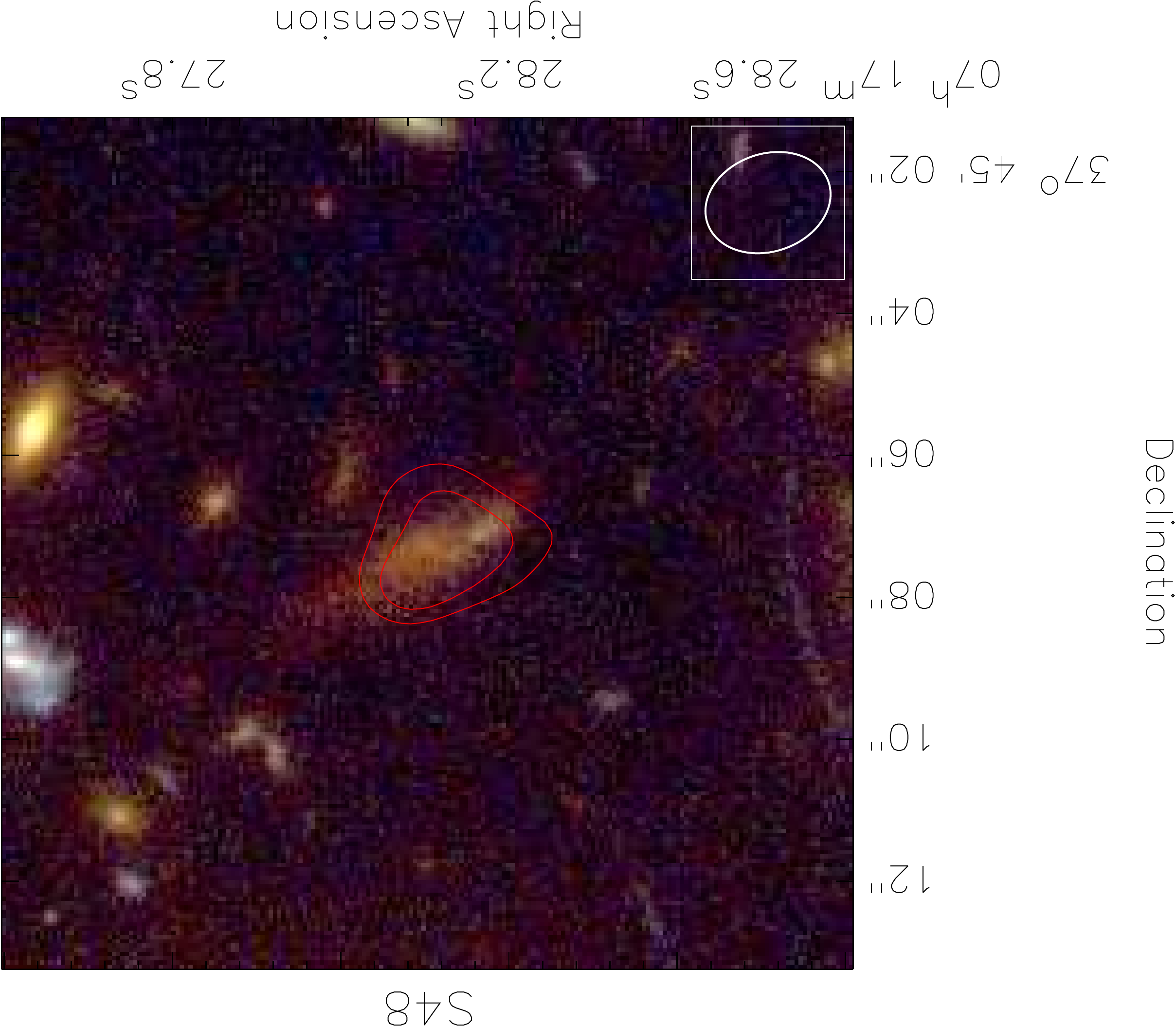} % N
\includegraphics[angle =180, trim =0cm 0cm 0cm 0cm,width=0.24\textwidth]{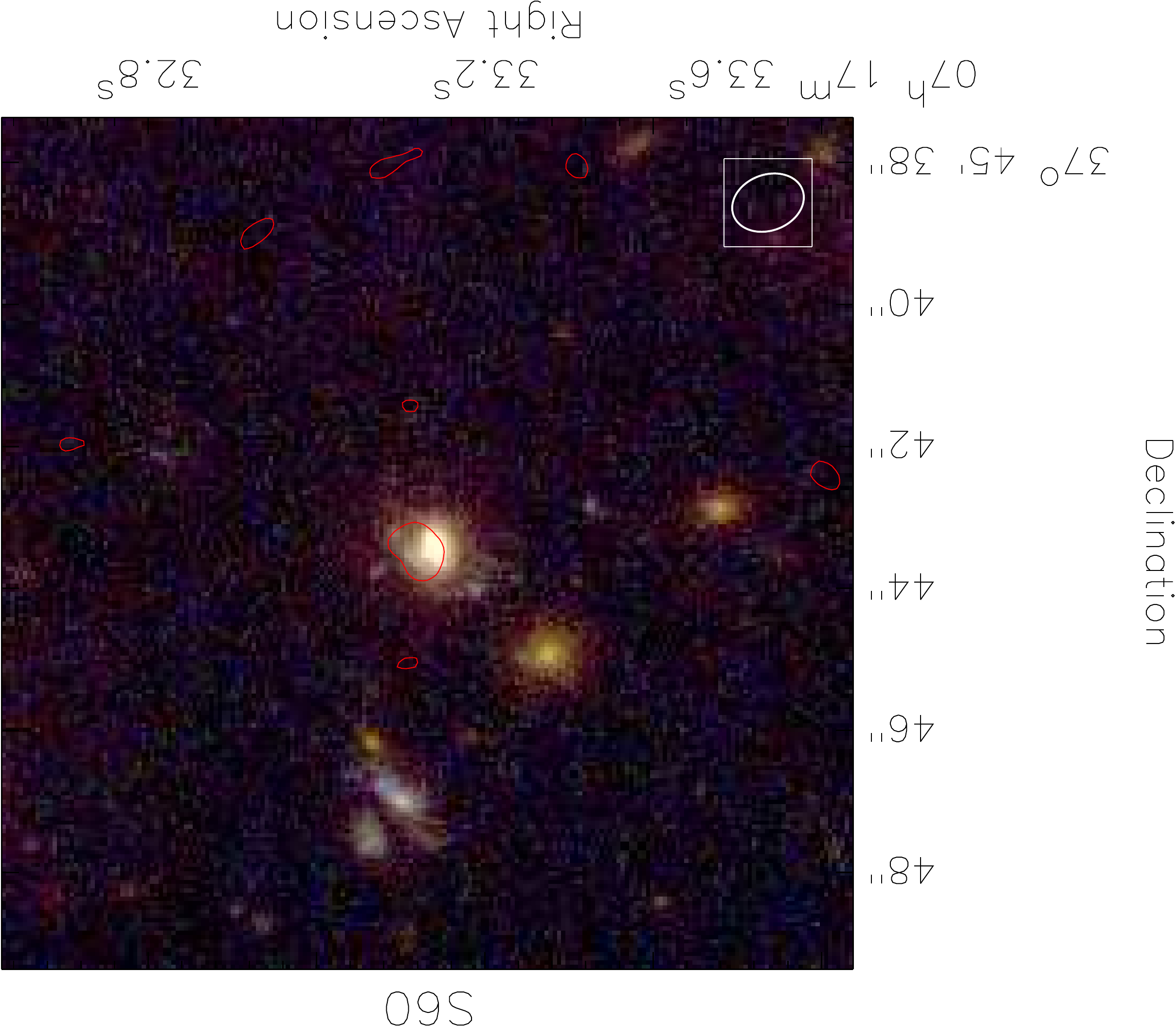} % O
\includegraphics[angle =180, trim =0cm 0cm 0cm 0cm,width=0.24\textwidth]{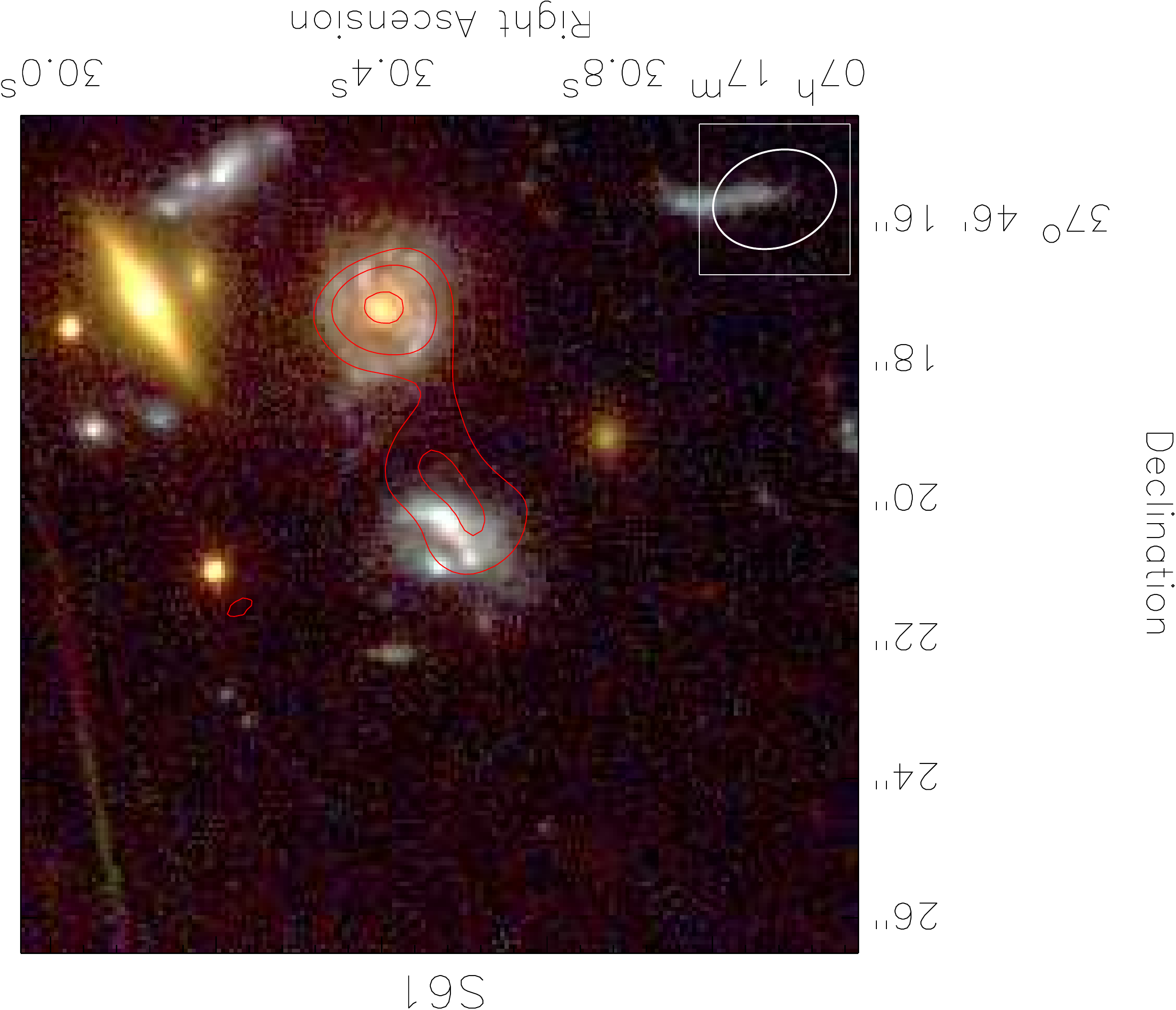} %P
\caption{HST F435, F606W, F814W postage stamp color images of the compact lensed radio sources in the MACS~J0717.3745 field. Similar postage stamp images, but for sources that are not lensed, are shown in Figure~\ref{fig:cutouts2}. The red radio contours are from the 2--4~GHz S-band image and drawn at levels $\sqrt{([1,2,4,\ldots])} \times 3\sigma_{\rm{rms}}$, with $\sigma_{\rm{rms}} = 1.8$~$\mu$Jy~beam$^{-1}$. The beam size is indicated in the bottom left corner. We draw white radio contours for some images to aid visibility (in case the area was not covered by all the HST filters). For sources without an S-band detection, we overlay contours from the C-band image (if detected there) or L-band image. We use contours from the combined L-, S-, and C-band image if the source is not detected in any of the three individual band images.  The values for $\sigma_{\rm{rms}}$ for the L and C-band images are listed in Table~\ref{tab:jvlaimages}. }
\label{fig:cutouts1}
\end{figure*}

MACS~J0717.5+3745 was observed with \chandra\ for a total of 243~ks between 2001 and 2013. A summary of the observations is presented in Table~\ref{tab:chandraobs}. The datasets were reduced with CIAO v4.7 and CALDB v4.6.5, following the same methodology that was described by \citet{2015ApJ...812..153O}. 

Point sources were detected with the CIAO script \emph{wavdetect} in the energy bands $0.5-2$ and $2-7$~keV, using wavelet scales of 1, 2, 4, 8, 16, and 32 pixels and ellipses with radii $5\sigma$ around the centers of the detected sources. {Due to the complicated morphology of MACS~J0717.5+3745, all sources found by  \emph{wavdetect} were visually inspected and some false detections associated with the extended ICM of the cluster were removed. }%All the point sources were also checked against the \jvla\ and \hst\ images. 
The local background around each point source was described using an elliptical annulus with an inner radius equal to the source radius and an outer radius approximately $\sim 3$~times larger than the source radius. To model the X-ray spectra of the point sources, the local background spectra were subtracted from the corresponding source spectra. The spectra were binned to a minimum of 1~count/bin, and modeled with \textsc{Xspec}~v12.8.2 using the extended C-statistics\footnote{http://heasarc.gsfc.nasa.gov/docs/xanadu/xspec/wstat.ps} \citep{Cash1979,Wachter1979}. All the source spectra were modeled as power-laws and included galactic absorption. The hydrogen column density in the direction of MACS~J0717.5+3745 was fixed to $8.4\times 10^{20}$~cm$^{-2}$, which is the sum of the weighted average atomic hydrogen column density from the Leiden-Argentine-Bonn \citep[LAB;][]{Kalberla2005} Survey and the molecular hydrogen column density determined by \citet{Willingale2013} from \emph{Swift} data. 

X-ray fluxes and luminosities were calculated in the energy band $2-10$~keV, with uncertainties quoted at the $90\%$ confidence level.

\section{Results}
\label{sec:results}

% overview of radio sources in the Field
The 2--4~GHz S-band image of the cluster region is shown in Figure~\ref{fig:Sband}. The most prominent source in the images is a large filamentary radio relic that is associated with the cluster MACS~J0717.5+3745. At the center of this relic, a narrow angle tail (NAT) galaxy is visible, which is associated with a cluster member at $z=0.5528$ \citep{2014ApJS..211...21E}. Another tailed radio source \citep[$z=0.5399$,][]{2014ApJS..211...21E} is visible at the far SE corner of the image. A bright linearly shaped FRI-type radio source \citep{1974MNRAS.167P..31F} is located to the SE. This source is associated with an elliptical foreground galaxy (\object{2MASX J07173724+3744224}) located at z=0.1546 \citep{2009A&A...503..707B}. The radio relics and tailed radio galaxies belonging to MACS~J0717.5+3745 will be discussed in a separate paper (van Weeren et al. in prep).

\subsection{Source detection}

We used  the {\tt PyBDSM}\footnote{http://dl.dropboxusercontent.com/u/1948170/html/index.html} source detection package to find and determine the integrated flux densities of the radio sources in the images.  {\tt PyBDSM} identifies ``islands'' of contiguous pixels above a detection threshold and fits each island with Gaussians. 
 
For detecting these islands, we took a threshold of $3\sigma_{\rm{rms}}$ and a pixel threshold of $4\sigma_{\rm{rms}}$, meaning that at least one pixel in each island needs to exceed the local background by $4\sigma_{\rm{rms}}$. We determined the local rms noise using a sliding box with a size of $300$ pixels to take the noise increase by the primary beam attenuation into account. We manually inspected the output source catalogs and removed any source associated with the radio relic and foreground FR-I source since these sources are larger than the $300$-pixel box size for the local noise determination. The source detection was run on all three frequency maps. The locations of the detected sources are indicated on Figure~\ref{fig:Sband}.

We then searched for optical counterparts to the radio sources within the area covered by the HST CLASH catalog and HST Frontier Fields observations. For the detected radio sources, we overlaid the radio contours on a HST Frontier Fields color (v1.0\footnote{https://archive.stsci.edu/pub/hlsp/frontier/macs0717/images/hst/}, F425W, F606W, and F814W band) image to verify that the correct counterparts were identified, see Figures~\ref{fig:cutouts1} (lensed sources) and~\ref{fig:cutouts2}. Counterparts were found for all radio sources. In a few cases, more than one optical counterpart was identified for the radio source in the CLASH catalog because of the complex morphology of the galaxy in the HST images. An overview of all the compact lensed radio sources that were found within the HST FoV is given in Table~\ref{tab:compactsources}. The properties of the sources that are not lensed, or for which we could not determine if they were located behind the cluster, are listed in Table~\ref{tab:compactsources2}.

To detect even fainter radio sources, we combined the individual L-, S-, and C-band images into one deep 1--6.5~GHz wideband continuum image, scaling with a spectral index\footnote{$F_{\nu} \propto \nu^{\alpha}$} ($\alpha$) of $-0.5$. This was done by convolving all maps to the resolution of the C-band image. With the help of this deep image, we identified about a dozen more sources below the  {\tt PyBDSM}  detection threshold in the individual maps, but with peak fluxes above the local $3\sigma_{\rm{rms}}$. We manually determined the flux densities of these sources in {\tt AIPS} with the task {\tt JMFIT}. In addition, we visually identified  five more sources in this deep image that were below the $3\sigma_{\rm{rms}}$ thresholds in the individual maps and thus not recognized there.
  These sources are also listed in Table~\ref{tab:compactsources} (and Table~\ref{tab:compactsources2}). We do not report integrated flux densities for these sources since they are not clearly detected in any of the individual maps. The CLASH photometric redshift and best fitting BPZ \citep{2000ApJ...536..571B} spectral template for each source are also listed. For more details on the photometric redshifts and spectral templates the reader is referred to \cite{2014MNRAS.441.2891M} and \cite{2014A&A...562A..86J}.
  For 23 sources, spectroscopic redshifts are available from \cite{2014ApJS..211...21E}. 
  
{For the sources that are located at redshifts larger than the cluster, we include the amplification factors in Table~\ref{tab:compactsources}, taking the average over all the  8 publicly available\footnote{https://archive.stsci.edu/prepds/frontier/lensmodels/} HST Frontier Fields lensing models for MACS~J0717.5+3745. The amplifications at a given redshift are calculated directly from the mass ($\kappa$) and shear ($\gamma$) maps. For more details on how these lensing models were derived we refer to reader to the references provided in Table~\ref{tab:compactsources}. The reported uncertainties  in the amplification factors for the individual models are smaller than the scatter between these models. Therefore, for the uncertainty in the amplification factor we take the standard deviation between these models, which should better reflect the actual uncertainties.}

In total we find 51 compact radio sources within the area covered by the HST imaging. In this sample, 16 sources are located behind the cluster, i.e., those where the 95\% confidence limit of the lower redshift bound places it beyond the cluster redshift of $z = 0.5458$. Of these sources, 7~have amplification factors larger than $2$. We plot the location of the lensed radio and X-ray sources on top of an amplification map for a $z=2$ source in Figure~\ref{fig:mag}. {We used the {\tt Zitrin-ltm-gauss\_v1} model amplification map as an example here.}
We discuss these lensed sources with amplification factors larger than $2$ in some more detail in Sect.~\ref{sec:lensedsources}. About a dozen radio sources are associated with cluster members (i.e., $ 0.5 < z_{\rm{phot}} < 0.6$).

We also search for the presence of compact X-ray sources within the HST FoV. In total we detect 7~X-ray sources. Five of these have radio counterparts. The X-ray sources are also included in Table~\ref{tab:compactsources}, with the measured X-ray fluxes. Two of these X-ray sources which have radio counterparts, are lensed by the cluster. {The other X-ray sources are foreground objects, cluster members, or have uncertain redshifts.}

\subsection{Lensed sources}
\label{sec:lensedsources}
In this section we discuss the radio and X-ray properties of the lensed sources with amplification factors $>2$.  These sources are listed on boldface in Table~\ref{tab:compactsources} and double circled in Figure~\ref{fig:Sband}. For the sources that are not obvious AGN, (i.e., those that do not have X-ray counterparts), we compute the star formation rate based on the measured radio luminosity. When converting from flux density to luminosity we used the amplification factors listed in Table~\ref{tab:compactsources}.
The radio luminosity of the galaxies (non-AGN) can be converted to mean star formation rate over the past $\sim 10^8$ yrs \citep[][]{2002MNRAS.330...17B}
 using
\begin{equation}
\frac{\rm{SFR}}{\rm{M}_\odot \: \rm{yr}^{-1}} \approx 4.5 \left(\frac{\rm{GHz}}{\nu} \right)^{\alpha} \frac{L_\nu} {10^{22} \rm{W} \: \rm{Hz}^{-1}}  ,
\label{eq:SFR}
\end{equation}
where $L_\nu$ represents the k-corrected rest-frame radio luminosity (which is also corrected for the amplification). The underlying assumptions are that cosmic rays from type II supernovae trace star forming regions and that the number of type II supernovae is directly proportional to the SFR. An advantage of these radio-derived SFRs is that they are not significantly affected by dust extinction. For our sources we use the 3~GHz measurement (unless stated otherwise), scaling with a spectral index of $\alpha = -0.5$. The uncertainty in the radio derived SFR is about a factor of 2 \citep{2003ApJ...586..794B}.

%Based on the IR-radio correlation \cite{2003ApJ...586..794B} give for 
%
%\begin{equation}
%\frac{\rm{SFR}}{\rm{M}_\odot \: \rm{yr}^{-1}} \approx =   \frac{ 5.52 \times L_{\rm{1.4~GHz}}} {10^{22} \rm{W} \: \rm{Hz}^{-1}}  . 
%\end{equation}
%
%which is valid for $L_{\rm{1.4~GHz}} > 6.4 \times 10^{21}$~W~Hz$^{-1}$ and valid for the radio luminosity of the lensed galaxies. 
%s8 8, s9 39, s45 13

We can also compute the specific SFRs (sSFR) by computing the stellar mass (M$_{\star}$) from the Spitzer 3.6 and 4.5~$\mu$m fluxes, {following the approach by \cite{2014MNRAS.442..196R}}. We corrected these fluxes for the amplification and computed the K-correction using the BPZ spectral templates. 
The Spitzer fluxes are taken from SEIP Source List (Enhanced Imaging Products from the Spitzer Heritage Archive,  {we took the 3.8\arcsec~diameter aperture flux densities}). To compute the stellar mass, we use the relation from \cite{2012AJ....143..139E}

\begin{equation}
\rm{M}_{*} \left[\rm{M}_\odot \right] = 10^{5.65} S^{2.85}_{3.6\mu\rm{m}}   S^{-1.85}_{4.5\mu\rm{m}}  \left( D_{\rm{L}}/0.05 \right)^2 \mbox{ ,}
\label{eq:stellarmass}
\end{equation}
where $S_\lambda$ is in units of Jy, and $D_{\rm{L}}$ the luminosity distance in Mpc. {This relation was derived for the Large Magellanic Cloud and assumes a \cite{1955ApJ...121..161S} initial mass function (IMF). It may break down for more strongly star-forming systems and might also vary with metallicity. Therefore Equation~\ref{eq:stellarmass} should be taken as an approximation to the stellar mass.}

\subsubsection{Comments on individual sources}

% S0

The source S0 has a spectroscopically measured redshift of $1.6852\pm0.001$ \citep{2014ApJS..211...21E} and an amplification factor of  $3.6\pm1.0$ The source is associated with a disk galaxy that has a central bright core in the HST image. The source is also detected with \chandra\ ($171 \pm 25$ net counts) with an unabsorbed  $2-10$~keV flux of $1.71_{-0.20}^{+0.16}\times 10^{-14}$~erg~s$^{-1}$~cm$^{-2}$. The photon index of the power-law of was determined at $0.89_{-0.42}^{+0.33}$. Together with the amplification factor {and associated uncertainty} this translates into a rest-frame luminosity of ${0.29_{-0.07}^{+0.11}\times 10^{44}}$ erg~s$^{-1}$, typical of an AGN. The optical spectrum of the galaxy contains emission lines and the best-fitting BPZ template is that of an Sbc type spiral galaxy.

%S6
Source S6 seems to be associated with a compact  star-like object. %, although another galaxy is located just 1.0\arcsec\ to the north of it. 
We also find an X-ray counterpart to the source ($150 \pm 12$ net counts).  The object has a $z_{\rm{phot}} = 1.89_{-0.09}^{+0.08}$ and an amplification factor of $2.3\pm0.7$. With this amplification factor and redshift, the unabsorbed $2-10$~keV flux of $8.48_{-1.08}^{+1.24}\times 10^{-15}$~erg~s$^{-1}$~cm$^{-2}$ translates to a rest-frame luminosity of
%$1.00_{-0.15}^{+0.13}\times 10^{44}$~erg~s$^{-1}$. 
${0.88_{-0.25}^{+0.41}\times 10^{44}}$~erg~s$^{-1}$.  
The photon index of the power-law of was determined at $2.05_{-0.23}^{+0.24}$. The best-fitting BPZ template was that of a spiral galaxy, {but the $\chi^2$ value of 29.3 indicated a very poor fit. Given the star-like nature of the object and poor fit,} this source could be a quasar\footnote{Note that the BPZ fitting did not include AGN or quasar galaxy templates}.

%S8
S8 is located at $z_{\rm{phot}} =  1.15_{-0.03}^{+0.04}$ and has a high amplification factor of about $6.4\pm1.8$. The best-fitting BPZ template is that of a starburst galaxy. Based on the S-band radio flux, we compute a SFR of ${15^{+12}_{-7}}$~M$_{\odot}$~yr$^{-1}$. With $\rm{M}_{*} = {2.4^{+1.4}_{-1.0}} \times 10^{10}$~M$_{\odot}$ we compute a sSFR of ${0.6}^{+0.4}_{-0.3}$~Gyr$^{-1}$. {The errors take into account the uncertainties in the radio and Spitzer flux density measurements, amplification factor (Table~\ref{tab:compactsources}), and redshift. For the radio spectral index we assumed an uncertainty of $\Delta\alpha=0.3$. Errors where estimated via a Monte Carlo approach, drawing $10^4$ realizations. 
Note that $\rm{M}_{*}$, SFR, and sSFR are computed under the assumption that Equations~\ref{eq:SFR} and \ref{eq:stellarmass} hold.}

%S9
S9 is associated with a faint red galaxy with $z_{\rm{phot}} = 1.69_{-0.06}^{+0.22}$ and is amplified with a factor of $3.4\pm1.0$. The best-fitting BPZ template corresponds to a Sbc/ES0 galaxy. The source has a relatively flat spectral index of $\alpha=-0.4 \pm 0.2$ between 1.5 and 5.5~GHz. Based on the integrated flux density, we compute a high SFR of ${49^{+46}_{-21}}$~M$_{\odot}$~yr$^{-1}$, scaling with $\alpha=-0.4$ . A blue galaxy is located about 1\arcsec~to the east of this galaxy at $z_{\rm{photo}} = 1.70_{-0.14}^{+0.12}$. % ID 6555
With $\rm{M}_{*} = {4.5^{+3.3}_{-1.9}} \times 10^{10}$~M$_{\odot}$ we compute a sSFR of ${1.1^{+0.8}_{-0.5}}$~Gyr$^{-1}$.

%S45
S45 is associated with a faint red galaxy with  $z_{\rm{phot}} = 1.41_{-0.19}^{+0.06}$. It has a high amplification factor of $8.7 \pm 4.1$ with a best-fitting spiral galaxy spectral template. We compute a SFR of ${17^{+29}_{-11}}$~M$_{\odot}$~yr$^{-1}$.

In addition to the above sources that were detected in the individual L-, S-, or C-band images, we also found two sources in the deep broad-band stacked radio image with amplifications $>2$. %S48
S48 is associated with a red galaxy at  $ z_{\rm{photo}} = 0.91_{-0.06}^{+0.07}$, with an amplification factor that is slightly less than 3. It is best fit by a spiral BPZ template.
% S61
S61 is associated with a blue irregular galaxy. Its amplification factor is $3.4\pm2.5$ and the best-fitting spectral template is that of a starburst galaxy. It is listed as two separate objects in the CLASH photometric catalog, {with component 2 (Table~\ref{tab:compactsources}) being a bright ``knot'' to the north, embedded within the overall emission from the galaxy (component 1).}
Thus, the HST images suggest that both components  belong the same galaxy with a complex morphology. This is also consistent with the two photometric redshifts that indicate $z_{\rm{photo}} \approx 1.6$.

\begin{figure}[h]
\begin{center}
\includegraphics[angle =180, trim =0cm 0cm 0cm 0cm,width=0.48\textwidth]{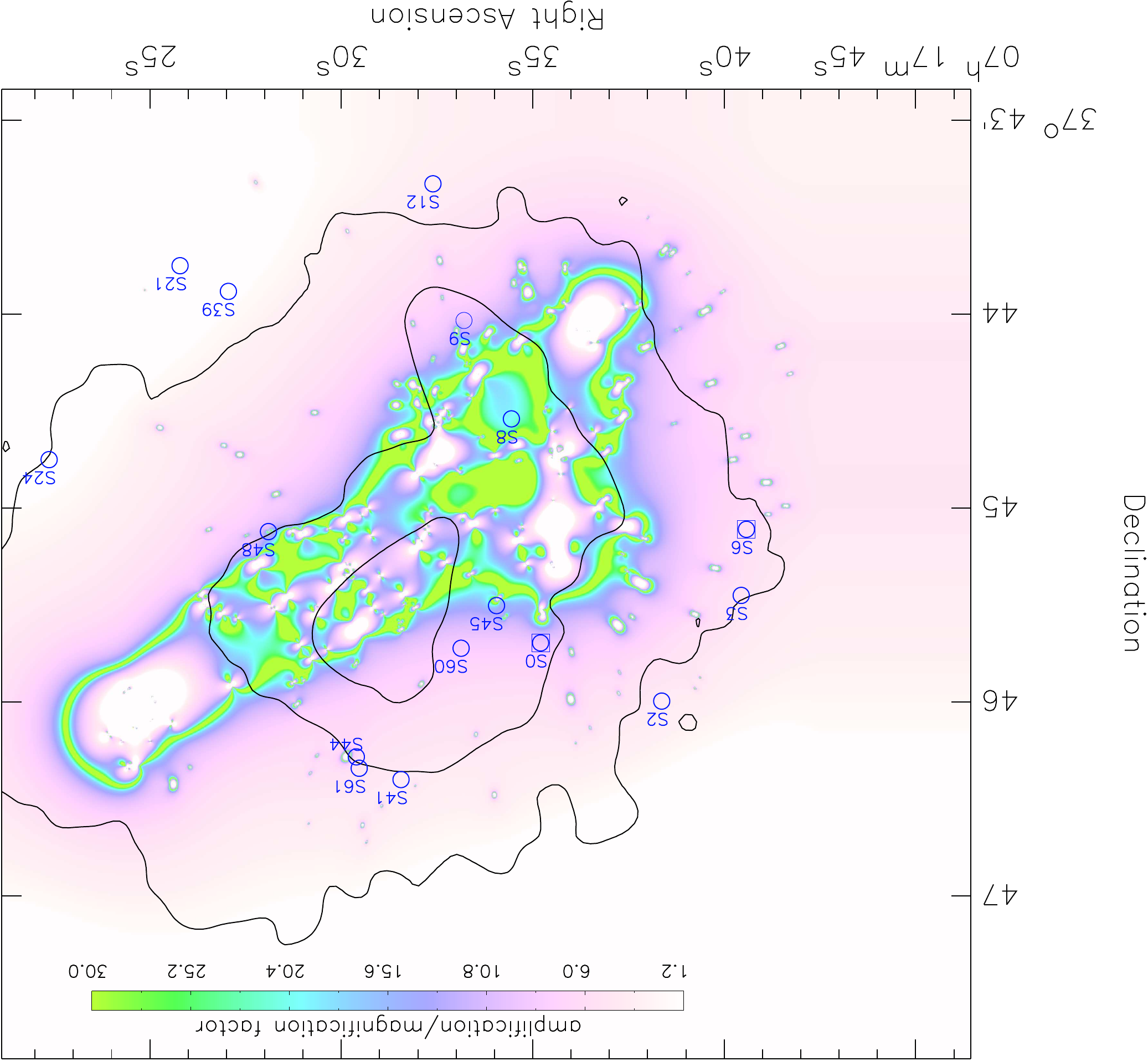}
\end{center}
\vspace{-3mm}
\caption{The amplification/magnification  map for a $z=2$ source from the {\tt Zitrin-ltm-gauss\_v1} lensing model. For details of the lens modeling see \cite{2013ApJ...762L..30Z} and \cite{2009MNRAS.396.1985Z}. {The location of the background compact sources are indicated as in Figure~\ref{fig:Sband}.} Black contours show the X-ray emission from \chandra, smoothed with a Gaussian with a FWHM of 10\arcsec. X-ray contours are drawn at levels of $[5,20,50] \times$ the X-ray background level as in Figure~\ref{fig:Sband}. \vspace{3mm}}
\label{fig:mag}
\end{figure}

It should be acknowledged that some of the sources we identify here, particularly those that lay behind the cluster according to their photometric redshifts, may be multiple images of multiply-lensed background sources. A brief search according to the predictions of the model seen in Figure~\ref{fig:mag} did not assign sources listed here to the same multiply imaged sources, nor were other counter images located. We also cross-checked the position of our lensed sources against the list of multiply lensed sources by \cite{2012A&A...544A..71L} but none of our sources appear in this list.  This means that  -- adopting the lens model in hand -- either the photometric redshift for these images significantly deviate than those listed in Table~\ref{tab:compactsources}, or that the predicted counter images's flux is below the detection limit given a possibly smaller magnification, for example. A more dedicated examination and search for multiple images among our sample will be performed elsewhere.

\begin{table*}[h!]
\begin{center}
\tiny
\caption{Properties of cluster background sources}
\begin{tabular}{lllllllllll}
name/ID & RA & DEC &  $S_{1.5}$  & $S_{3.0}$ & $S_{5.5}$  & $z_{\rm{phot}}$  & $t_{b}$ & amplification & Xray flux & $z_{\rm{spec}}$ \\
                 & degr (J2000) &  degr (J2000) & $\mu$Jy & $\mu$Jy & $\mu$Jy & &&&$10^{-15}$~erg~cm$^{-2}$~s$^{-1}$& \cite{2014ApJS..211...21E}  \\
\hline
\hline
\textbf{S0  3358} &  109.3967560  &  37.7615846 &  $48.0 \pm 11.1$ &  $25.9  \pm 3.3$  & $11.7\pm 3.9$ &  $1.59_{-0.06}^ {+0.09}$ & 5.90 & $3.6\pm1.0$ & $17.1^{+0.16}_{-0.20}$ & $1.6852\pm0.001$ (E)\\
S2  2759 &   109.4097441  &  37.7666944 & \ldots &  $21.2 \pm 3.2$& $16.4 \pm 3.8$ &  $0.74_{-0.03}^{+0.07}$ &    5.10 & $1.25 \pm 0.8$ & \ldots & \ldots \\
S3 3866  &   109.4184262  &   37.7576298 & $24.4 \pm  9.2^a$ & $19.4 \pm 3.1$ & $11.7 \pm 3.6$ & $1.52_{-0.07}^{+0.10}$ &  9.40 & $1.8\pm0.4$ & \ldots& \ldots\\
\textbf{S6 4614}  &   109.4189706  &  37.7517924  &  \ldots& $10.1 \pm 1.9^{\star b}$ &\ldots  & $1.89_{-0.09}^{+0.08}$ & 6.40& $2.3\pm0.7$ & $8.48^{+1.24}_{-1.08}$ & \ldots \\
\textbf{S8 5637}  & 109.3935136   & 37.7423563  & \ldots & $24.0 \pm 5.5^{a}$ & \ldots & $ 1.15_{-0.03}^{+0.04}$ & 7.70 & $6.4\pm1.7$ & \ldots & \ldots\\ 
\textbf{S9 6554}  &  109.3882566   &  37.7338089 & $43.7 \pm 7.9$ &  $20.0 \pm 3.0$& $24.9 \pm 3.5$ & $1.69_{-0.06}^{+0.22}$ & 5.60 &$3.4\pm1.0$ & \ldots & \ldots\\
S12 7885 &  109.3850184 &  37.7221324  & $96.6 \pm 10.5$  &  $54.5 \pm 3.3$ & $36.9 \pm 3.8$ & $2.32_{-0.05}^{+0.14} $ &5.10 & $1.8\pm0.5$ & \ldots & \ldots\\
S21 7107 & 109.3575070 &  37.7291308    &$86.4 \pm 8.8$  &$75.7 \pm 3.0$  & $87.1 \pm 4.6$ & $1.02_{-0.06}^{+0.15}$ & 4.70& $1.4\pm0.3$ & \ldots & \ldots\\
S24$_1$ 5304  & 109.3430929 &  37.7459519  & \ldots & $19.1 \pm 3.0$ & \ldots & $ 0.9_{-0.4}^{+0.3}$ & 7.90 &$1.4\pm0.2^{c,d}$& \ldots & \ldots\\
S24$_2$ 5307  & 109.3432347 &  37.7456667  & \ldots & $19.1 \pm 3.0$ &  \ldots& $0.7_{-0.3}^{+0.3}$  & 8.80 & $1.4\pm0.2^{c,d}$& \ldots & \ldots\\ 
S24$_3$ 5305  & 109.3431472 &  37.7458181  & \ldots & $19.1 \pm 3.0$ &\ldots  & $0.9_ {-0.2}^{+0.3}$ &8.80 & $1.4\pm0.2^{c}$ & \ldots & \ldots\\
S24$_4$ 5306  & 109.3434442 &  37.7458095 & \ldots &  $19.1 \pm 3.0$ &  \ldots& $0.8_{-0.6}^{+0.3}$ & 6.80 & $1.4\pm0.2^{c,d}$ & \ldots & \ldots\\
S39 6832  & 109.3626639 &  37.7313808    & $26.7 \pm  8.6^a$ & $22.8\pm3.2$ & \ldots & $1.1_{-0.1}^{+1.5}$ &6.50& $1.4\pm0.2$ & \ldots & \ldots\\  
S41 1901 & 109.3813749  & 37.7733429  & \ldots & \ldots & $22.6 \pm 3.0$  & $1.02_{-0.08}^{+0.07}$ & 6.00& $1.7\pm0.3$ & \ldots & \ldots\\ 
S44 2104 & 109.3766785  & 37.7714612  & \ldots & \ldots &\ldots  & $0.82_{-0.06}^{+0.06}$ & 6.50& $1.6\pm0.3$ & \ldots & \ldots\\
\textbf{S45 3811} & 109.3919138  & 37.7583808  &\ldots  & $22.7\pm 2.6$ & \ldots & $ 1.41_{-0.19}^{+0.07}$ & 6.30& $8.7\pm4.1$ & \ldots & \ldots\\ 
\textbf{S48 4567} & 109.3670705  & 37.7520604   &  &  &  &  $0.91_{-0.06}^{+0.07}$ & 5.40&$2.7\pm0.9$ & \ldots & \ldots\\
S60 3367  &109.3880803  & 37.7620638  & \ldots & $8.3 \pm 3.3^a$   & \ldots &  $0.80_{-0.05}^{+0.07}$ & 6.80&$1.7\pm0.2$ & \ldots & \ldots\\
\textbf{S61$_1$ 2016}  &109.3770241  & 37.7723431  & \ldots & \ldots & \ldots &   $1.73_{-0.14}^{+0.09}$ & 7.60&$3.4\pm2.5^{c}$ & \ldots & \ldots\\
\textbf{S61$_2$  2015} & 109.3769567 &   37.7724821   &  \ldots& \ldots & \ldots &   $1.6_{-0.2}^{+0.2}$  & 8.30&$3.4\pm2.5^{c,e}$ & \ldots& \ldots \\
\hline
\hline
\end{tabular}
\label{tab:compactsources}
\end{center}
$^{\star}$ flux density measurement could be affected by radio emission from other sources\\
$^{a}$ manually measured \\
$^{b}$ very faint source, only peak flux was measured \\
$^{c}$ the complex morphology of the galaxy likely caused it to be fragmented and listed as separate objects in the catalog\\
$^{d}$ we assume that these source components have the same redshift as  S24$_3$\\
$^{e}$ {we assume that this source component has the same redshift as  S61$_1$}\\
The source ID and positions are taken from the CLASH catalog. The photometric  redshifts ($z_{\rm{phot}}$) and best  fitting spectral template ($t_b$) are also taken from the CLASH catalog. The photometric redshifts are given at a 95\% confidence level. The spectral templates are described in \cite{2014MNRAS.441.2891M}. In total there are 11 possible spectral templates, five  for elliptical galaxies (1--5), two for spiral galaxies (6, 7) and four for starburst galaxies (8--11), along with emission lines and dust extinction. Non-integer values indicate interpolated templates between adjacent templates. Boldface source IDs indicate amplification factors $> 2$, these sources are discussed in Sect.~\ref{sec:lensedsources}. For the radio flux density errors we include a 2.5\% uncertainty from bootstrapping the flux density scale. {The listed amplification factors are the mean values from the models: {\tt CATS\_v1} \citep{2012MNRAS.426.3369J,2014MNRAS.444..268R}, {\tt Sharon\_v2} \citep{2014ApJ...797...48J}, {\tt Zitrin-ltm\_v1}, {\tt Zitrin-ltm-gauss\_v1} \citep[e.g.,][]{2013ApJ...762L..30Z,2009MNRAS.396.1985Z}, {\tt GLAFIC\_v3} \citep{2015ApJ...799...12I}, {\tt Williams\_v1} \citep[e.g.,][]{2006MNRAS.367.1209L}, {\tt Bradac\_v1} \citep{2005A&A...437...39B,2009ApJ...706.1201B}, {\tt Merten\_v1} \citep[e.g.,][]{2011MNRAS.417..333M}}. \\
For the spectroscopic redshifts ($z_{\rm{spec}}$) the spectral classification is as follows: A =	absorption-line spectrum; E =	emission-line spectrum.\\
\end{table*}

\subsection{Radio luminosity function}
It is expected that the number of star forming galaxies increases with redshift, peaking at $z \sim 2$ \citep[e.g.,][]{2014ARA&A..52..415M}.
We compute the radio luminosity function from the sources detected in our S-band image that have a 3~GHz flux density above 18~$\mu$Jy ($10 \times \sigma_{\rm{rms}}$). Above this flux density we should be reasonably complete, also for sources that are resolved. 

For the luminosity function, we determine the volume behind the cluster in which we could detect a hypothetical source above the flux limit for a given luminosity. We then divide the number of detected sources (for that luminosity range) by the obtained volume. To compute the volume, we take the varying magnification (as a function of position and redshift) into account using the {publicly available lensing models and provided {\tt Python} code}.

We excluded the regions covered by the radio relic. We scaled our 3~GHz S-band luminosities to 1.4~GHz taking $\alpha=-0.5$, to facilitate a comparison with the literature results from \cite{2005MNRAS.362....9B}. We limited ourselves to the range $0.6 < z < 2.0$, because we only detect a single source above $z > 2$. Restricting the redshift range also limits the effects of the redshift evolution of the radio luminosity function. 

\begin{figure}[h]
\begin{center}
\includegraphics[angle =180, trim =0cm 0cm 0cm 0cm,width=0.48\textwidth]{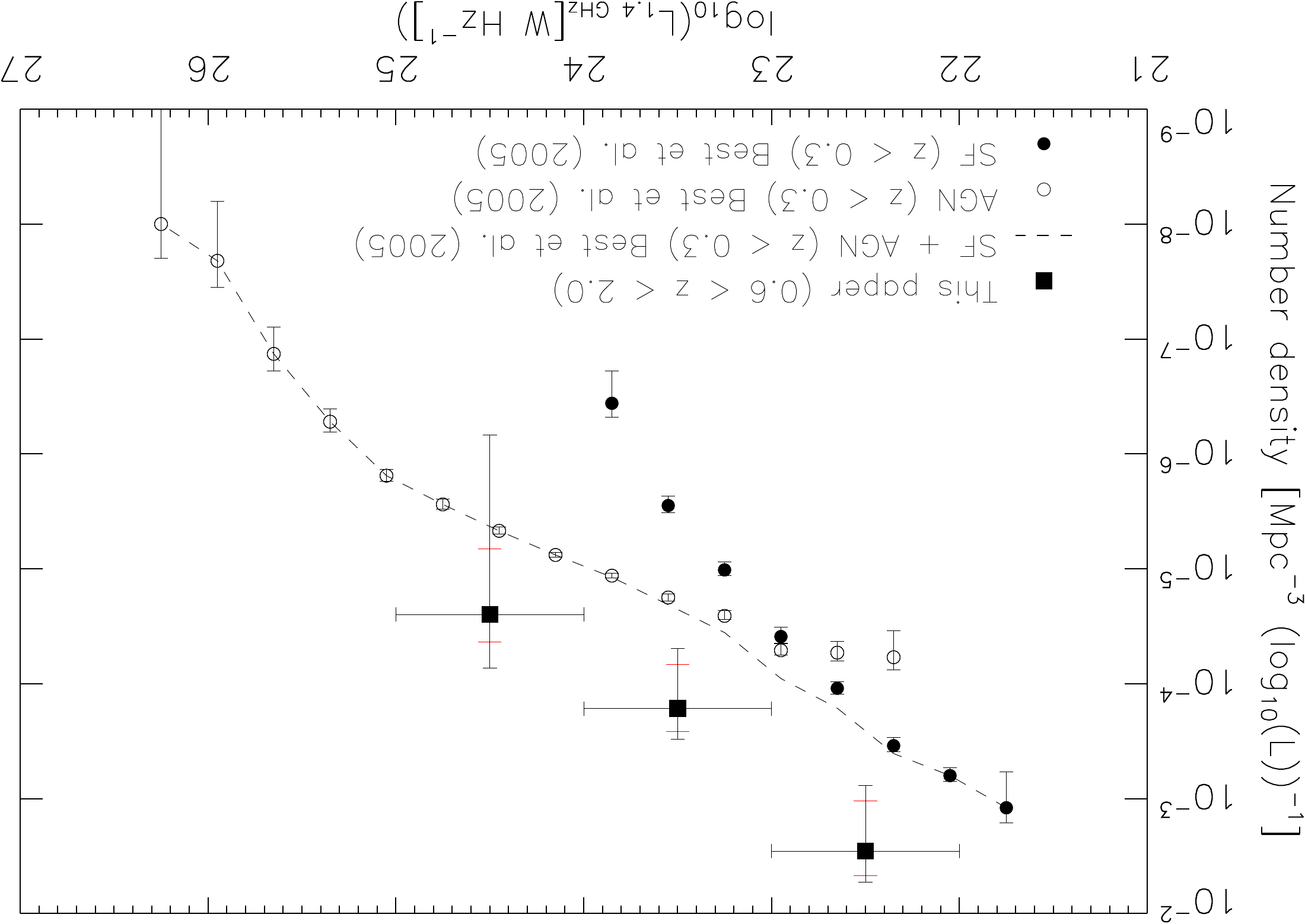}
\end{center}
\caption{The radio luminosity function (number density of sources) at $0.6<z<2.0$ derived from our detected S-band sources. {The luminosities were scaled to 1.4~GHz and averaged over all 8 lensing models. The uncertainties in black are shown at $1\sigma$ and include the Poisson errors and scatter between the 8 different lensing models. For reference, the red error bars only show the scatter (standard deviation) in the luminosity function from the 8 different lensing models.}
%. A $1\sigma$ upper limit would correspond to a mean occurrence rate of 1.15 sources and a 90\% upper limit confidence range corresponds to a mean occurrence rate of 2.3 sources. 
Results from \cite{2005MNRAS.362....9B} for a $z<0.3$ sample are also shown. The \citeauthor{2005MNRAS.362....9B} sample is divided into SF galaxies and AGN.}
\label{fig:lum}
\end{figure}

{The differences between the lensing models are larger than the uncertainties provided for the individual models. We therefore compute the volume for each for the 8 models listed in Table~\ref{tab:compactsources} and take the average. The same is done for counting the number of sources in each luminosity bin (as this also depends on the amplification factors). In Figure~\ref{fig:lum} we plot the luminosity function averaged over these 8 different lensing models. The red error bars represent the standard deviation over the 8 lensing models. The black error bars show the combined uncertainty from the lensing models and Poisson errors on the galaxy number counts. These two  errors were added in quadrature. We find that the uncertainties in the redshifts and flux density measurements do not contribute significantly to the error budget. 

Another uncertainty for the derived luminosity function is related to  cosmic variance. Based on the computed volume we probe for the three luminosity bins and the number of objects in each bin, we compute the cosmic variance using \cite{2008ApJ...676..767T}. From this computation we find that cosmic variance  introduces an extra uncertainty between 22\% and 30\%. Note however that  this is a factor of a few smaller the Poisson errors and scatter due to the different lensing models.

In Figure~\ref{fig:lum} we also plot the low-redshift ($z < 0.3$) luminosity function derived by \cite{2005MNRAS.362....9B}. Although the uncertainties in our luminosity function are substantial, we find evidence for an increase in the number density of sources of a factor between $4$ and $10$ compared to the \citeauthor{2005MNRAS.362....9B} low-redshift sample.

With a larger sample (for example including all six Frontier Fields clusters) it should become possible to map out the luminosity function more accurately, as the Poisson errors can be reduced by a factor of $\sim \sqrt{6}$. Surveys covering a larger area will be needed to map out the high-luminosity end. These surveys can be shallower and do not require the extra amplification by lensing. A major limitation of the precision that can be achieved for the faint-end of the luminosity function, which can only be accessed with the power of lensing, is the accuracy of the lensing models. This highlights the importance of  further improving the precision of the Frontier Fields lensing models \citep[see also][]{2015arXiv151008077L}.

}

\section{Discussion and Conclusions}
\label{sec:discussion}

In our \jvla\ radio images of MACS~J0717.5+3745 we discovered 7~lensed sources with {expected amplification factors} larger than 2. This makes  MACS~J0717.5+3745 the cluster with the largest number of known lensed radio sources. Two of these radio sources are also detected in our \chandra\ X-ray image. To our knowledge only two other lensed X-ray sources behind a galaxy cluster are known \citep[Abell~370,][]{2000ApJ...543L.119B}. 
All of the lensed sources (with amplification factors~$> 2$) are located at $1 \lesssim z \lesssim 2$ and most seem to be star forming galaxies with SFR of~$\sim$~10--50 M$_\odot$~yr$^{-1}$ based on their radio continuum fluxes. Our search for lensed radio sources is different from previous radio studies performed by e.g., \cite{1997ApJ...490L...5S, 2005A&A...431L..21G,2010MNRAS.404..198I,2010A&A...518L..35I,2010A&A...509A..54B} which targeted previously known lensed submillimetre galaxies that have SFR $\sim 10^2$--$10^3$~M$_\odot$~yr$^{-1}$. The two lensed sources that are also detected in our \chandra\ image have 2--10~keV X-ray luminosities of $\sim 10^{43-44}$~erg~s$^{-1}$. We therefore classify these sources as AGN \citep[e.g.,][]{2004AJ....128.2048B}. 

From the derived luminosity function we find evidence for an increase in the number density of  $0.6<z<2.0$ radio sources compared to the $z < 0.3$ sample from \cite{2005MNRAS.362....9B}.  The increase is expected given the cosmic evolution of AGN and SF galaxies between these two redshift ranges \citep[e.g.,][]{2014ARA&A..52..415M,2014MNRAS.445..955B}. {Besides the Poisson errors, we find that the scatter between the available lensing models contributes significantly to the uncertainty in our derived luminosity function.}

From our \jvla\ observations we conclude that, as expected, lensing by massive (merging) clusters enables studying star forming galaxies at moderate to high redshifts, with the advantage of not being affected by extinction. Some of these radio sources have flux densities that are below the detection limits of typical radio observations without the amplification by lensing. In the case of MACS~J0717.5+3745 our highest amplification factor is about~9. To detect a source with a similar signal-to-noise ratio in the S-band, without the lensing amplification, would have required  about $10 \times 8^{2} \approx 600$~hrs of \jvla\ integration time. Practically this means that  the Square Kilometre Array (SKA) would be the only instrument that could achieve such a detection without the help of lensing. This example is similar to the lensed radio source found by \cite{2011ApJ...739L..28J}. {Radio observations also nicely complement  far-infrared and submillimeter observations that can detect strongly star-forming galaxies behind clusters \citep[e.g.,][]{2010A&A...518L..12E,2015arXiv150800586R}.}

Besides the amplification of the integrated flux density, the lensing magnification offers a chance to study these lensed sources at high spatial resolution. In the case of MACS~J0717.5+3745, all lensed radio sources are not or only slightly resolved at the current $\sim 1\arcsec$ resolution. 
A larger sample of lensing clusters is needed to increase the chances of finding a rare bright, and more highly magnified source, that would allow a detailed spatially resolved study. Based on our  MACS~J0717.5+3745 result, for a massive lensing cluster (for example from the CLASH sample) we expect to detect about a handful of lensed radio sources with a pointed $\sim10$~hr \jvla\ observation. 

Detection of background lensed X-ray sources suffer from the increased X-ray background from the cluster's ICM, in particular in regions of high magnification (see Figure~\ref{fig:mag}). This makes X-ray observations of lensed sources comparatively less efficient than optical or radio observations. 
In principle, by choosing a harder energy band (i.e., 2--10~keV) the contrast between a typical AGN and the ICM emission can be increased.
However, for massive lensing clusters this does not work so well because they are generally quite hot ($\sim 5-10$~keV), producing significant ICM emission in the hard X-ray band. In addition, the number density of X-ray sources on the sky that could potentially be lensed by a cluster and detected with current instruments with reasonable exposure times is typically lower than that of radio observations and thus decreases the chance of finding an object at high magnification \citep[e.g.,][]{2012ApJ...758...23C,2008A&A...479..283B}. Extrapolating from our MACS~J0717.5+3745 result, we expect to find of the order one lensed X-ray object (amplification $\gtrsim 2$) per massive lensing cluster for a $\sim 10^2$~ks \chandra\ observation.

\acknowledgments
{\it Acknowledgments:}
We would like to thank the anonymous referee for useful comments. We thank Megan Gralla for a discussion on the lensed radio sources.
The National Radio Astronomy Observatory is a facility of the National Science Foundation operated under cooperative agreement by Associated Universities, Inc. Support for this work was provided by the National Aeronautics and Space Administration through Chandra Award Number GO4-15129X issued by the Chandra X-ray Observatory Center, which is operated by the Smithsonian Astrophysical Observatory for and on behalf of the National Aeronautics Space Administration under contract NAS8-03060.

R.J.W. is supported by NASA through the Einstein Postdoctoral
grant number PF2-130104 awarded by the Chandra X-ray Center, which is
operated by the Smithsonian Astrophysical Observatory for NASA under
contract NAS8-03060. G.A.O. acknowledges support by NASA through a Hubble Fellowship grant HST-HF2-51345.001-A awarded
by the Space Telescope Science Institute, which is operated by the Association of Universities
for Research in Astronomy, Incorporated, under NASA contract NAS5-26555. M.B acknowledge support by the research group FOR 1254 funded by the Deutsche Forschungsgemeinschaft: ``Magnetisation of interstellar and intergalactic media: the prospects of low-frequency radio observations''. W.R.F., C.J., and F.A-S. acknowledge support from the Smithsonian Institution. F.A-S. acknowledges support from Chandra grant GO3-14131X. A.Z. is supported by NASA through Hubble Fellowship grant HST-HF2-51334.001-A awarded by STScI. This research was performed while T.M. held a National Research Council Research Associateship Award at the Naval Research Laboratory (NRL). Basic research in radio astronomy at NRL by T.M. and T.E.C. is supported by 6.1 Base funding. M.D. acknowledges the support of STScI grant 12065.007-A. P.E.J.N. was partially supported by NASA contract NAS8-03060. E.R. acknowledges a Visiting Scientist Fellowship of the Smithsonian Astrophysical Observatory, and the hospitality of the Center for Astrophysics in Cambridge. Part of this work performed under the auspices of the U.S. DOE by LLNL under Contract DE-AC52-07NA27344.

Part of the reported results are based on observations made with the NASA/ESA Hubble Space Telescope, obtained from the Data Archive at the Space Telescope Science Institute. STScI is operated by the Association of Universities for Research in Astronomy, Inc. under NASA contract NAS 5-26555. 
This work utilizes gravitational lensing models produced by PIs Brada{\v c}, Ebeling, Merten \& Zitrin, Sharon, and Williams funded as part of the HST Frontier Fields program conducted by STScI. The lens models were obtained from the Mikulski Archive for Space Telescopes (MAST).
This research has made use of the NASA/ IPAC Infrared Science Archive, which is operated by the Jet Propulsion Laboratory, California Institute of Technology, under contract with the National Aeronautics and Space Administration


%% in parentheses, after the keyword, but they will not be verified.

{\it Facilities:} \facility{VLA}, \facility{CXO}, \facility{HST}



\clearpage
\appendix

\section{Compact sources: cluster and foreground objects}
In Table~\ref{tab:compactsources2} we list the properties of all the radio and X-ray sources that are {cluster members, foreground objects, or sources with uncertain redshifts (so that we could not determine if they are lensed or not).} We also provide HST color postage stamp images around these radio sources in Figure~\ref{fig:cutouts2}.

\begin{table*}[h!]
\begin{center}
\tiny
\caption{Source properties}
\begin{tabular}{lllllllllll}
name/ID & RA & DEC &  $S_{1.5}$  & $S_{3.0}$ & $S_{5.5}$  & $z_{\rm{phot}}$  & $t_{b}$ &  Xray flux & $z_{\rm{spec}}$ \\
                 & degr (J2000)&  degr (J2000) & $\mu$Jy & $\mu$Jy & $\mu$Jy & &&$10^{-15}$~erg~cm$^{-2}$~s$^{-1}$& \cite{2014ApJS..211...21E}  \\
\hline
\hline
S1 3774   &  109.4073141  &  37.7583435 &  $47.5 \pm 11.2$ &  $22.6 \pm  3.5$  & \ldots &  $0.46_{-0.04}^{+0.07}$ &  6.80 &  \ldots &  \ldots\\
S4 2833  &   109.4232228  &  37.7654146  & $105.7 \pm 7.2$ & $60.4 \pm 2.2 $  &$24.7 \pm 3.5$  & $0.29_{-0.07}^{+0.06}$ & 5.90 &  \ldots &  \ldots\\ 
S5  6341 &  109.4018584   &  37.7340578 & $442.2 \pm 9.4$ &  $578.1 \pm3.2$ & $446.2 \pm 5.1$ & $0.69_{-0.10}^{+0.04}$ &5.50 &  \ldots&  $0.5426\pm0.0003$ (E) \\ 
S7 5353  &  109.3982487   &  37.7457351   & $157.0 \pm 9.2$ & $57.6\pm3.0$ & $16.3 \pm 3.5$ & $0.54_{-0.02}^{+0.11}$& 1.00& \ldots  & $0.5408\pm0.0003$ (A)   \\  
S10 7521 & 109.3823125  &  37.7260450 &  $64.4 \pm 8.2$&  $40.8 \pm 2.5$ & $19.1 \pm 3.7$ & $0.56_{-0.04}^{+0.06}$ & 6.60& \ldots  & $0.5315\pm0.0005$ (E)\\
S11 7828 & 109.3816839  &  37.7225923  & $132.2 \pm 9.9$ &  $85.3 \pm 3.1$& $51.2 \pm 4.3$ & $0.30_{-0.07}^{+0.05}$ & 6.00 &  \ldots &  \ldots\\ 
S13 7314 & 109.4023607  &  37.7268590  & \ldots & \ldots & $12.7 \pm  5.0^a$ & $ 0.57_{-0.04}^{+0.07}$ &8.20 &  \ldots & $0.5332\pm0.0003$ (E) \\  
S14 1407 & 109.4268701  & 37.7779977  & $43.9 \pm  13.5^a$  & \ldots & \ldots & $0.6_{-0.4}^{+3.0}$ & 7.90 &  \ldots \\
S15 3462 &  109.4122491  &  37.7606844 & $35.7 \pm 13.4^a$ & \ldots & \ldots & $0.50_{-0.04}^{+0.04}$ &6.80 &   \ldots&  $0.4992\pm0.0003$ (E) \\
S16 1096 & 109.4278576  & 37.7808548   & $66.2 \pm 8.7$ & $24.3\pm 3.1$ & $26.9 \pm 3.6$ & $1.0_{-0.7}^{+3.1}$  & 8.50 &  \ldots &  \ldots\\  
S17  952   & 109.4095733 &  37.7806377  & $1685.9 \pm 12.4$ & $1048.8 \pm 4.0$ & $847.8 \pm 7.9$ & $0.64_{-0.05}^{+0.07}$ &5.50&  $5.64_{-0.15}^{+0.18}$ & $0.5613\pm0.0003$ (A) \\ 
%S18 345   &  109.4204105 &  37.7896360   & $50.3 \pm 8.3$  & $18.1 \pm 2.8$ & $14.4 \pm 4.0$  & $0.462_{0.387}^{3.259}$ & 3.00\\ 
% NO CLASH COVERAGE
%S19 \ldots & \ldots                &  \ldots               & $138.1 \pm8.7$  & $88.5 \pm 3.2$  & $50.6 \pm 3.7$ &  & & &  $0.5374\pm0.0003$ (E) \\
S20 7679 & 109.3522480 &  37.7248602   & $38.8 \pm 8.7$ & $25.4 \pm 3.2$ & $20.1 \pm  6.0^a$ & $0.9_{-0.5}^{+3.4}$ & 7.90 &  \ldots &  \ldots\\
S22 6064 & 109.3537889  & 37.7355689   & $26.7 \pm 10.2^a$ & $16.5 \pm 2.9$  &  \ldots& $0.49_{-0.05}^{+0.04}$  &6.00&   \ldots &$0.5357\pm0.0003$ (E) \\ 
S23 5574 & 109.3647573  & 37.7448496   & $90.6 \pm 9.3$ & $48.6 \pm 3.0$ & $26.8 \pm 3.2$ & $0.55_{-0.04}^{+0.03}$  &7.50&  \ldots & $0.5261\pm0.0003$ (E)\\ 
% NO CLASH COVERAGE
%S25 \ldots            & \ldots               & \ldots               & $31.8 \pm 9.2$ & $23.6 \pm 3.1$ &  $28.2 \pm 7.5^a$ \\
S26 4123 & 109.3470266  & 37.7555482   & $85.0 \pm 8.8$ & $50.6 \pm 3.2$ & $27.0 \pm 3.9$ & $0.5_{-0.2}^{+0.2}$ &6.50&  \ldots & $0.4212\pm0.0003$ (E)\\
S27 3789  & 109.3478763 &  37.7578892  & $2013.0 \pm 16.1$ & $1274.5 \pm 7.8$ & $811.2 \pm 9.2$ & $0.65_{-0.37}^{+0.07}$ & 7.00 &  \ldots &  \ldots \\  
S28 3852  & 109.3739700 &  37.7568547  & $45.4 \pm 10.1$ & $24.6 \pm 2.7$ &$15.9 \pm 3.5$  & $0.58_{-0.03}^{+0.03}$ &5.60&   \ldots& $0.5565\pm0.0003$ (E) \\
S29 3402  & 109.3598798 &  37.7621055   & \ldots  & $19.7 \pm 4.6^a$ & $17.0 \pm  5.7^a$  & $0.55_{-0.03}^{+0.04}$ & 5.40&  \ldots& $0.5442\pm0.0003$ (A)  \\ 
S30 2446  & 109.3634032 &  37.7683109    & $33.5 \pm  12.5^a$  & $31.5 \pm 4.6$  & $19.6 \pm 3.2$ & $0.15_{-0.05}^{+0.03}$ & 6.80 &  \ldots &  \ldots\\ 
S31 1660  &  109.3568926 &  37.7756024   & \ldots &  \ldots& $14.4 \pm 5.8^a$  & $0.9_{-0.7}^{+2.9}$ & 7.90 &  \ldots  &  \ldots\\
%% NO CLASH COVERAGE
%S32 \ldots &   \ldots                & \ldots                & $163.2 \pm 12.9$ & $90.6\pm 4.5$    & $55.7 \pm 3.5$ & \\
S33 2974 &   109.3936629  &  37.7644686  &\ldots  &\ldots  & \ldots & $0.34_{-0.03}^{+0.06}$ & 6.80 &  \ldots\\ 
S34 4426  & 109.3981162  &  37.7514879   & $17797  \pm 29 $ & $7058 \pm 14$  & $3149 \pm 9$ & $0.56_{-0.02}^{+0.03}$ &5.50 & \ldots &$0.5528\pm0.0003$ (A) \\
%S34 4426  & 109.3981162  &  37.7514879   & $17796.53 (31beams)$ & $7057.8696 (57beams)$  & $3149.387 (24beams)$ & $0.558_{0.534}^{0.585}$ &5.50& & &$0.5528\pm0.0003$ (A) \\
S35 4020  &  109.3985608 &  37.7547888   & $59.79 \pm  5.1^{\star a}$ &  $41.8\pm8.0^{\star}$ & $25.2 \pm 2.4^{\star a}$& $0.54_{-0.02}^{+0.02}$ &4.30&  \ldots &  $0.5443\pm0.0003$ (A)\\
%S36  5598 &  109.4050484 &  37.7397125  & $10504.55 (189beams )$ & $8355.724 (beams409)$  &  $7132.200 (162beams)$  & $0.159_{0.137}^{0.218}$ &1.30& &X \\
S36  5598 &  109.4050484 &  37.7397125  & $10505 \pm 71$ & $8356 \pm 36$  &  $7132 \pm 24$  & $0.16_{-0.02}^{+0.06}$ &1.30& $5.88_{-0.14}^{+0.15}$ &  \ldots \\
%S37 \ldots  & \ldots                 & \ldots               &  $257.7 \pm  4.9^\star (peak)$  & $326.9\pm33.6^{\star}$  &\\
S38 6779 & 109.3951581  & 37.7315723    & $38.4 \pm  4.9^{\star a,b} $ &  $29.1\pm6.7$ &  $29.5\pm  2.4^{\star a,b}$ & $0.53_{-0.05}^{+0.03}$ &4.40&  \ldots & $0.5366\pm0.001$ (A)\\
%% NO CLASH COVERAGE
%S40 \ldots & \ldots                & \ldots                &    $30.7 \pm  10.6^a$  &  $12.7 \pm  3.9^a$  & $15.0 \pm  5.5^a$  &  \\
S42 5335 & 109.3813319  & 37.7437654  & $31.6 \pm  11.2^a$ & $13.2 \pm 3.1$ &  \ldots & $0.48_{-0.03}^{+0.03}$ & 5.50&    \ldots   & $0.4919\pm0.0003$ (E) \\
S43 4992 & 109.3970833  & 37.7464913  &\ldots  & $19.8 \pm 3.1$ & $8.5 \pm 3.3$ & $0.54_{-0.03}^{+0.03}$ & 6.20&    \ldots   & $0.1779\pm0.0003$ (E)\\  
S46 5274  & 109.4179222 &  37.7459331  & $36.3 \pm  13.2^a$  &$23.0\pm 2.6$  &  & $0.50_{-0.05}^{+0.04}$& 9.40& \ldots & $0.5490\pm0.0003$  (E) \\ % correct spectroscopic position ?
S47 5076  & 109.4220234 &  37.7478226  & $33.0 \pm  12.3^a$ & $21.1 \pm 6.8$ & \ldots & $0.54_{-0.05}^{+0.06}$ &6.30 &  \ldots & $0.5660\pm0.0003$ (E)\\
S49 7693 & 109.3710599 &  37.7222104   & \ldots &  $7.5 \pm  1.9^{a,b}$  & \ldots & $0.30_{-0.18}^{+0.09}$& 6.40& \ldots & $0.2288\pm0.0003$ (E)   \\
%S50 6667 & 109.3736725  & 37.7315984   & \ldots & \ldots & \ldots &  $0.529_{0.505}^{0.554}$ \\  % just too faint in combined image
%S51 6070 & 109.3687591  & 37.7379402   & \ldots & \ldots & \ldots & $0.011_{0.010}^{0.027}$ \\  % just too faint in combined image
S52 5299 & 109.3654725  & 37.7433269  & $105.0 \pm 6.2$ & \ldots &\ldots  &  $0.22_{-0.03}^{+0.03}$ &7.00 &  \ldots &  \ldots\\
%% NO CLASH COVERAGE
%S53 \ldots & \ldots                & \ldots               &  \ldots& \ldots &\ldots & &&& $0.5285\pm0.0003$ (E) \\
%% NO CLASH COVERAGE
%S54 \ldots & \ldots                 & \ldots               & $54.6 \pm 9.0$ & $35.4\pm  3.3$  & \ldots&&&& $0.5382\pm0.0003$ (E)  \\
S55$_1$$^c$ 1585 & 109.4246322 &  37.7763075  & \ldots & \ldots & \ldots &  $0.4_{-0.3}^{+2.5}$  & 8.00 &  \ldots &  \ldots\\
S55$_2$$^c$ 1584 & 109.4245383 &  37.7763555  & \ldots & \ldots &  \ldots  &  $2.5_{-2.2}^{+0.6}$ & 7.90 &  \ldots &  \ldots\\
%S56 982 & 109.4326983 &  37.7804454   &$40.4 \pm 8.6$ + $26.2 \pm 8.7$ &  &  &  $0.496_{0.102}^{1.366} $ & & & $0.3859\pm0.0003$ (E) \\
%S56 982 & 109.4326983 &  37.7804454   &$66.6\pm12.2$ &  \ldots& \ldots &  $0.496_{0.102}^{1.366} $ &7.40& & & $0.3859\pm0.0003$ (E) \\
%S57  1404 & 109.3990868  & 37.7751503 & \ldots &\ldots  &  \ldots&  $0.154_{0.108}^{0.336}$ \\ % just too faint
S58  792   & 109.3976281  & 37.7831987 &  \ldots&\ldots  & $12.1 \pm 5.2^a$ &  $0.8_{-0.4}^{+1.1}$ & 7.60 & \ldots &  \ldots\\ 
S59 2444  &109.3621323  &  37.7692569 & \ldots & $13.5 \pm  4.3^a$  & \ldots &  $0.56_{-0.05}^{+0.08}$ &6.50& \ldots & $0.4984\pm0.0005$ (A) \\
S62 2977 & 109.3963967 &  37.7646847  &  \ldots& $19.9 \pm 5.4^a$   & \ldots &   $0.65_{-0.05}^{+0.03} $ &6.80& $3.09_{-0.84}^{+0.98}$ & $0.5490\pm0.0005$ (E) \\ 
X63 1956 & 109.4139432   &  37.7728903  & \ldots & \ldots & \ldots &   $0.50_{-0.06}^{+0.03} $ &5.20&   $7.37_{-2.87}^{+3.89}$  & $0.4897\pm0.0003$ (A) \\
X64$_1$$^c$ 2220 & 109.4099881 &  37.7711734    & \ldots & \ldots & \ldots &   $0.8_{-0.5}^{+2.4} $ &8.00&   $6.20_{-4.71}^{+3.09}$ & \ldots \\
X64$_2$$^c$ 2222 & 109.4095458 &  37.7711695  & \ldots & \ldots & \ldots &   $0.6_{-0.2}^{+3.1} $ &8.70&  $6.20_{-4.71}^{+3.09}$ &  \ldots \\
\hline
\hline
\end{tabular}
\label{tab:compactsources2}
\end{center}
$^{\star}$ flux density measurement could be affected by radio emission from other sources\\
$^{a}$ manually measured \\
$^{b}$ very faint source, only peak flux was measured \\
$^{c}$ the complex morphology of the galaxy likely caused it to be fragmented and listed as separate objects in the catalog\\
The source ID and positions are taken from the CLASH catalog. The photometric  redshifts ($z_{\rm{phot}}$) and best  fitting spectral template ($t_b$) are also taken from the CLASH catalog. The photometric redshifts are given at a 95\% confidence level. The spectral templates are described in \cite{2014MNRAS.441.2891M}. In total there are 11 possible spectral templates, five  for elliptical galaxies (1--5), two for spiral galaxies (6, 7) and four for starburst galaxies (8--11), along with emission lines and dust extinction. Non-integer values indicate interpolated templates between adjacent templates. For the radio flux density errors we include a 2.5\% uncertainty from bootstrapping the flux density scale.\\
For the spectroscopic redshifts ($z_{\rm{spec}}$) the spectral classification is as follows: A =	absorption-line spectrum; E =	emission-line spectrum.\\
\end{table*}

\begin{figure*}[h]
\centering
\includegraphics[angle =180, trim =0cm 0cm 0cm 0cm,width=0.24\textwidth]{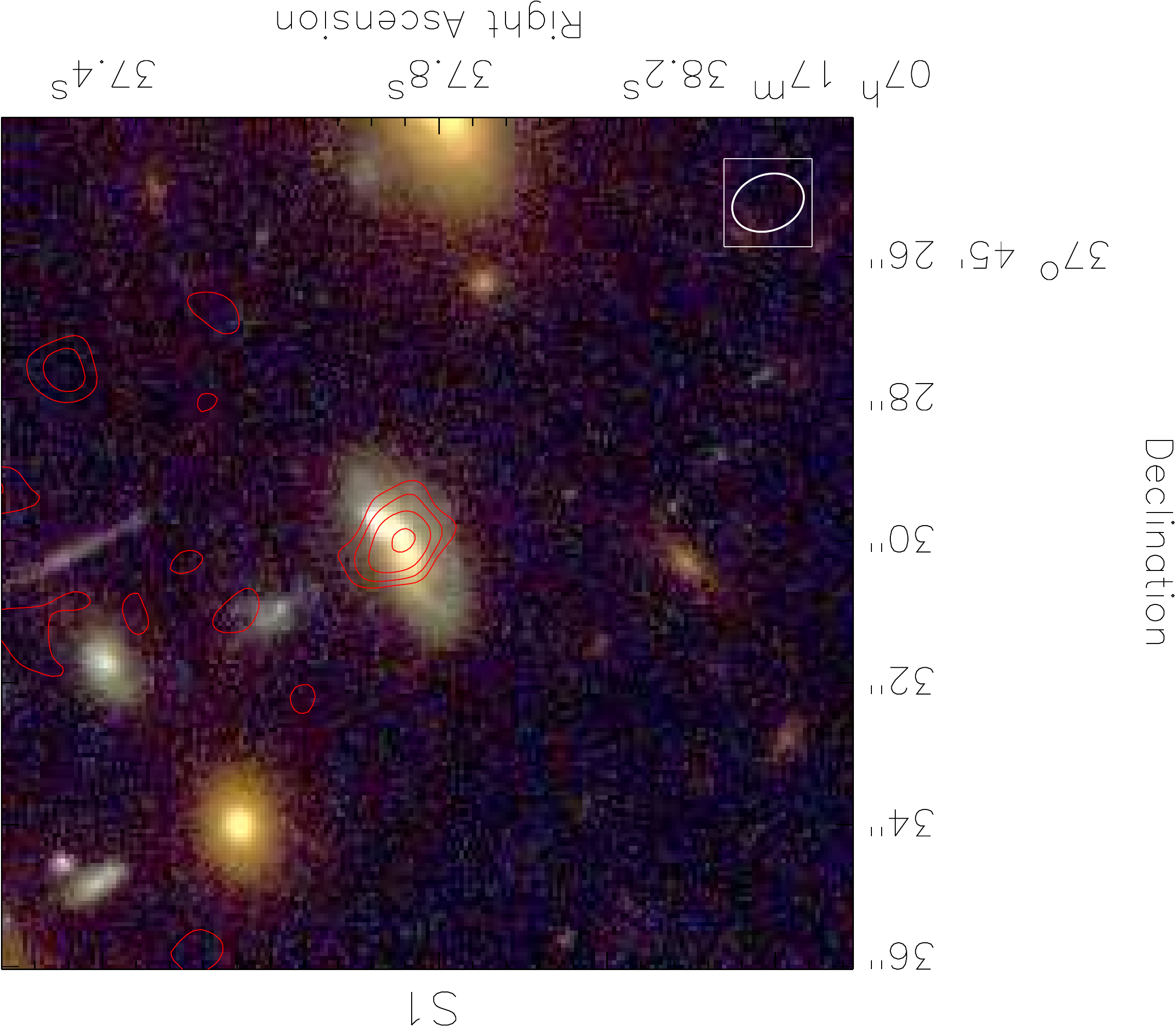} % 1
\includegraphics[angle =180, trim =0cm 0cm 0cm 0cm,width=0.24\textwidth]{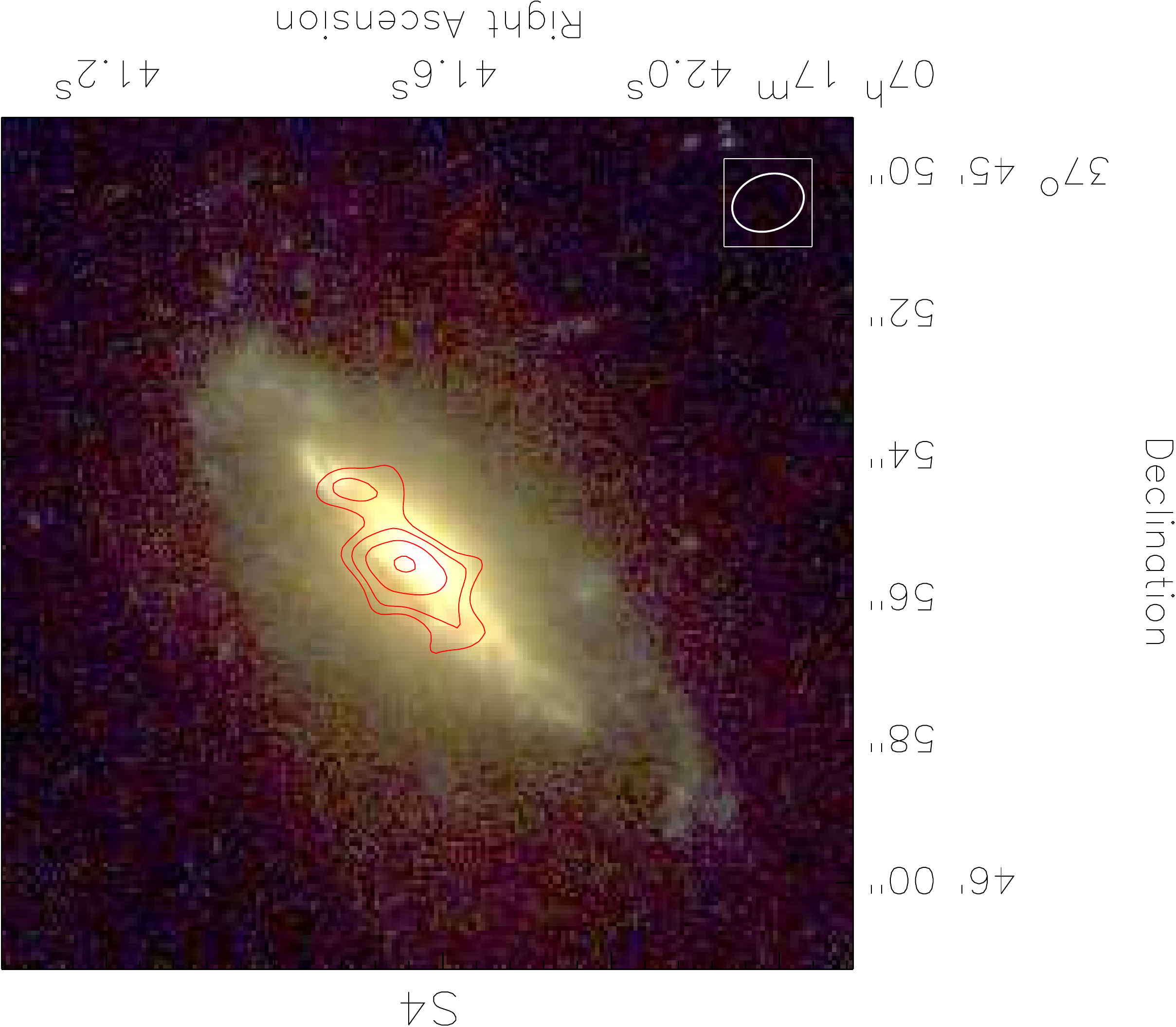} % 2
\includegraphics[angle =180, trim =0cm 0cm 0cm 0cm,width=0.24\textwidth]{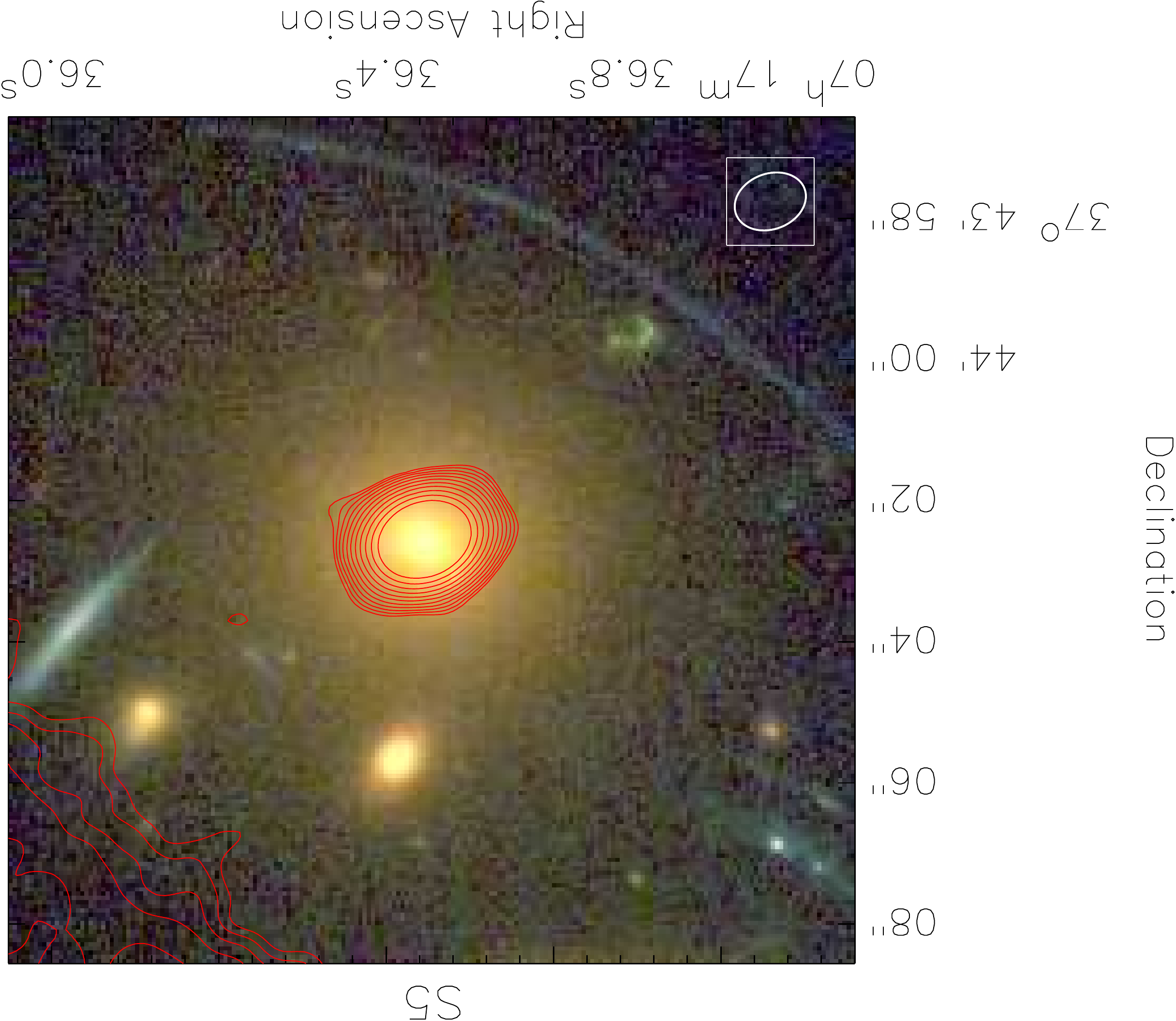} % 3
\includegraphics[angle =180, trim =0cm 0cm 0cm 0cm,width=0.24\textwidth]{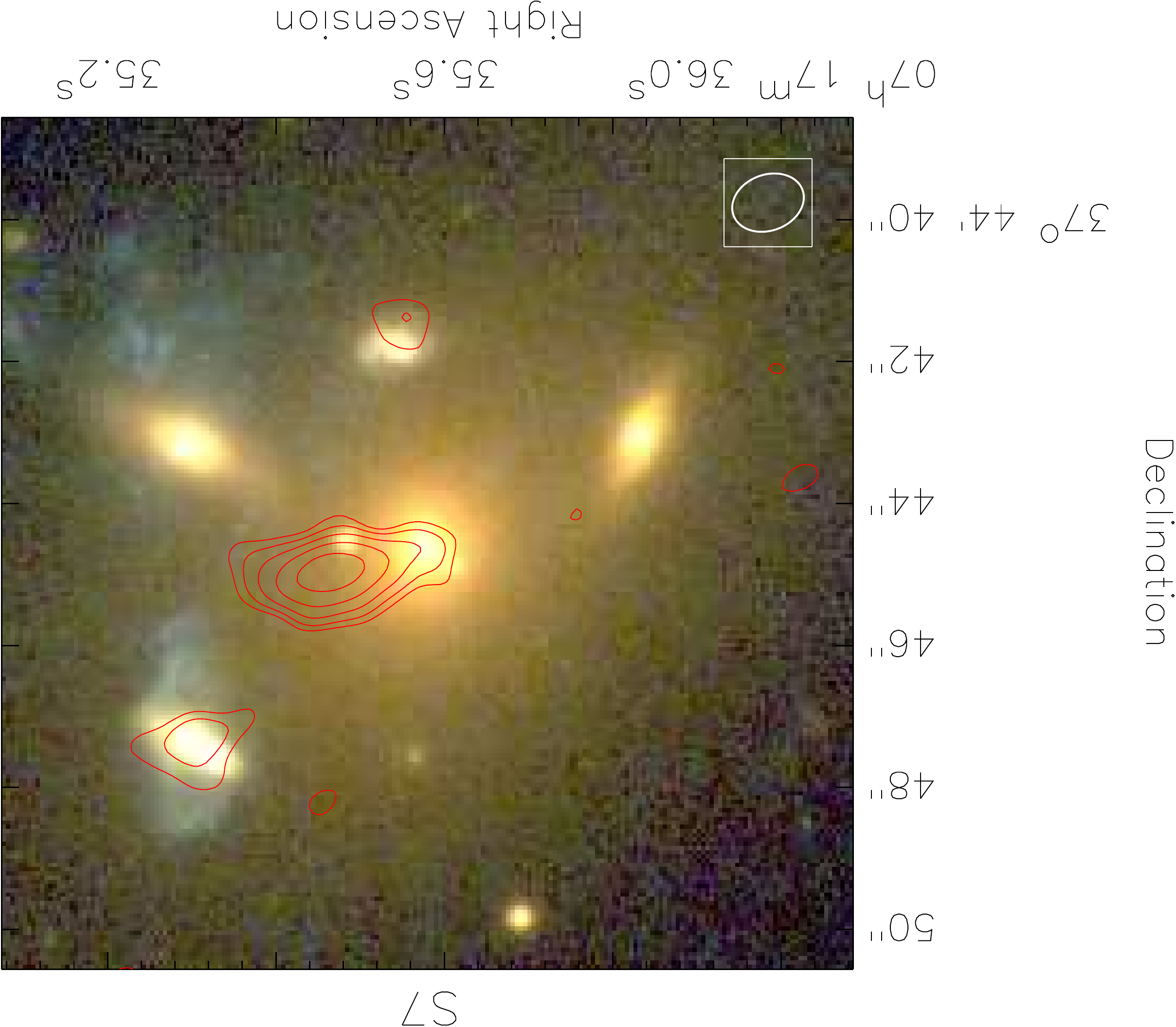} % 4
\includegraphics[angle =180, trim =0cm 0cm 0cm 0cm,width=0.24\textwidth]{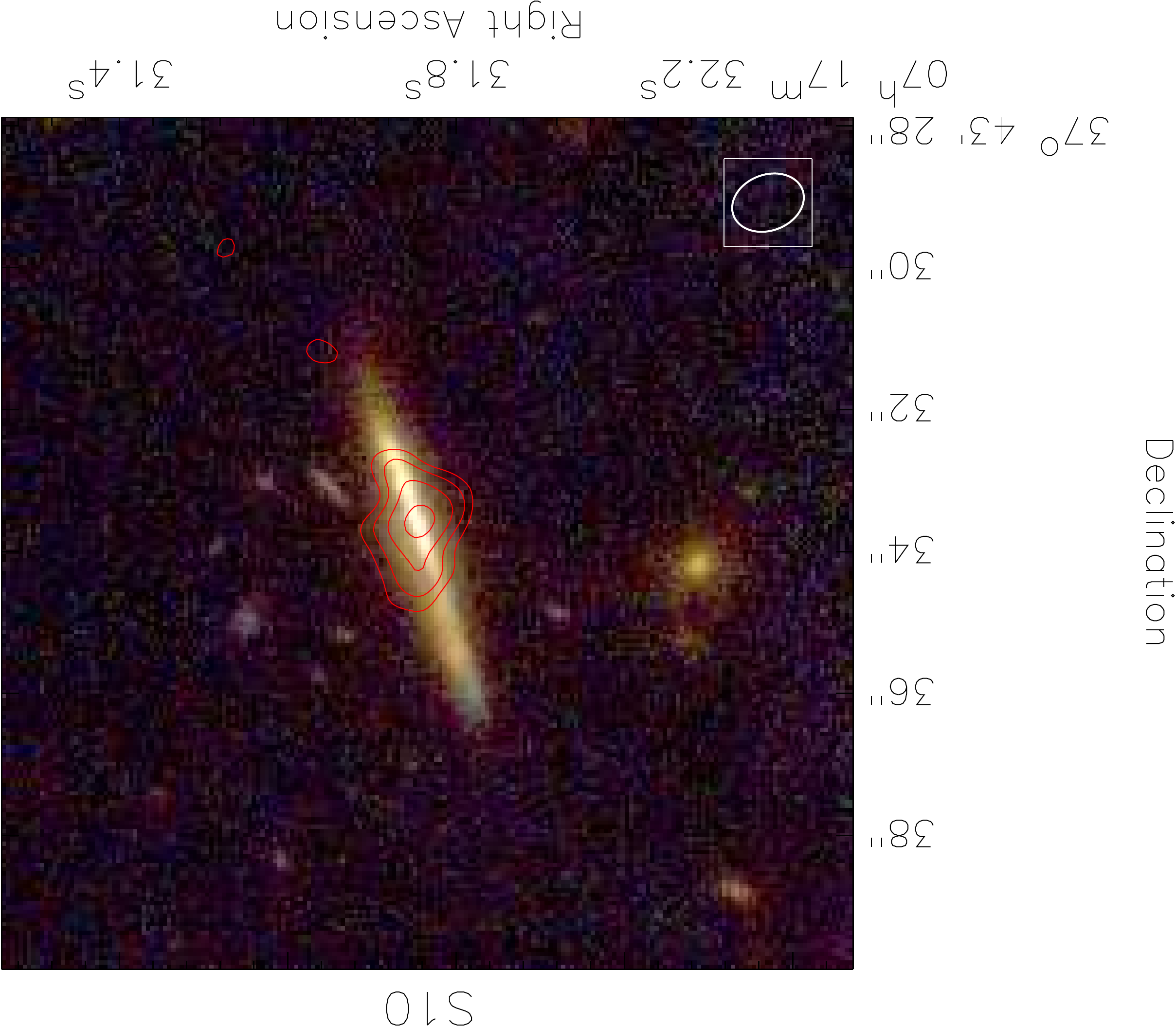} %5
\includegraphics[angle =180, trim =0cm 0cm 0cm 0cm,width=0.24\textwidth]{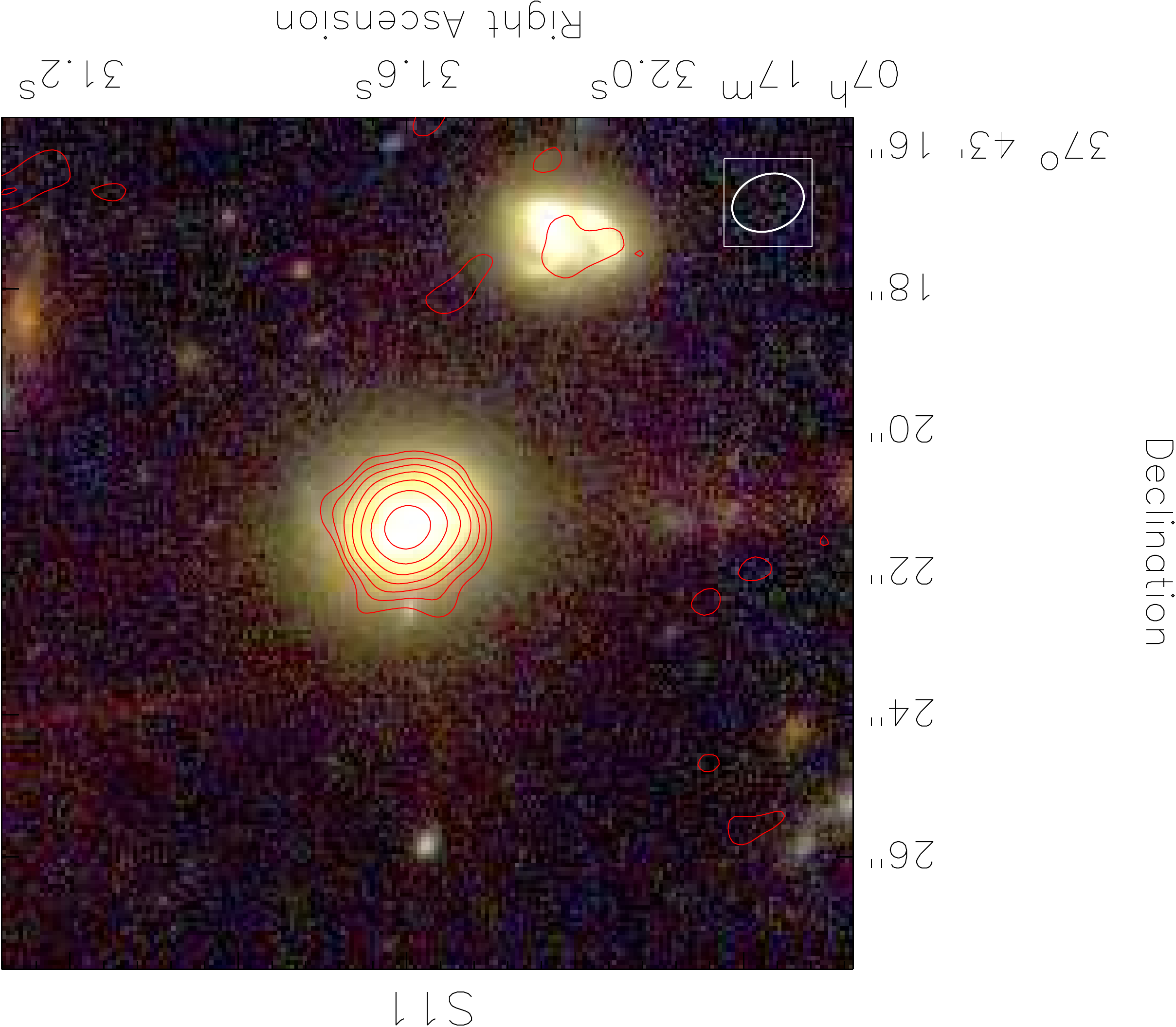}
\includegraphics[angle =180, trim =0cm 0cm 0cm 0cm,width=0.24\textwidth]{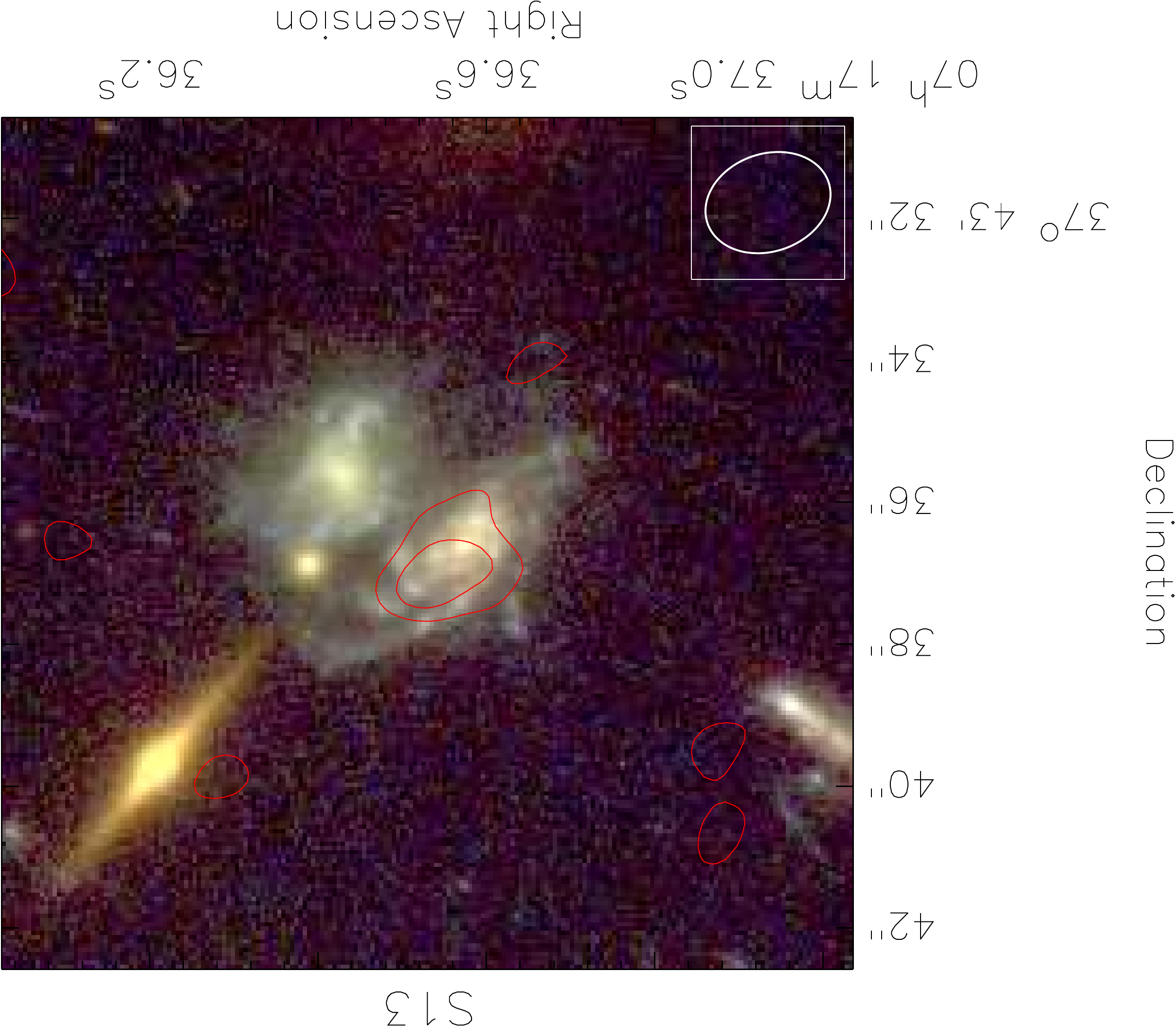} % 7
\includegraphics[angle =180, trim =0cm 0cm 0cm 0cm,width=0.24\textwidth]{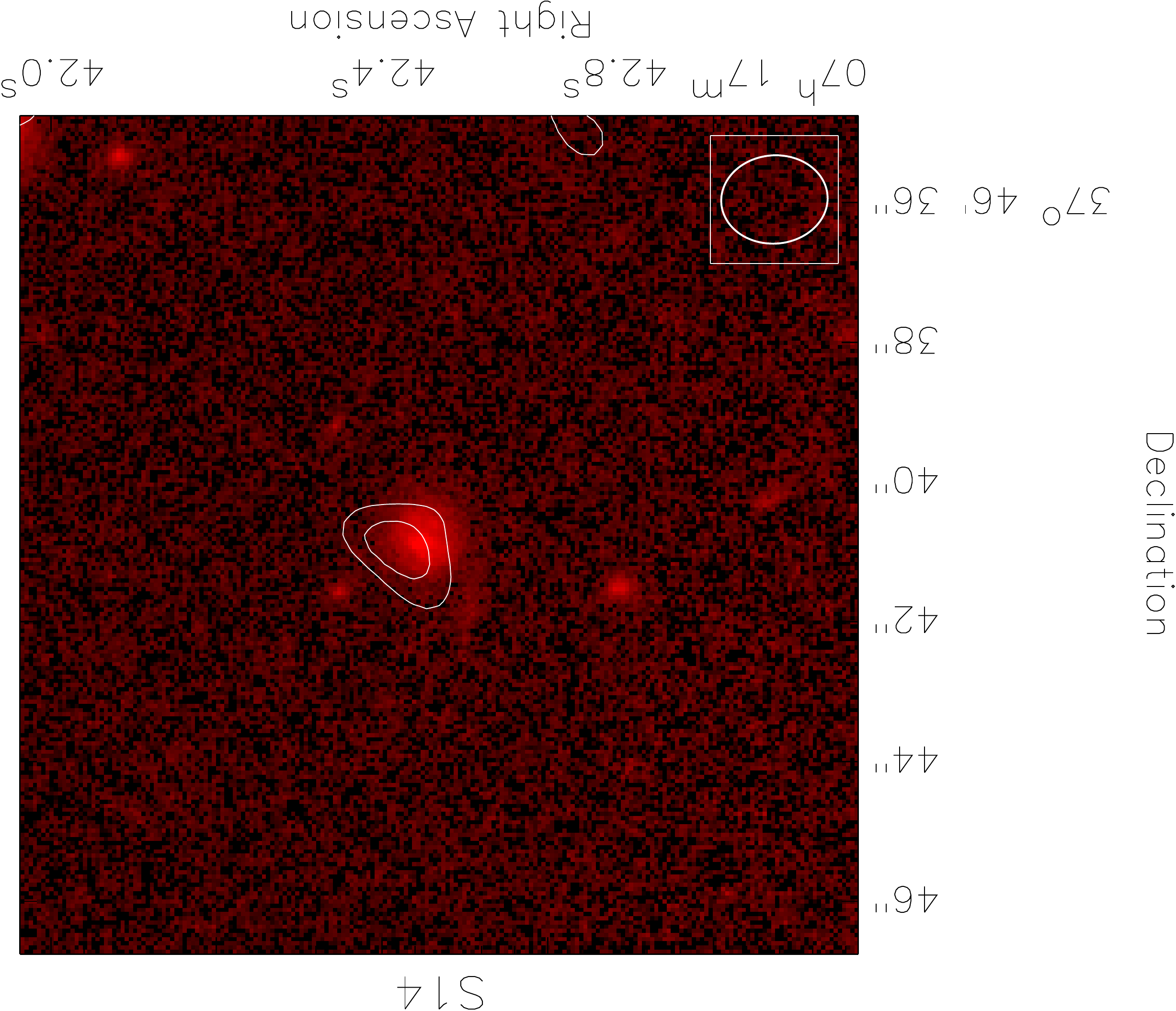}
\includegraphics[angle =180, trim =0cm 0cm 0cm 0cm,width=0.24\textwidth]{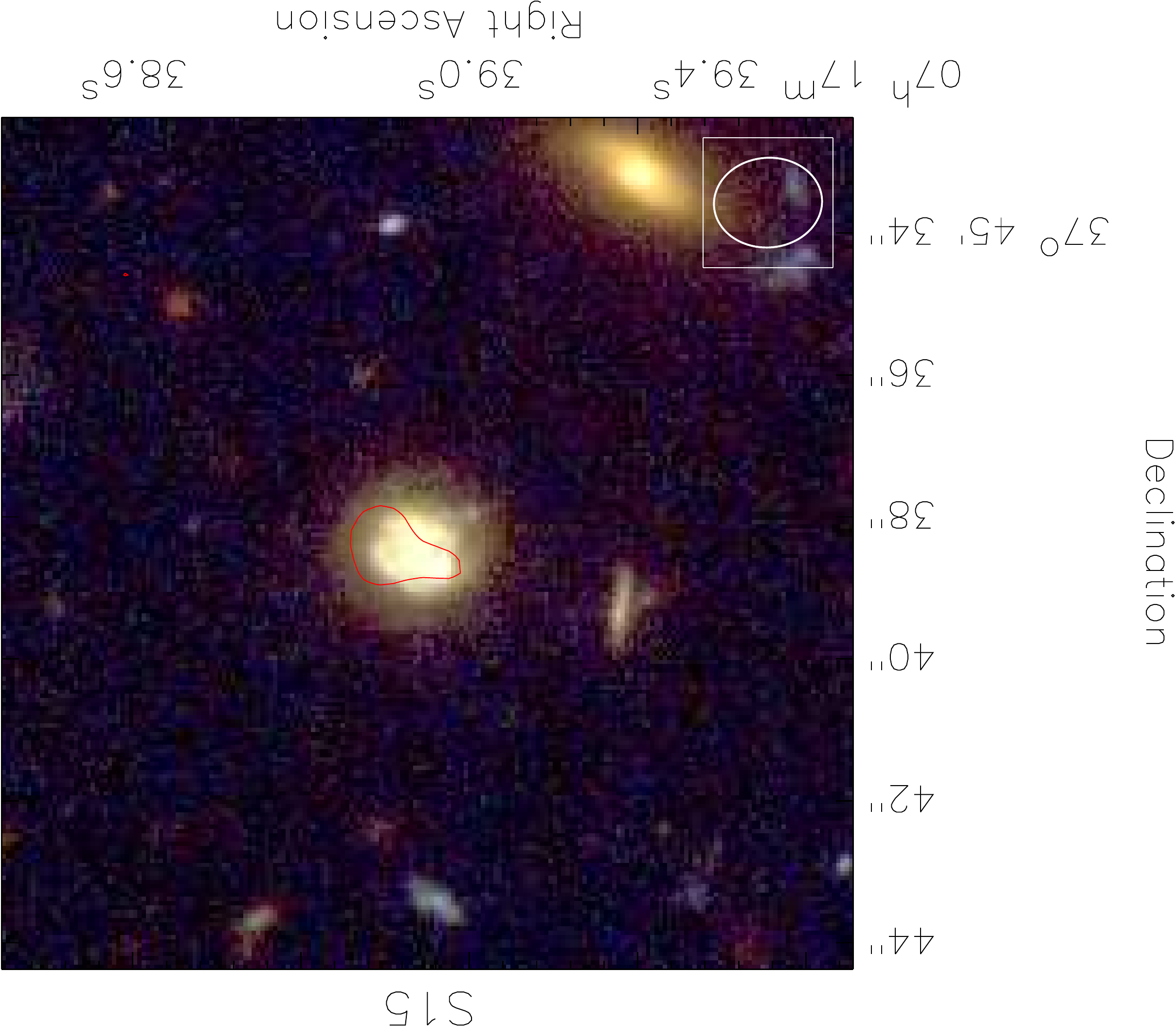}
\includegraphics[angle =180, trim =0cm 0cm 0cm 0cm,width=0.24\textwidth]{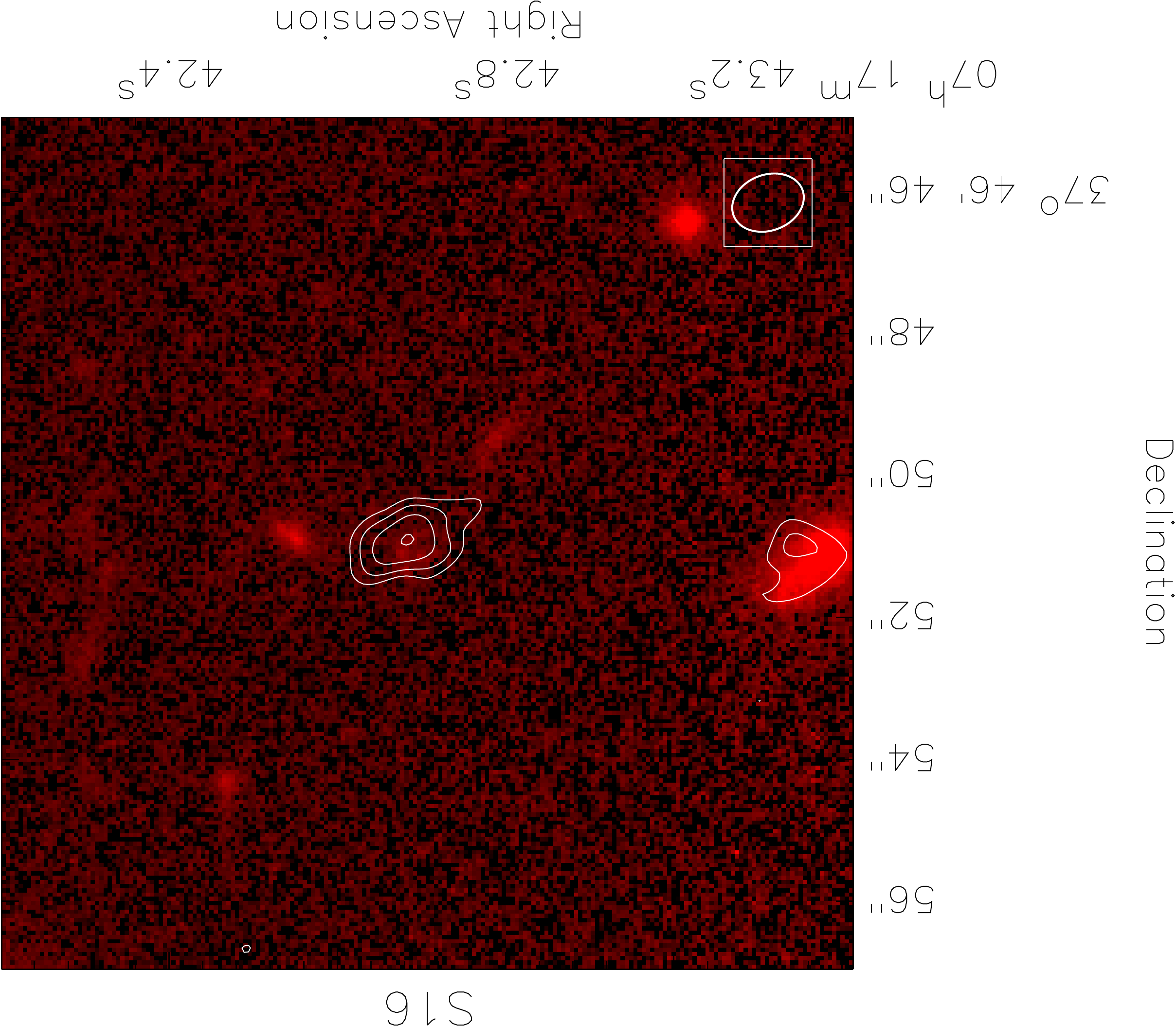}
\includegraphics[angle =180, trim =0cm 0cm 0cm 0cm,width=0.24\textwidth]{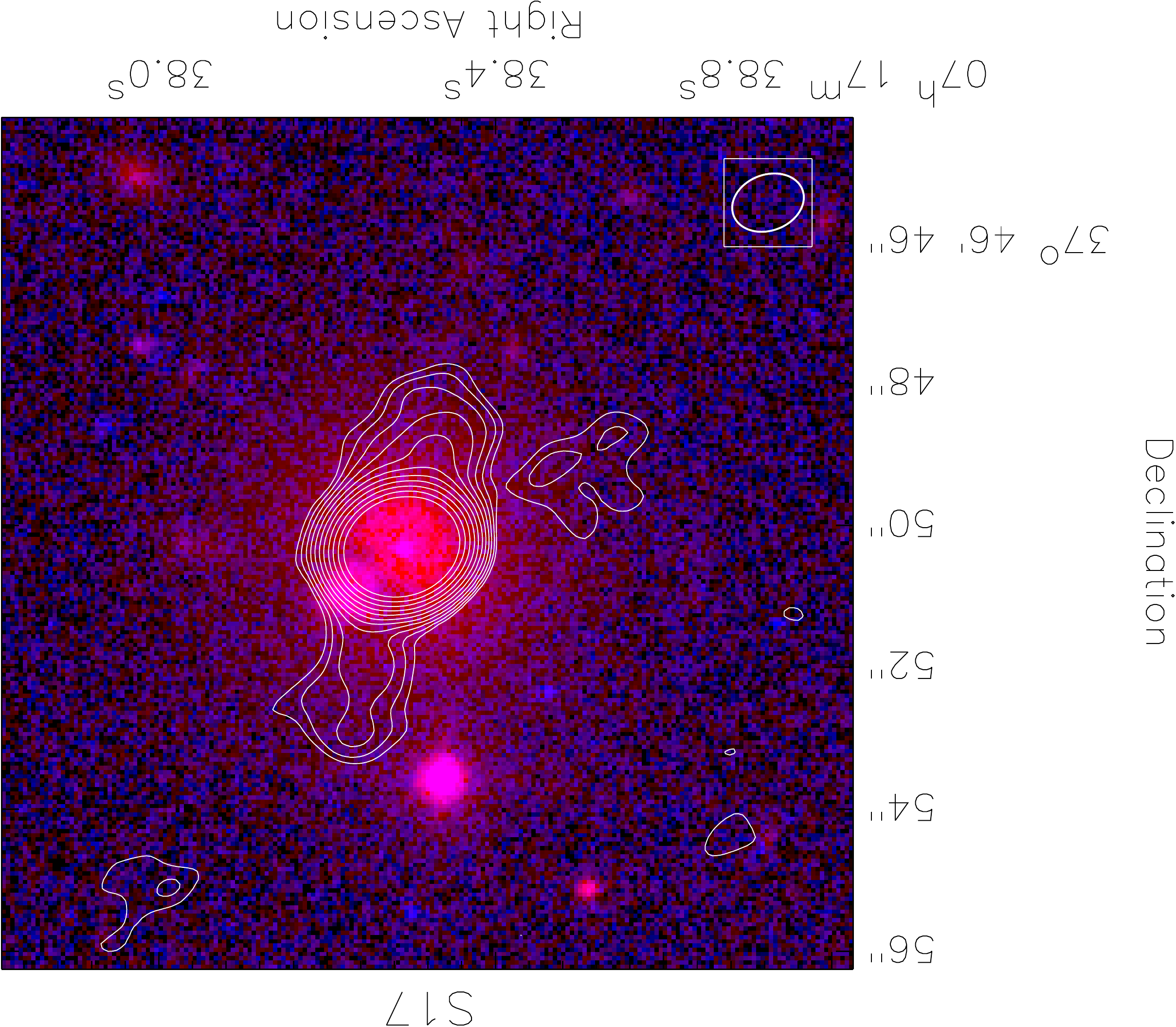} % 11
\includegraphics[angle =180, trim =0cm 0cm 0cm 0cm,width=0.24\textwidth]{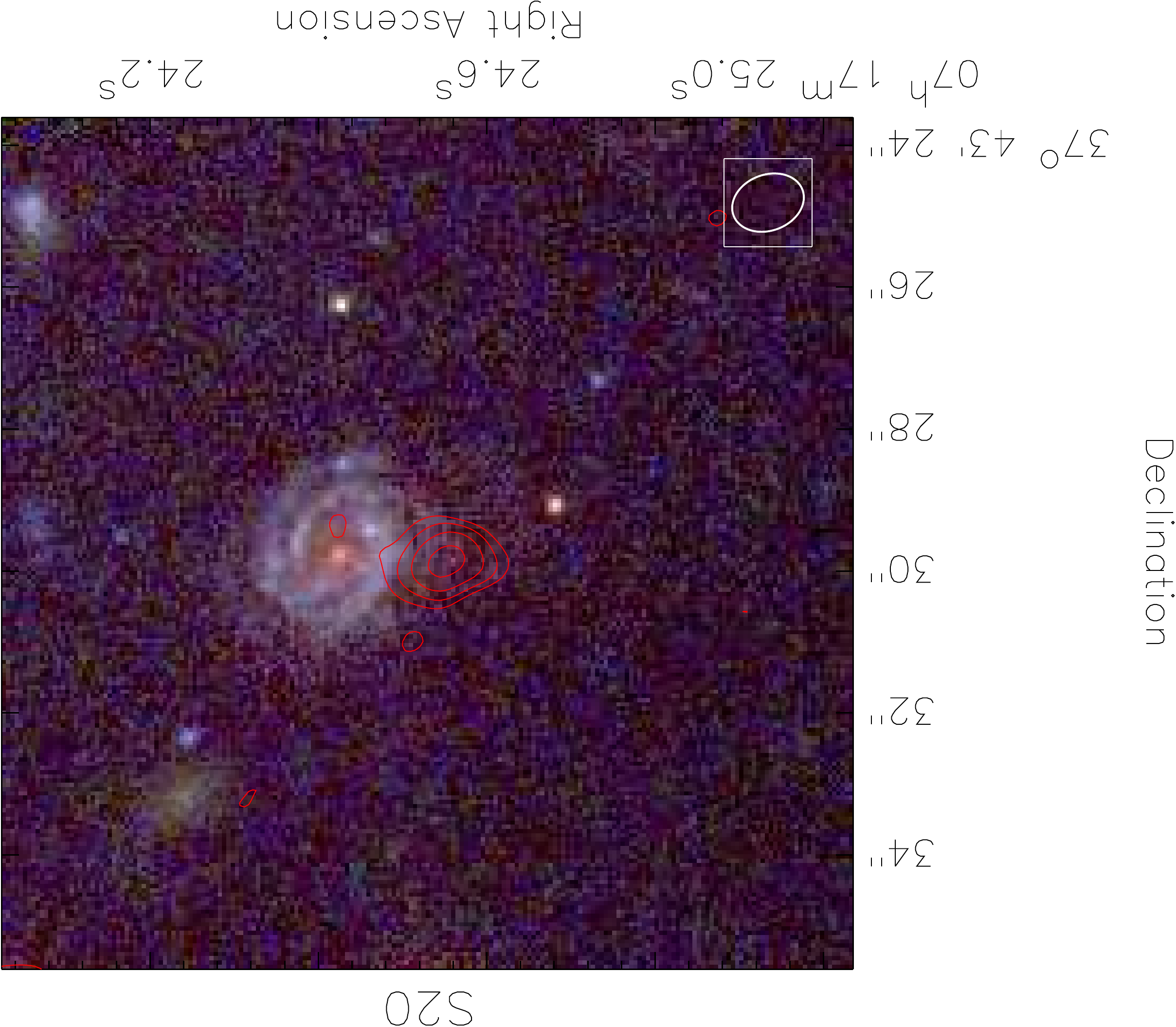}
\includegraphics[angle =180, trim =0cm 0cm 0cm 0cm,width=0.24\textwidth]{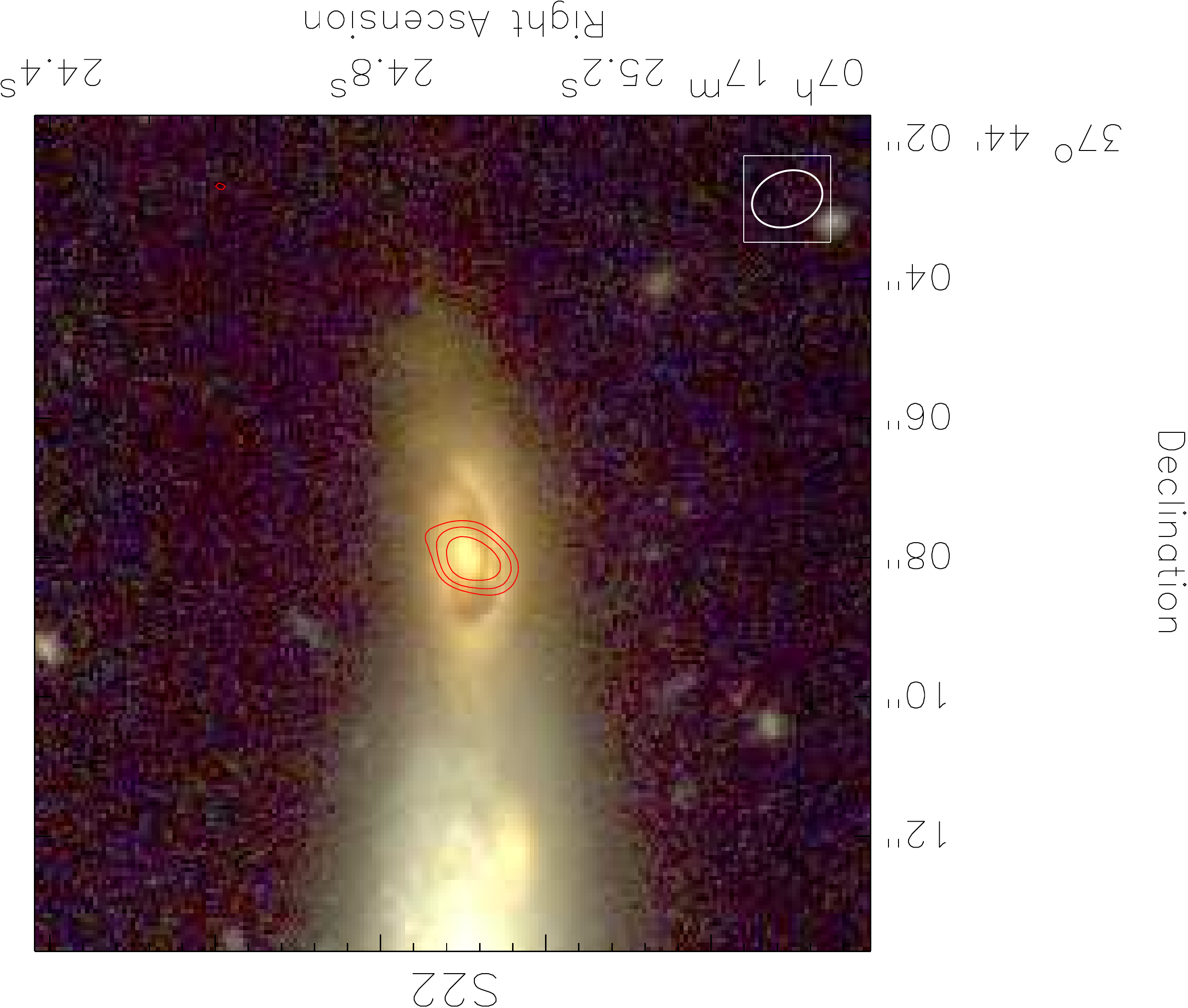} % 13
\includegraphics[angle =180, trim =0cm 0cm 0cm 0cm,width=0.24\textwidth]{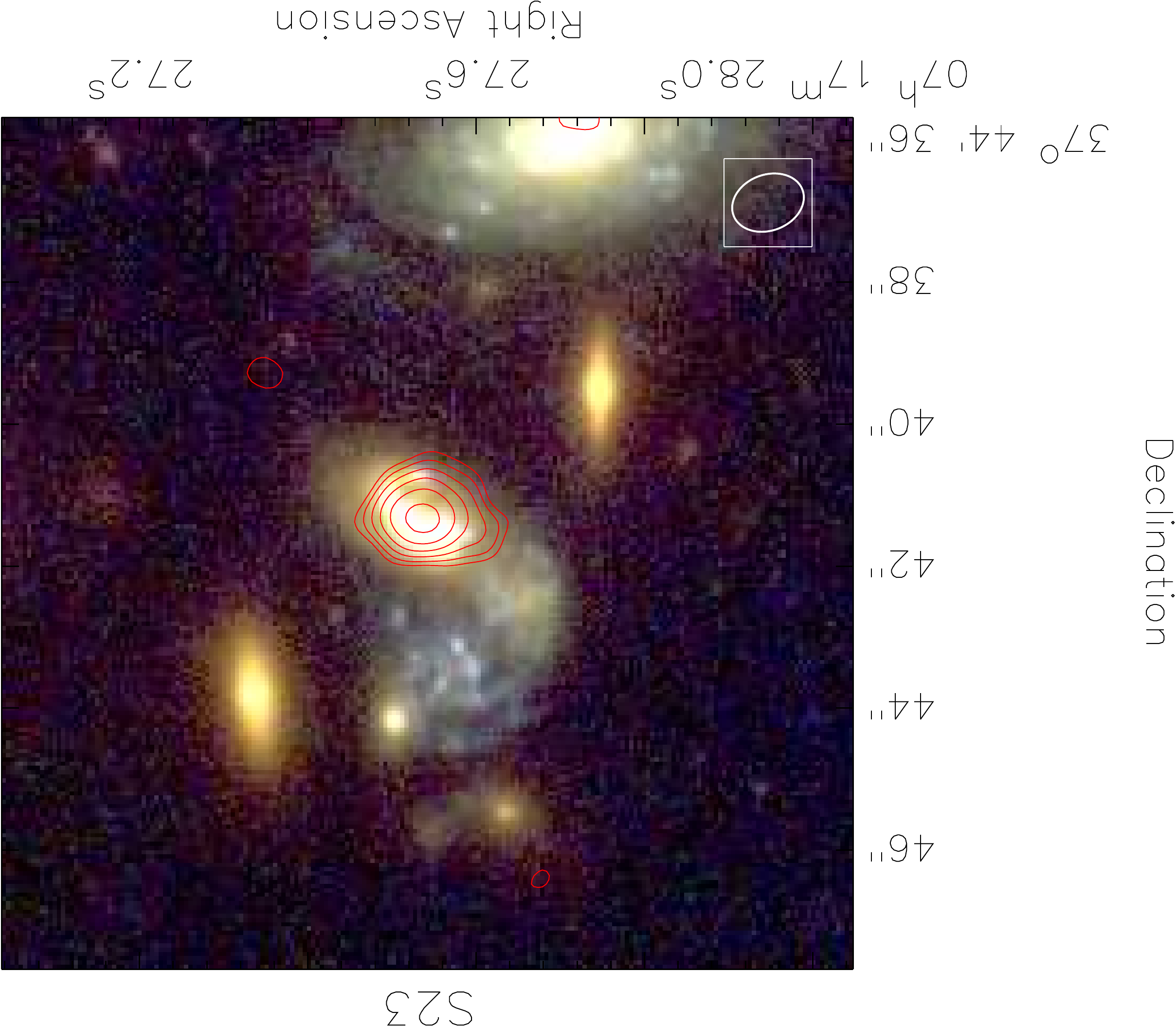}
\includegraphics[angle =180, trim =0cm 0cm 0cm 0cm,width=0.24\textwidth]{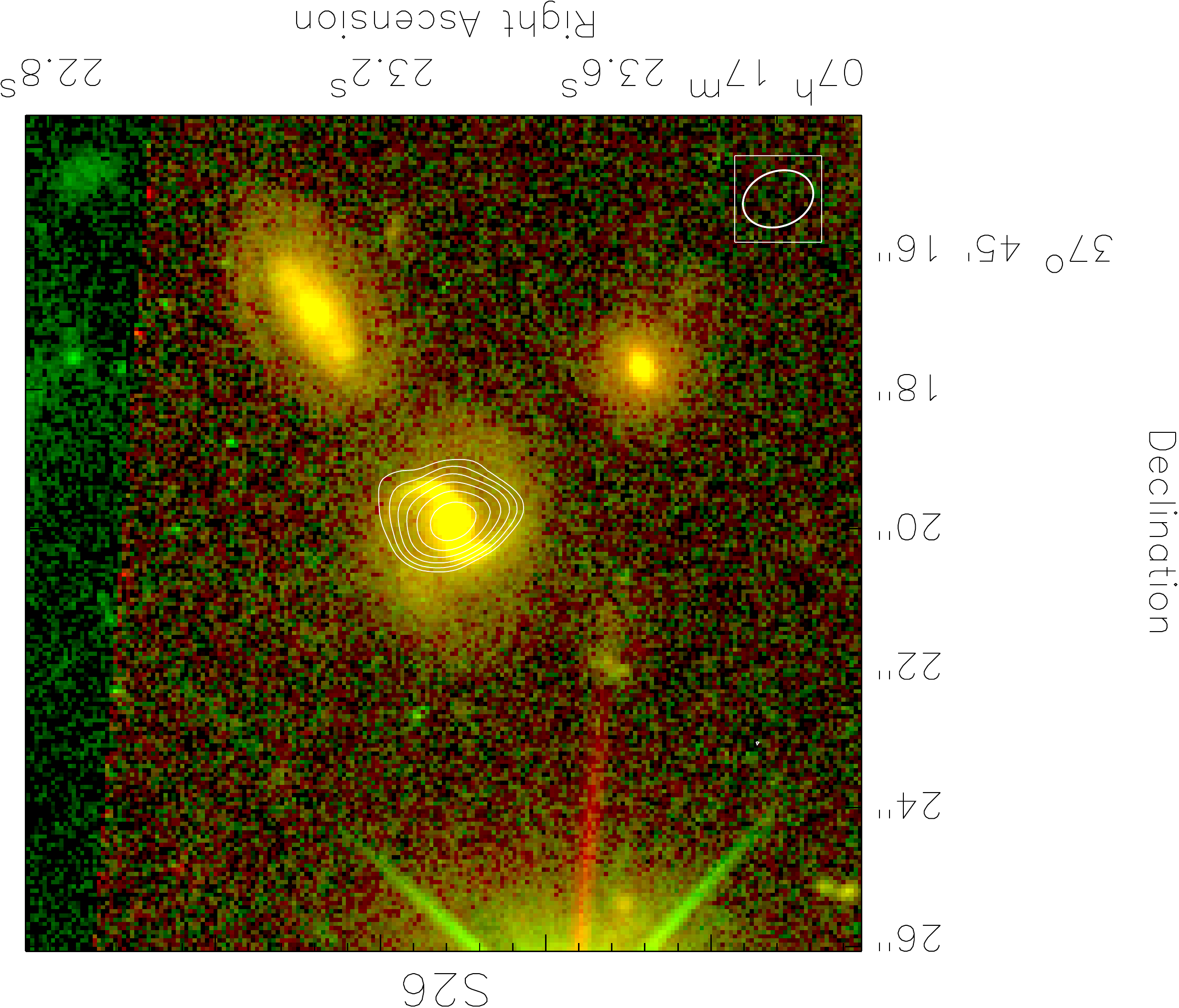}
\includegraphics[angle =180, trim =0cm 0cm 0cm 0cm,width=0.24\textwidth]{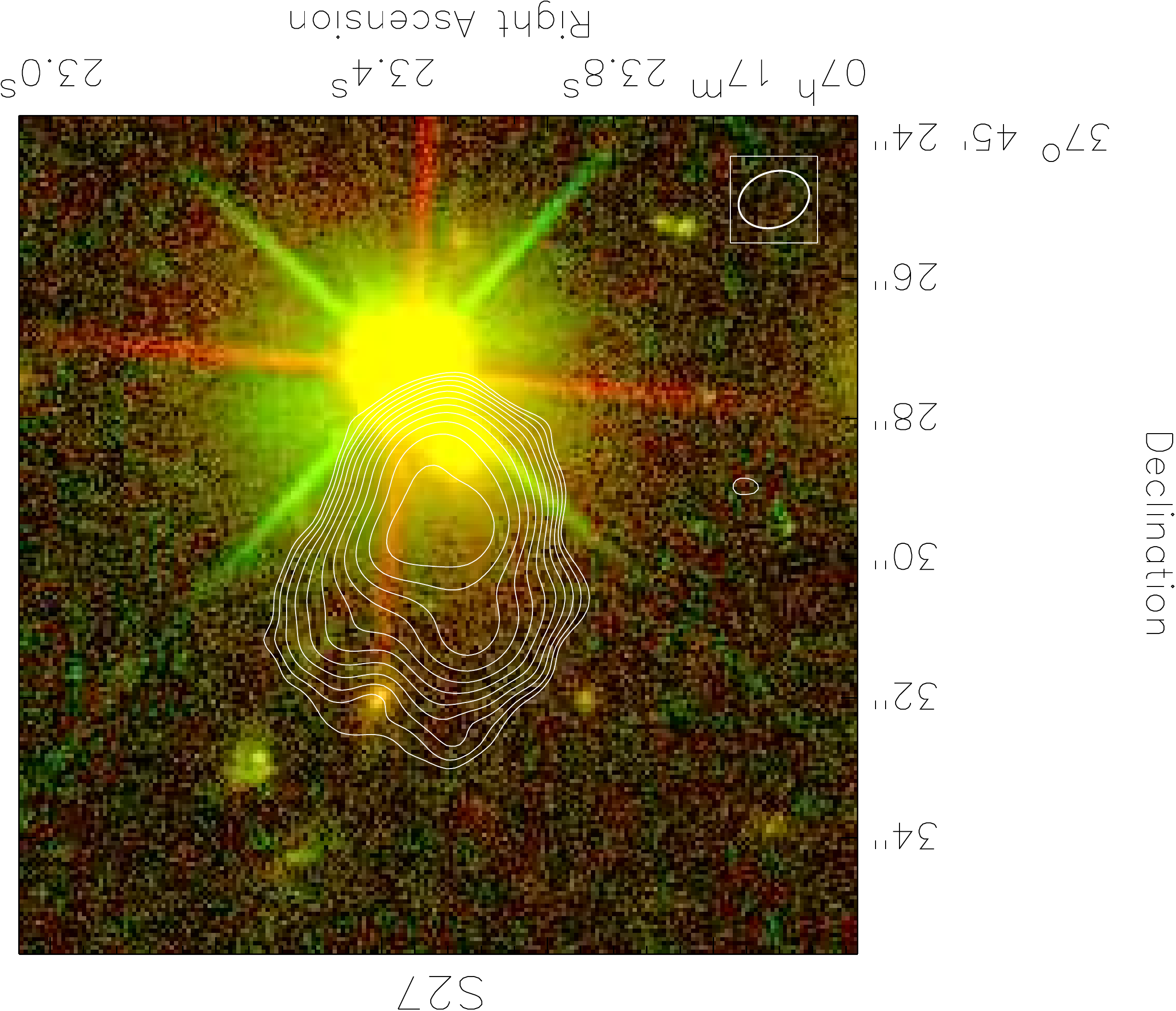}
\includegraphics[angle =180, trim =0cm 0cm 0cm 0cm,width=0.24\textwidth]{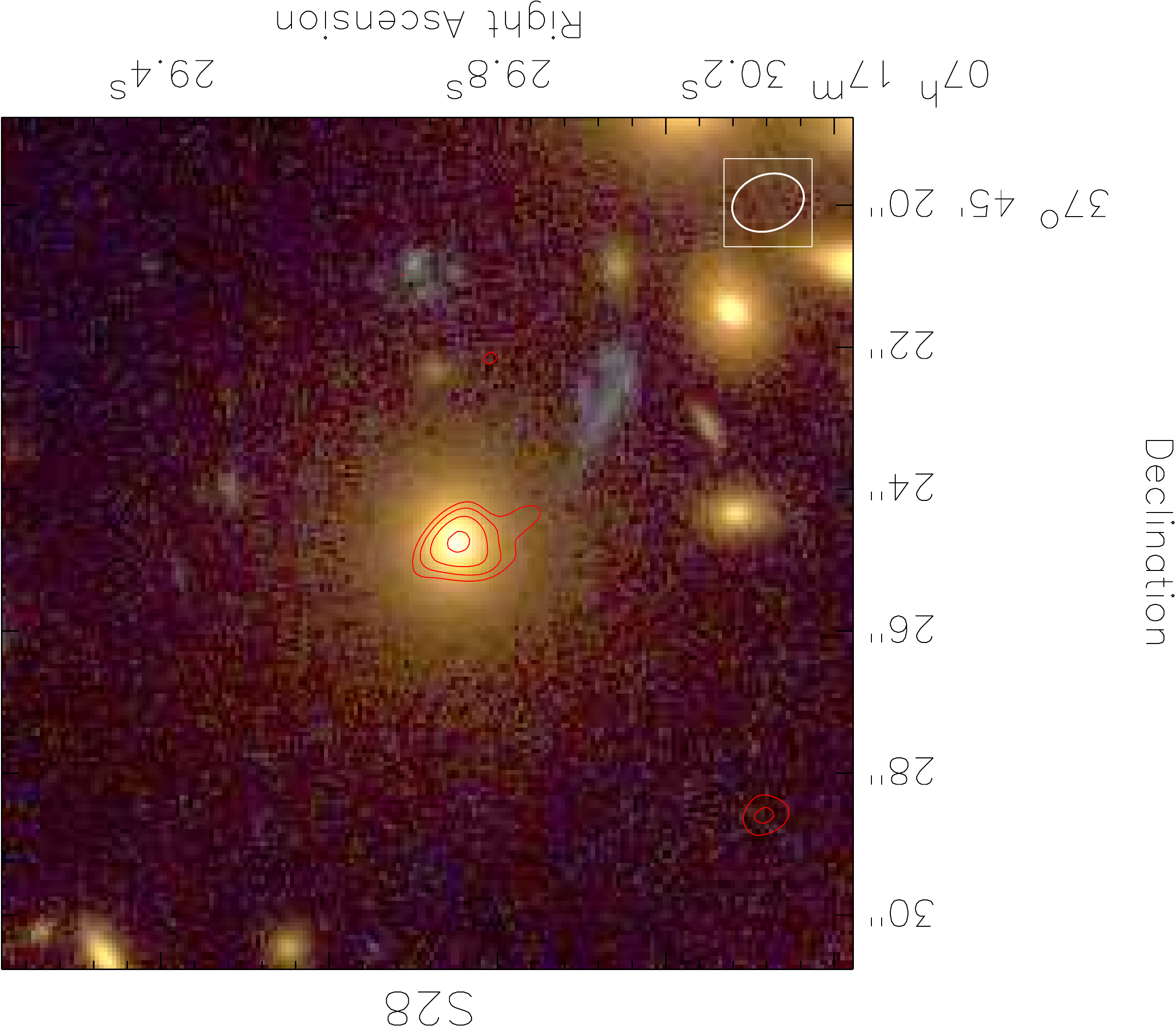}
\includegraphics[angle =180, trim =0cm 0cm 0cm 0cm,width=0.24\textwidth]{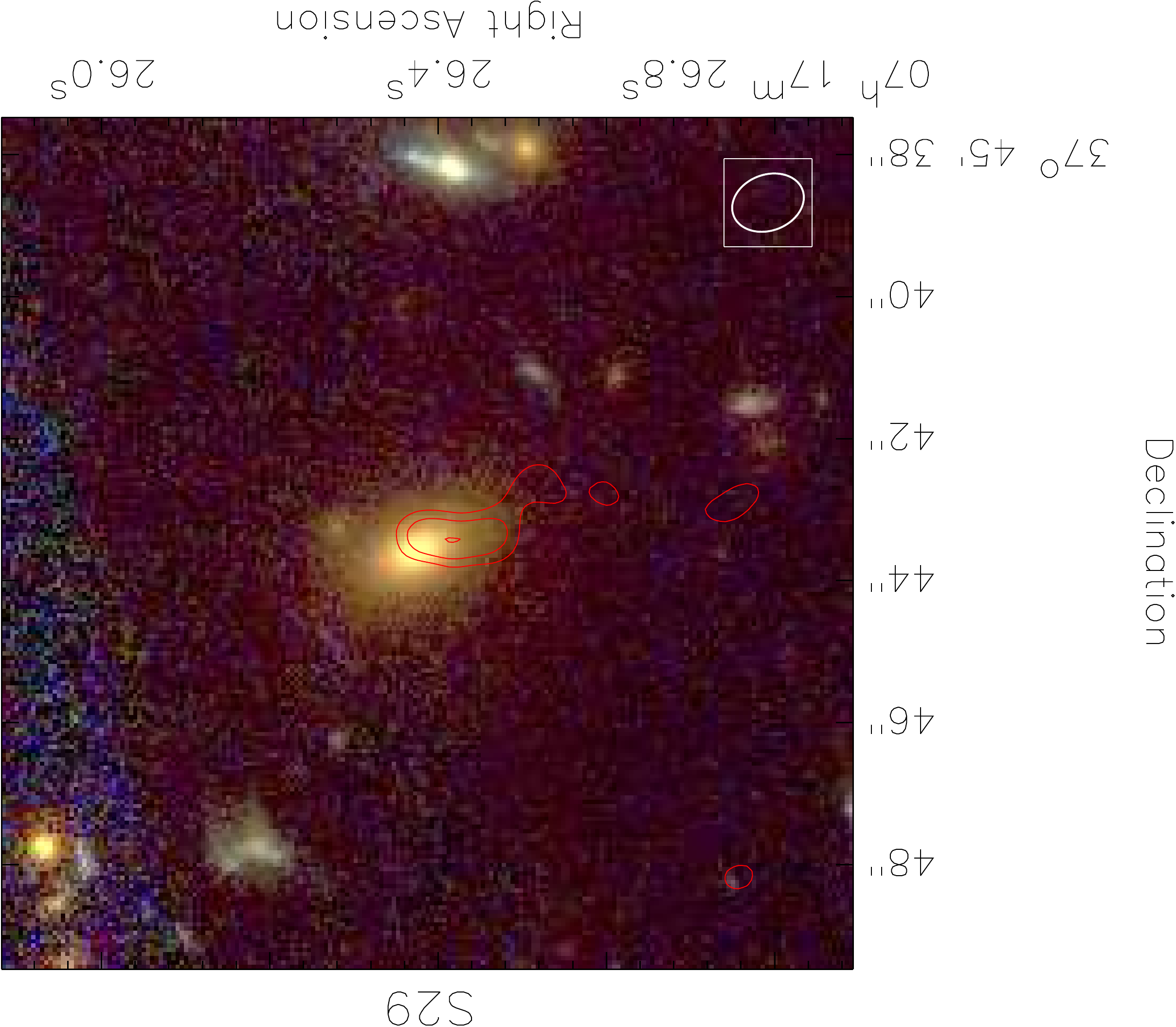} % 18
\includegraphics[angle =180, trim =0cm 0cm 0cm 0cm,width=0.24\textwidth]{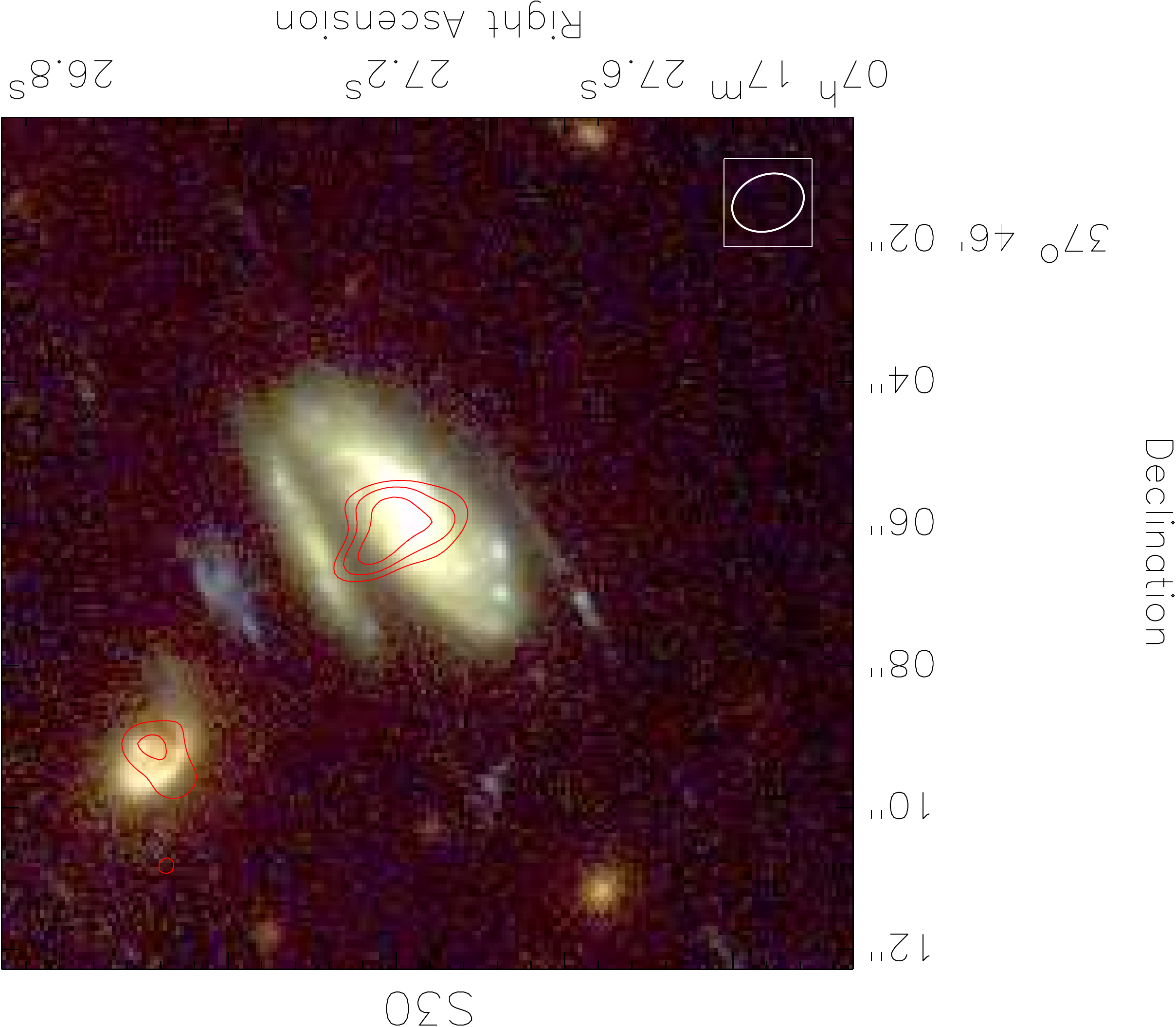}
\includegraphics[angle =180, trim =0cm 0cm 0cm 0cm,width=0.24\textwidth]{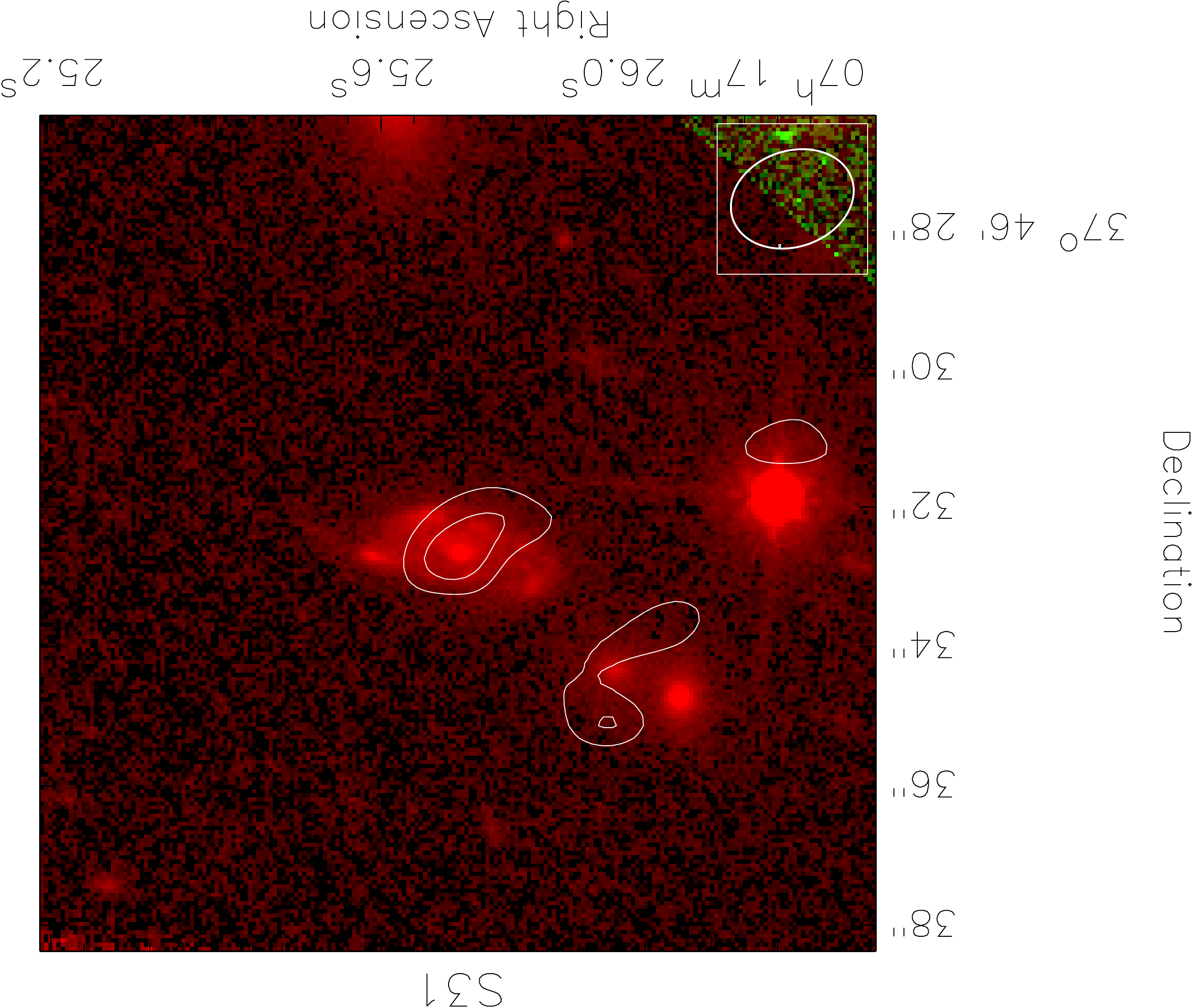} % 20
\caption{HST F435, F606W, F814W postage stamp color images around the compact radio sources in the MACS~J0717.3745 field. {The sources in this panel are cluster members, foreground objects, or sources with uncertain redshifts (so that we could not determine if they are lensed or not).} Postage stamp images for the lensed sources are shown in Figure~\ref{fig:cutouts1}.  The red radio contours are from the 2--4~GHz S-band image and drawn at levels $\sqrt{([1,2,4,\ldots])} \times 3\sigma_{\rm{rms}}$, with $\sigma_{\rm{rms}} = 1.8$~$\mu$Jy~beam$^{-1}$. The beam size is indicated in the bottom left corner. We draw white radio contours for some images to aid visibility (in case the area was not covered by all the HST filters). For sources without an S-band detection, we overlay contours from the C-band image (if detected there) or L-band image. We use contours from the combined L-, S-, and C-band image if the source is not detected in any of the three individual band images. The values for $\sigma_{\rm{rms}}$ for the L and C-band images are listed in Table~\ref{tab:jvlaimages}. Note that for source S20 there is a very faint CLASH counterpart located precisely at the radio position.} 
\label{fig:cutouts2}
\end{figure*}
\begin{figure*}[h]
%\ContinuedFloat
\centering
\includegraphics[angle =180, trim =0cm 0cm 0cm 0cm,width=0.24\textwidth]{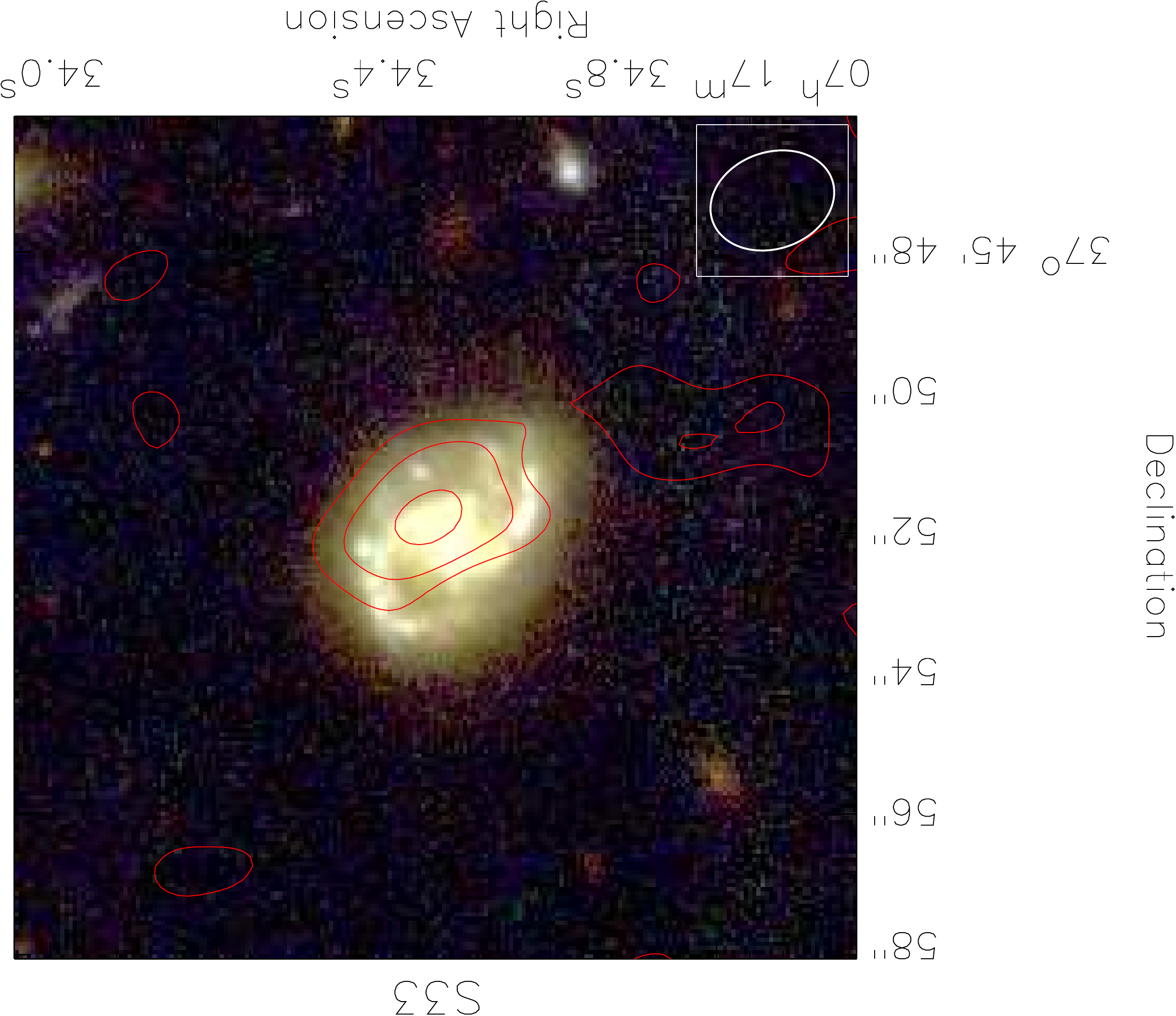} %21
\includegraphics[angle =180, trim =0cm 0cm 0cm 0cm,width=0.24\textwidth]{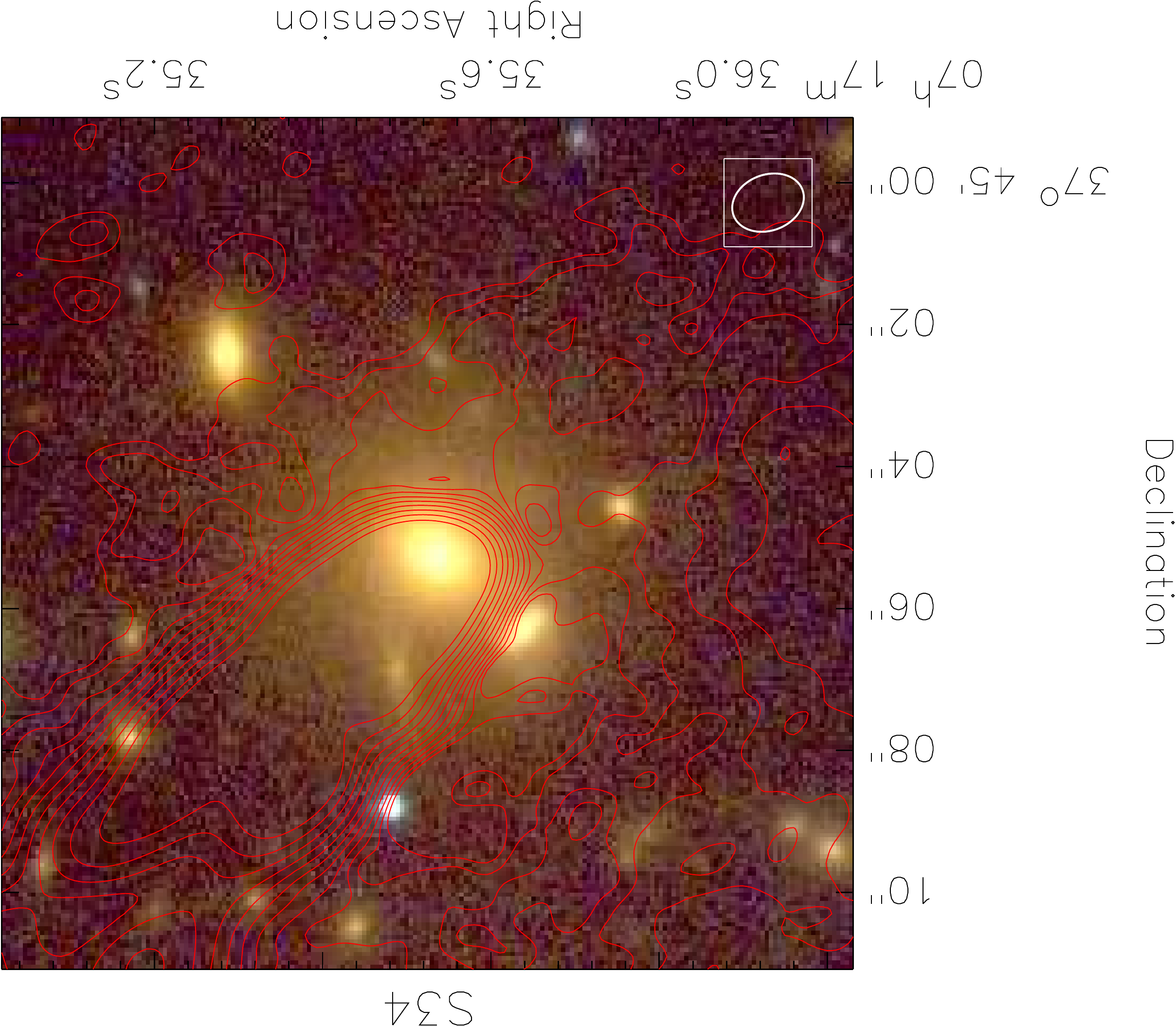}
\includegraphics[angle =180, trim =0cm 0cm 0cm 0cm,width=0.24\textwidth]{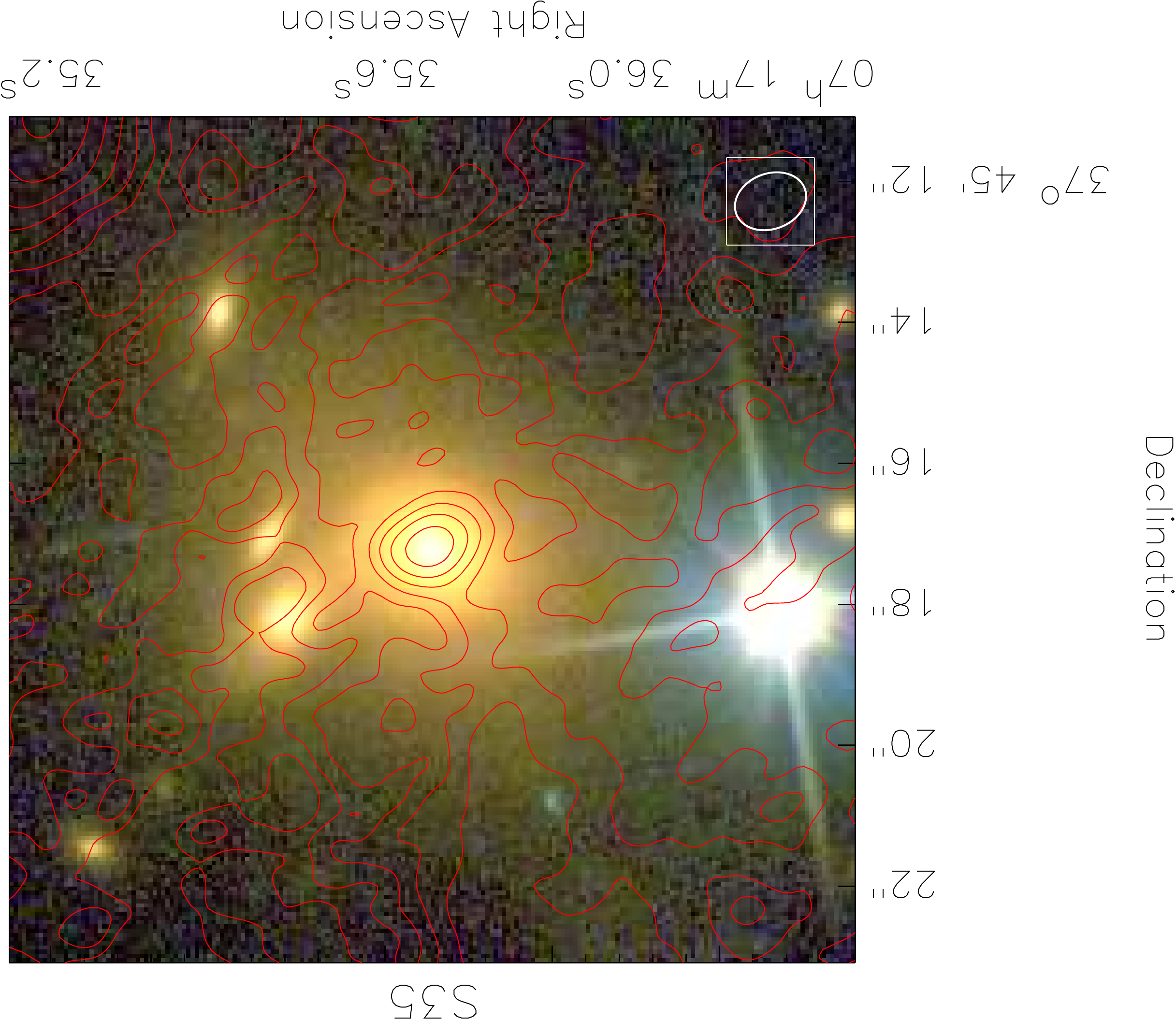}
\includegraphics[angle =180, trim =0cm 0cm 0cm 0cm,width=0.24\textwidth]{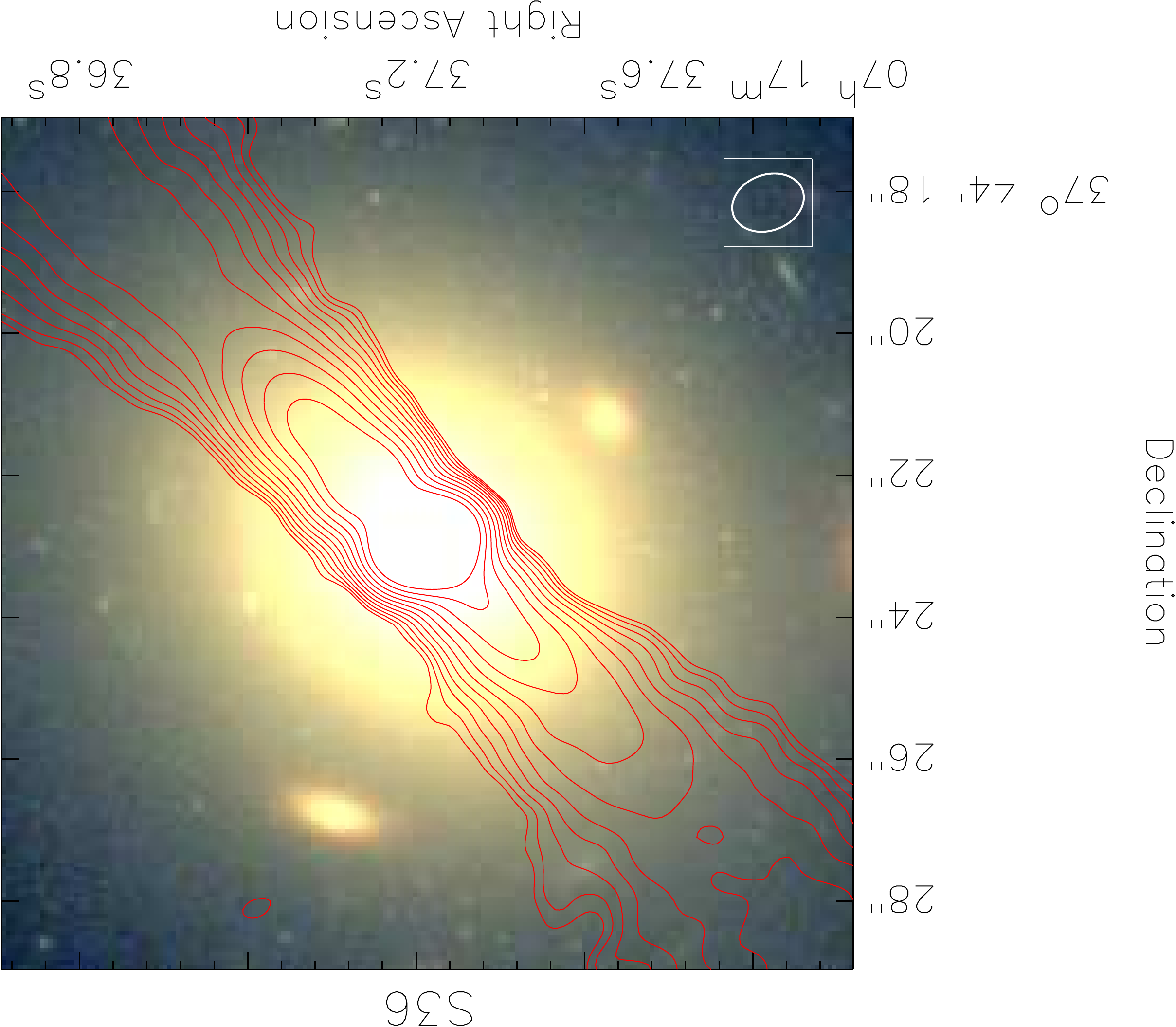}
\includegraphics[angle =180, trim =0cm 0cm 0cm 0cm,width=0.24\textwidth]{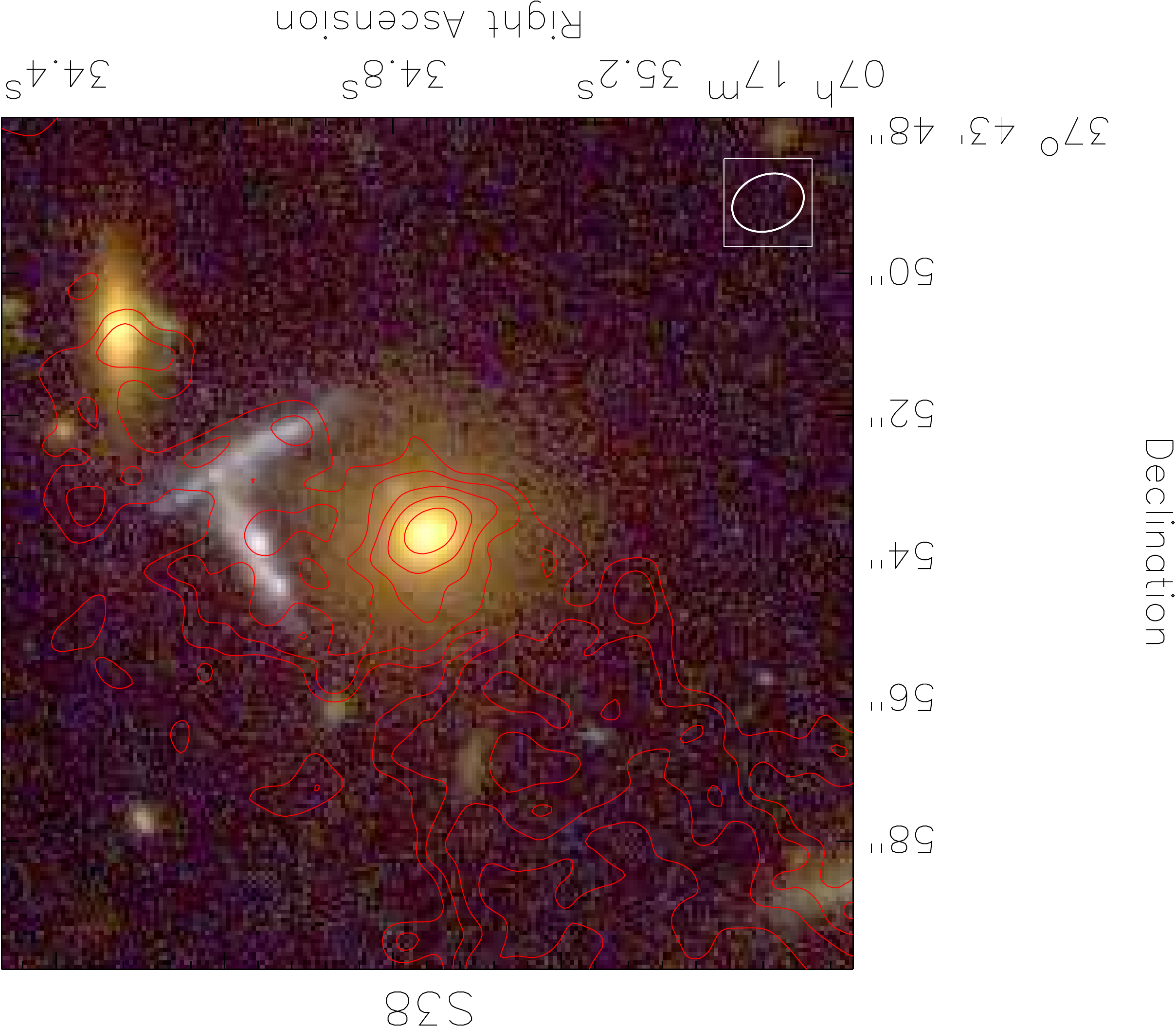} % 25
\includegraphics[angle =180, trim =0cm 0cm 0cm 0cm,width=0.24\textwidth]{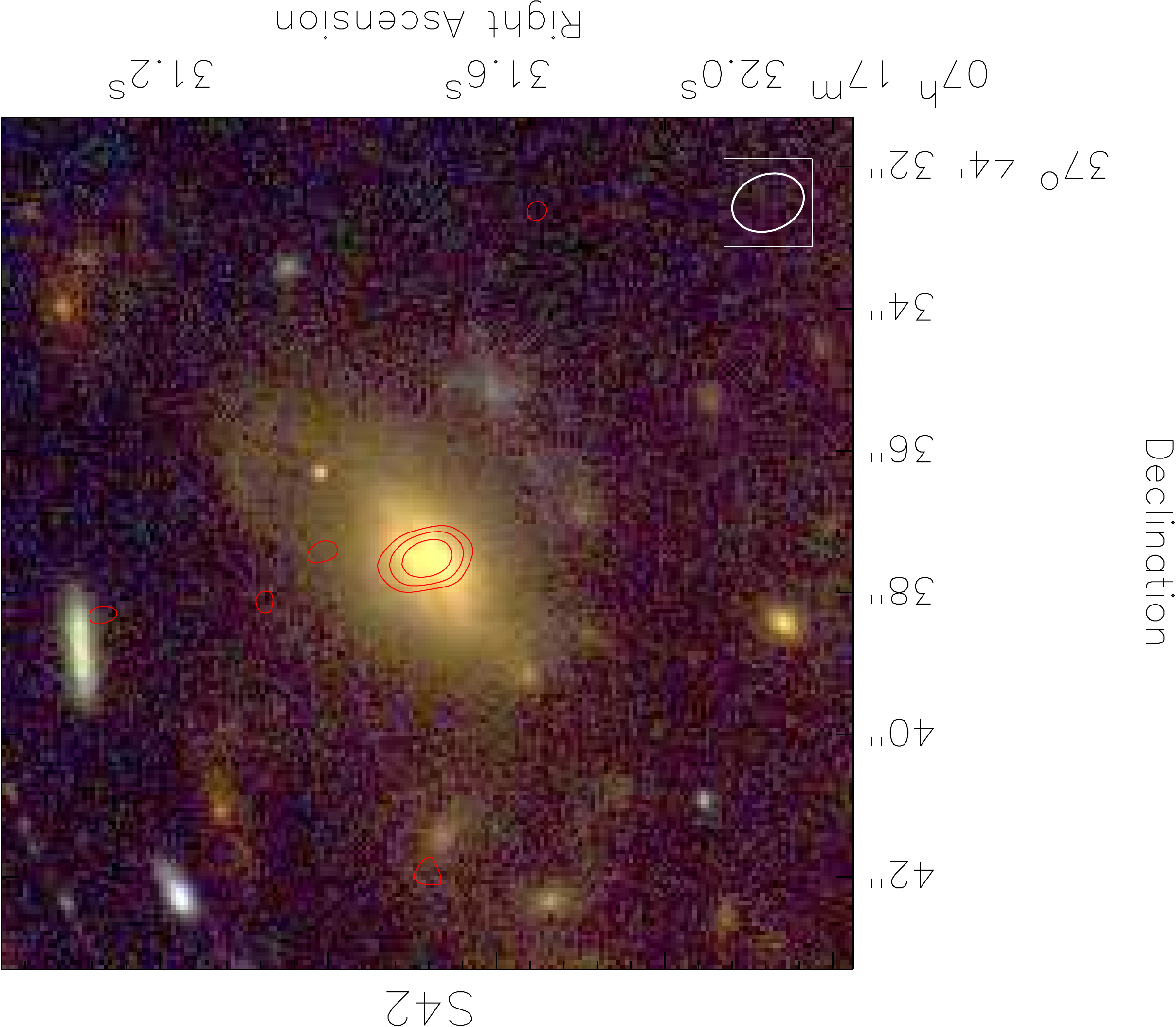}
\includegraphics[angle =180, trim =0cm 0cm 0cm 0cm,width=0.24\textwidth]{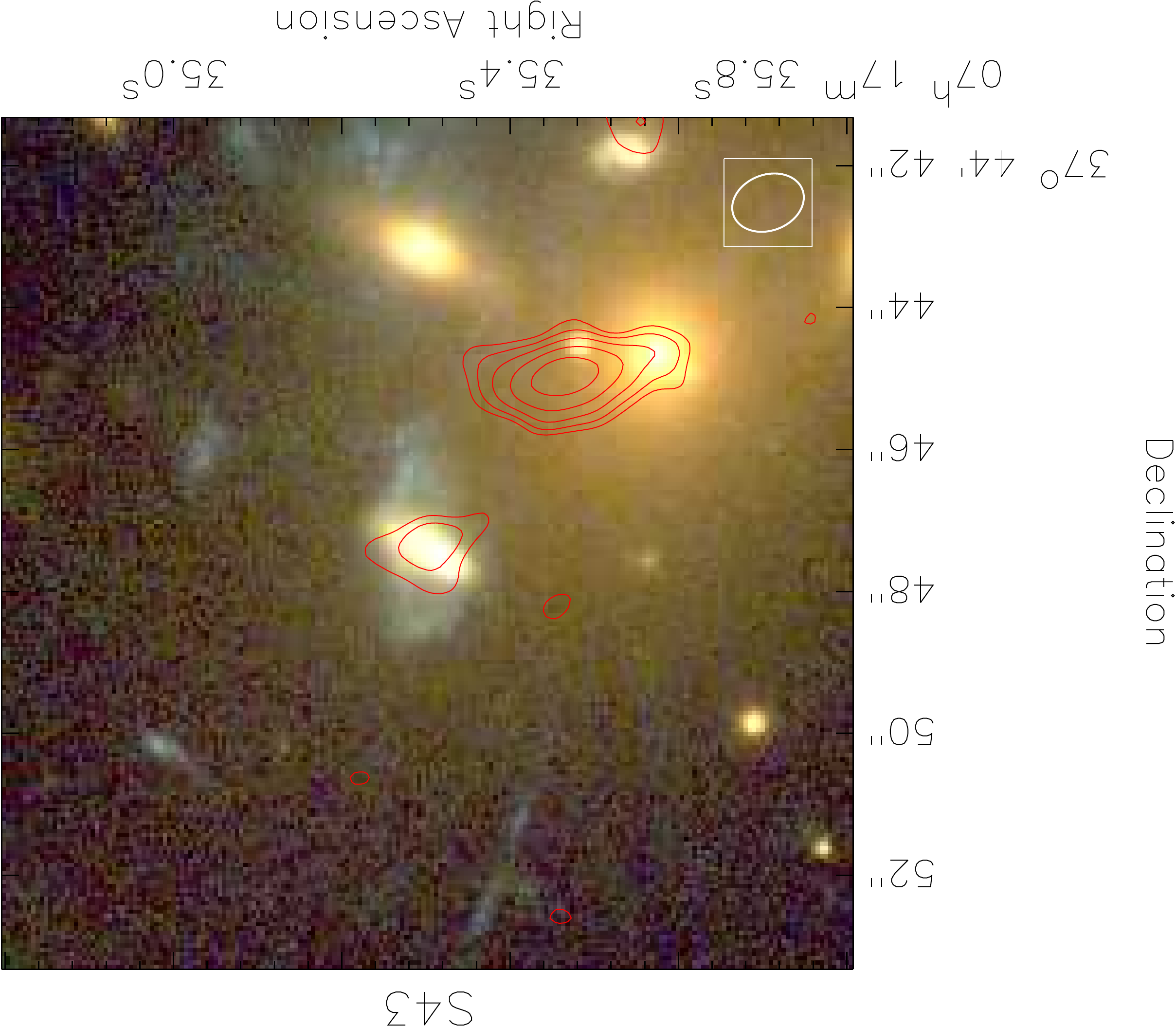}
\includegraphics[angle =180, trim =0cm 0cm 0cm 0cm,width=0.24\textwidth]{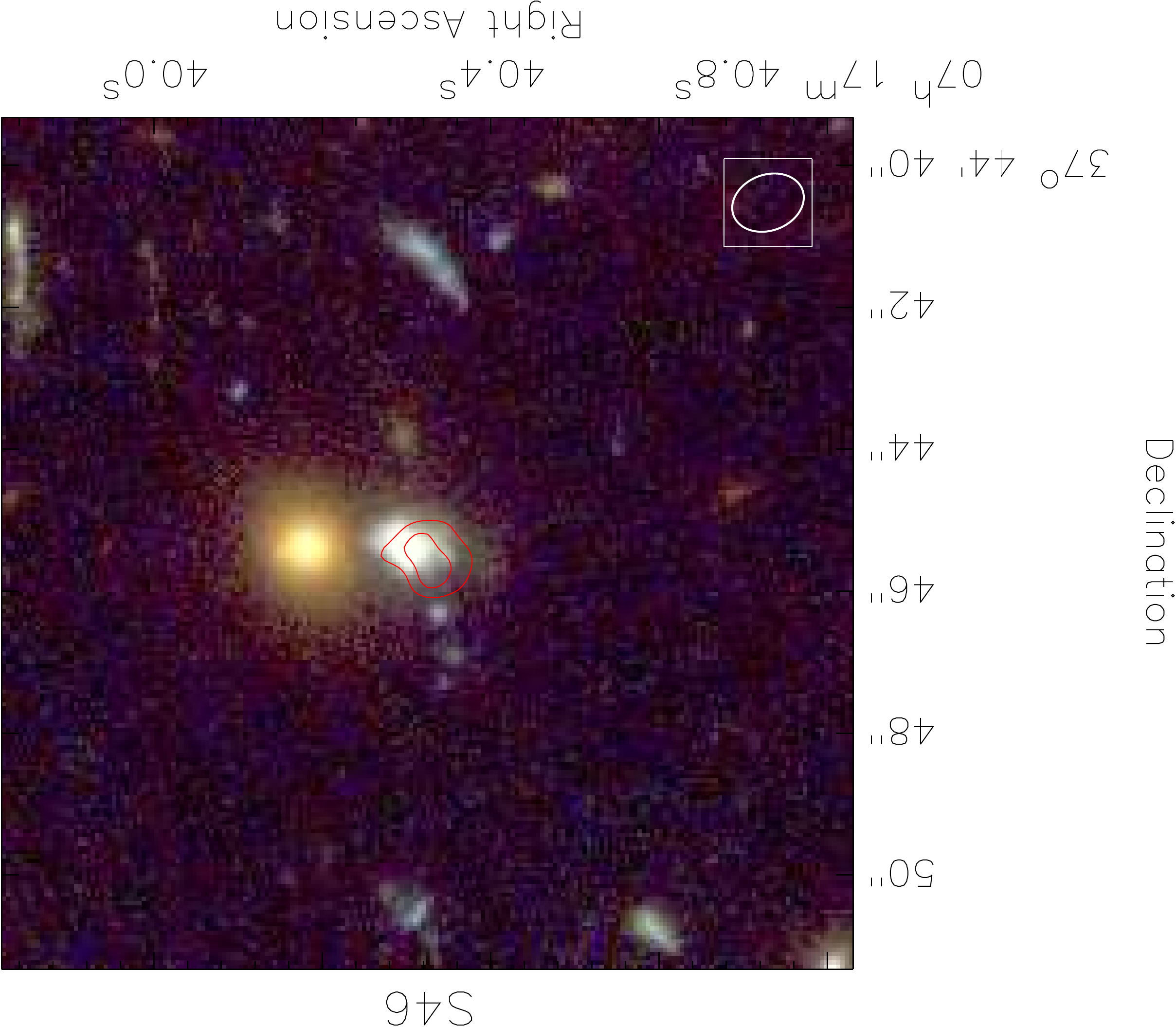} % 28
\includegraphics[angle =180, trim =0cm 0cm 0cm 0cm,width=0.24\textwidth]{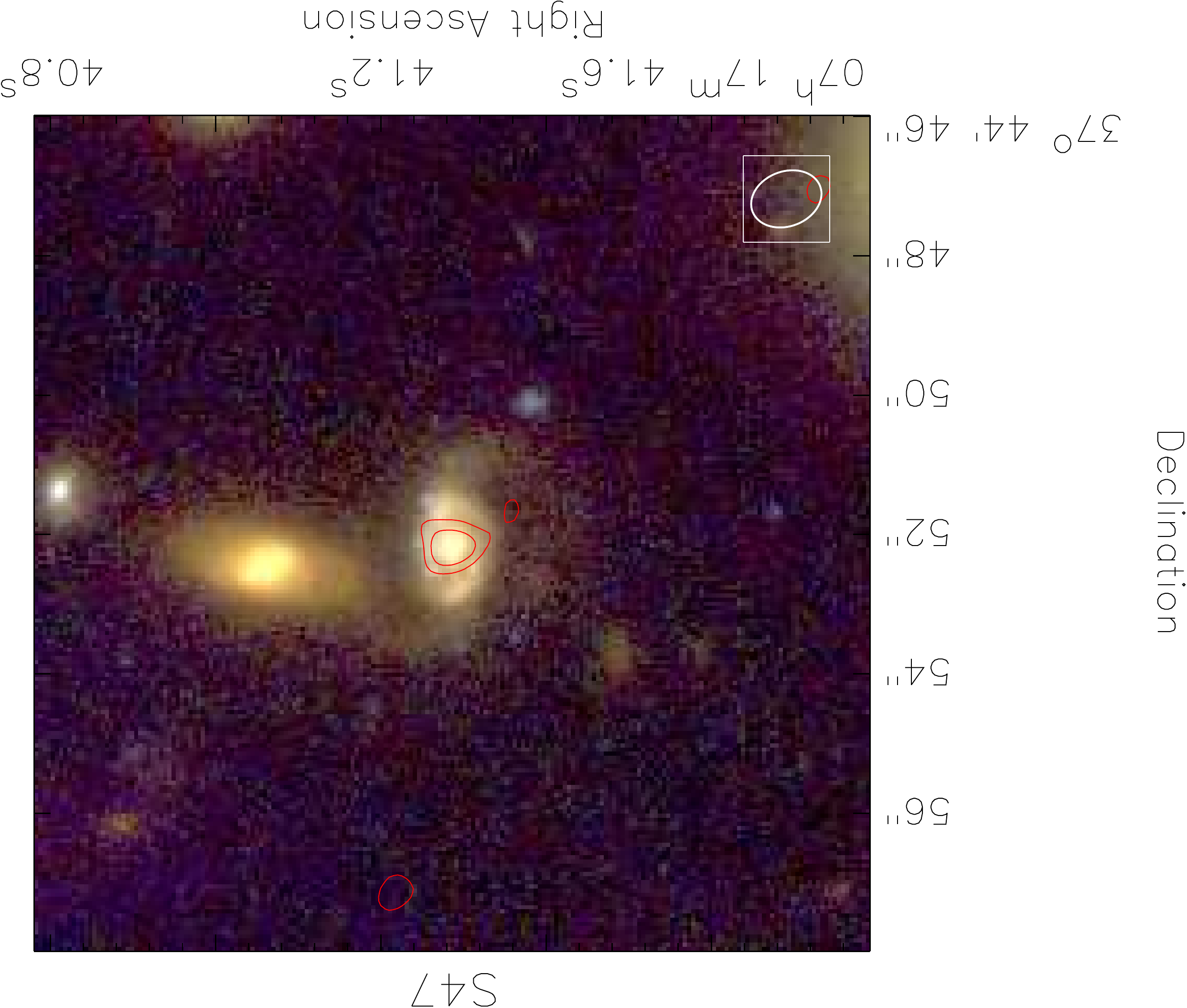}
\includegraphics[angle =180, trim =0cm 0cm 0cm 0cm,width=0.24\textwidth]{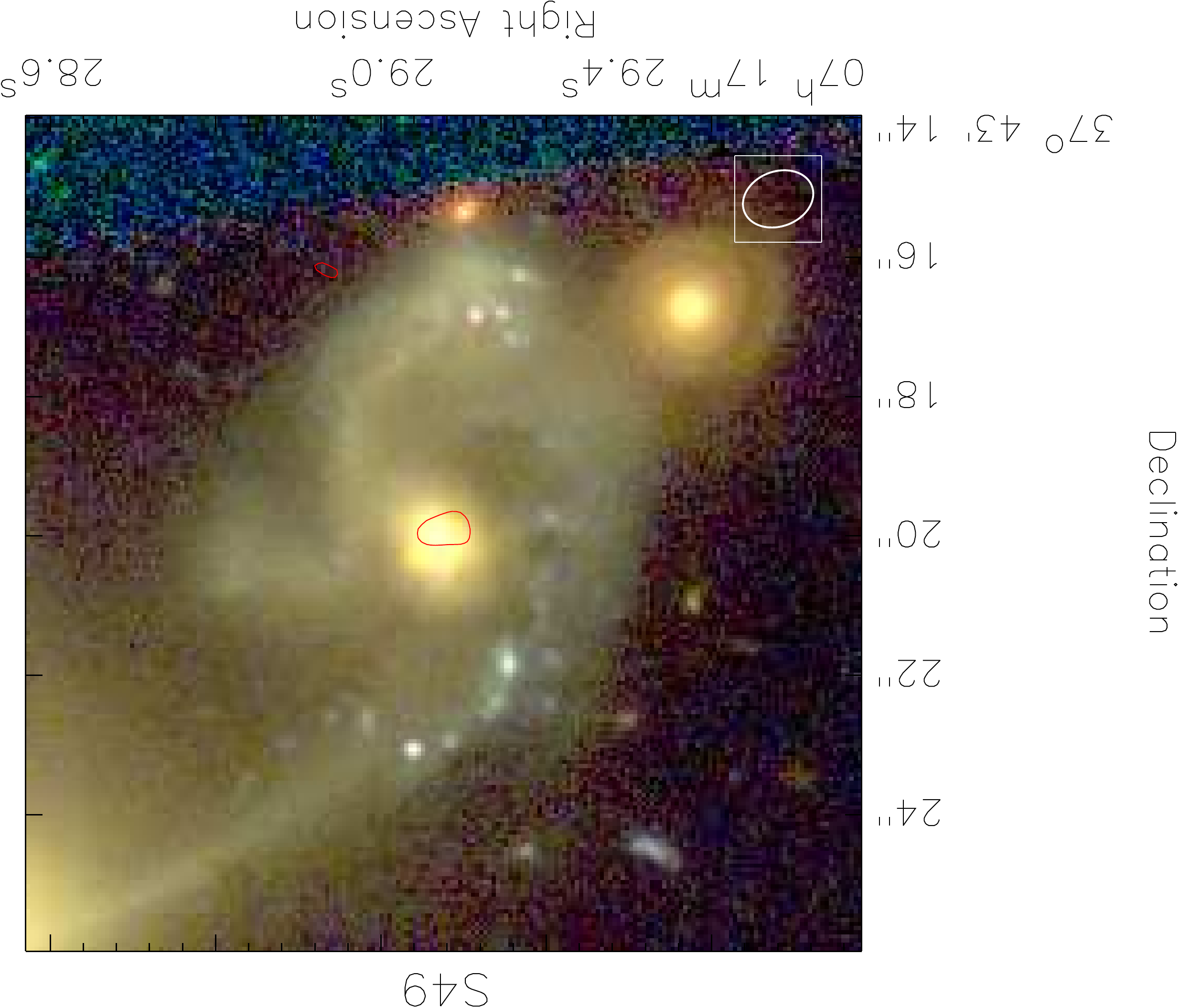} % 30
\includegraphics[angle =180, trim =0cm 0cm 0cm 0cm,width=0.24\textwidth]{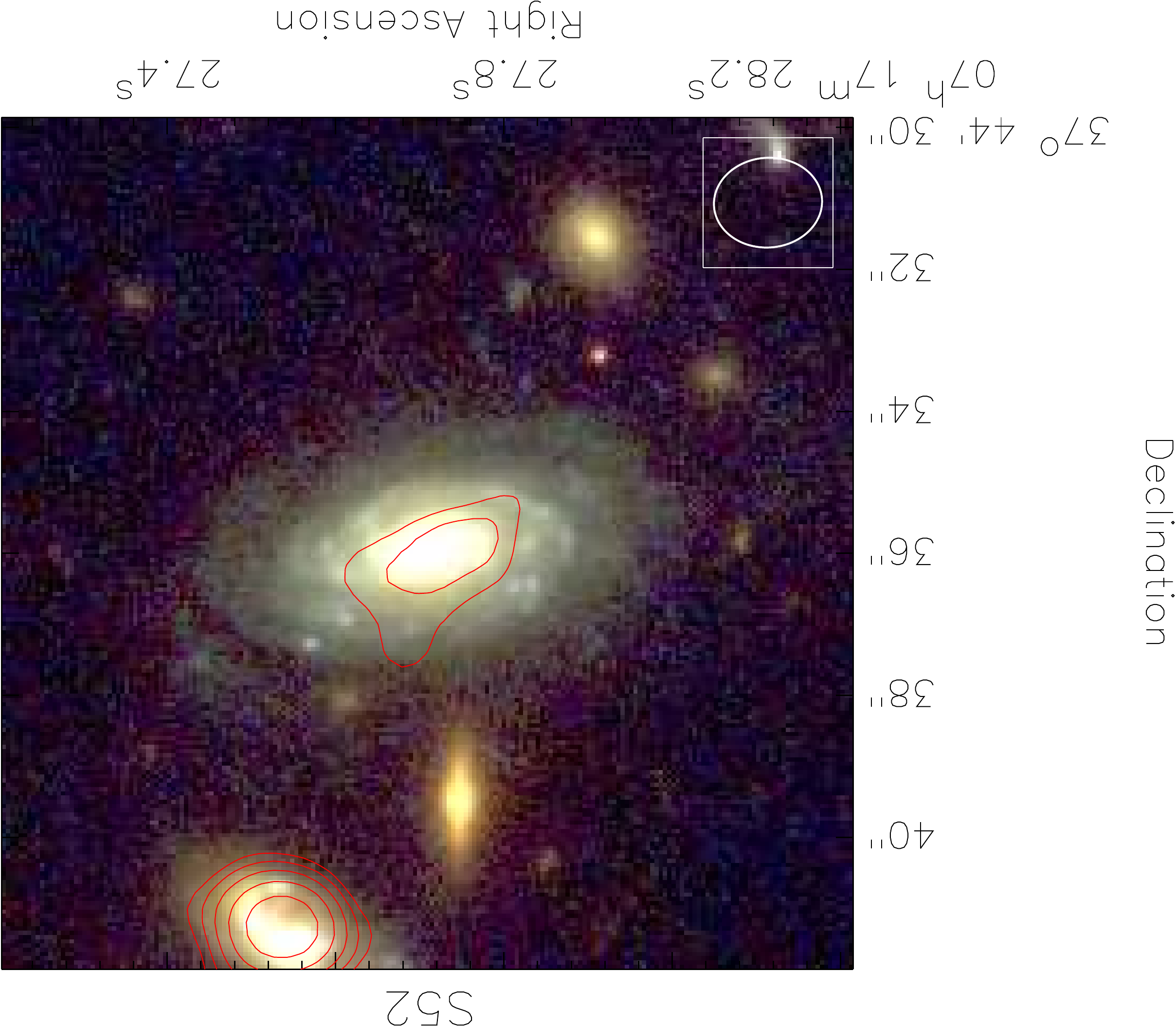}
\includegraphics[angle =180, trim =0cm 0cm 0cm 0cm,width=0.24\textwidth]{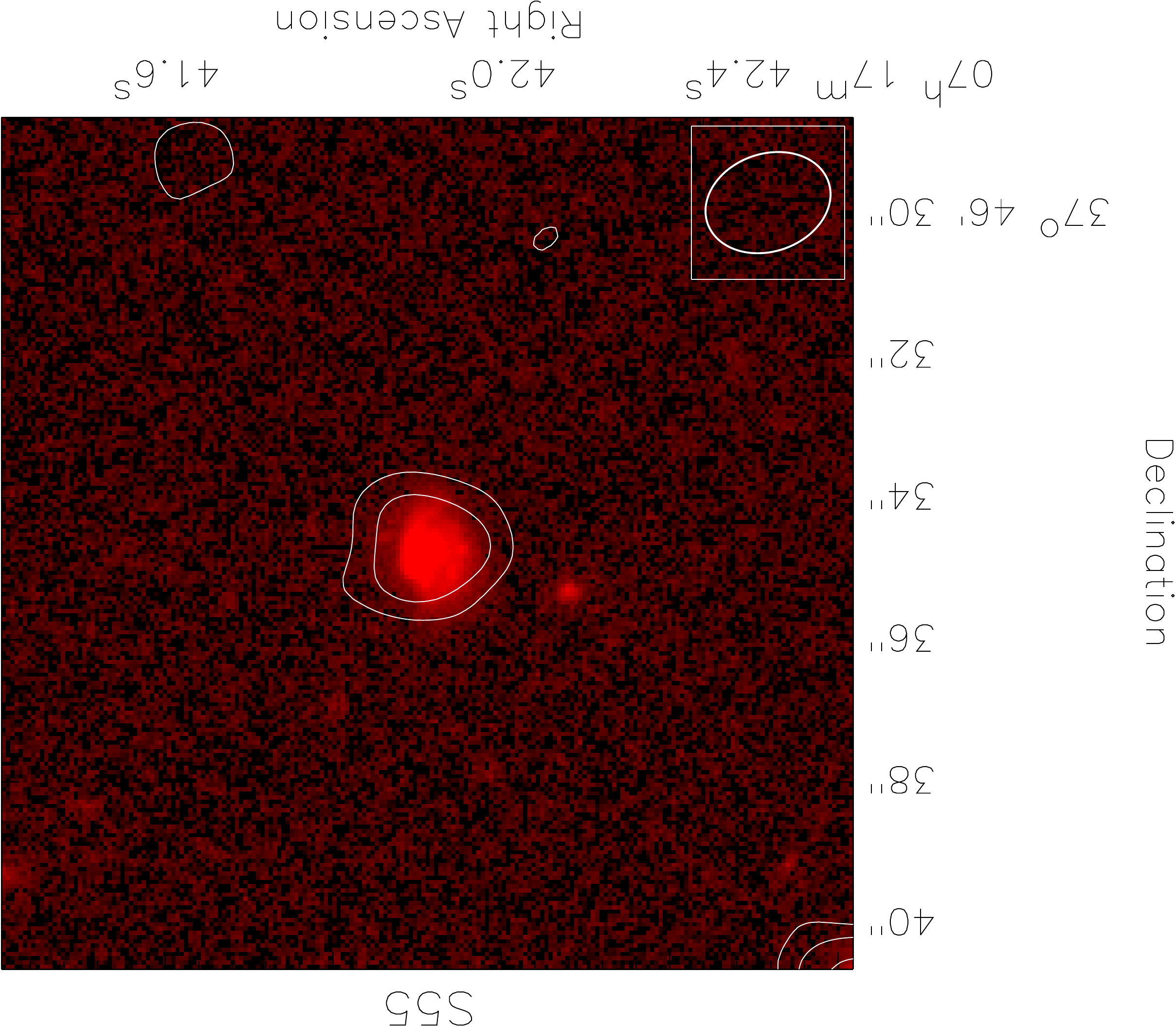}
\includegraphics[angle =180, trim =0cm 0cm 0cm 0cm,width=0.24\textwidth]{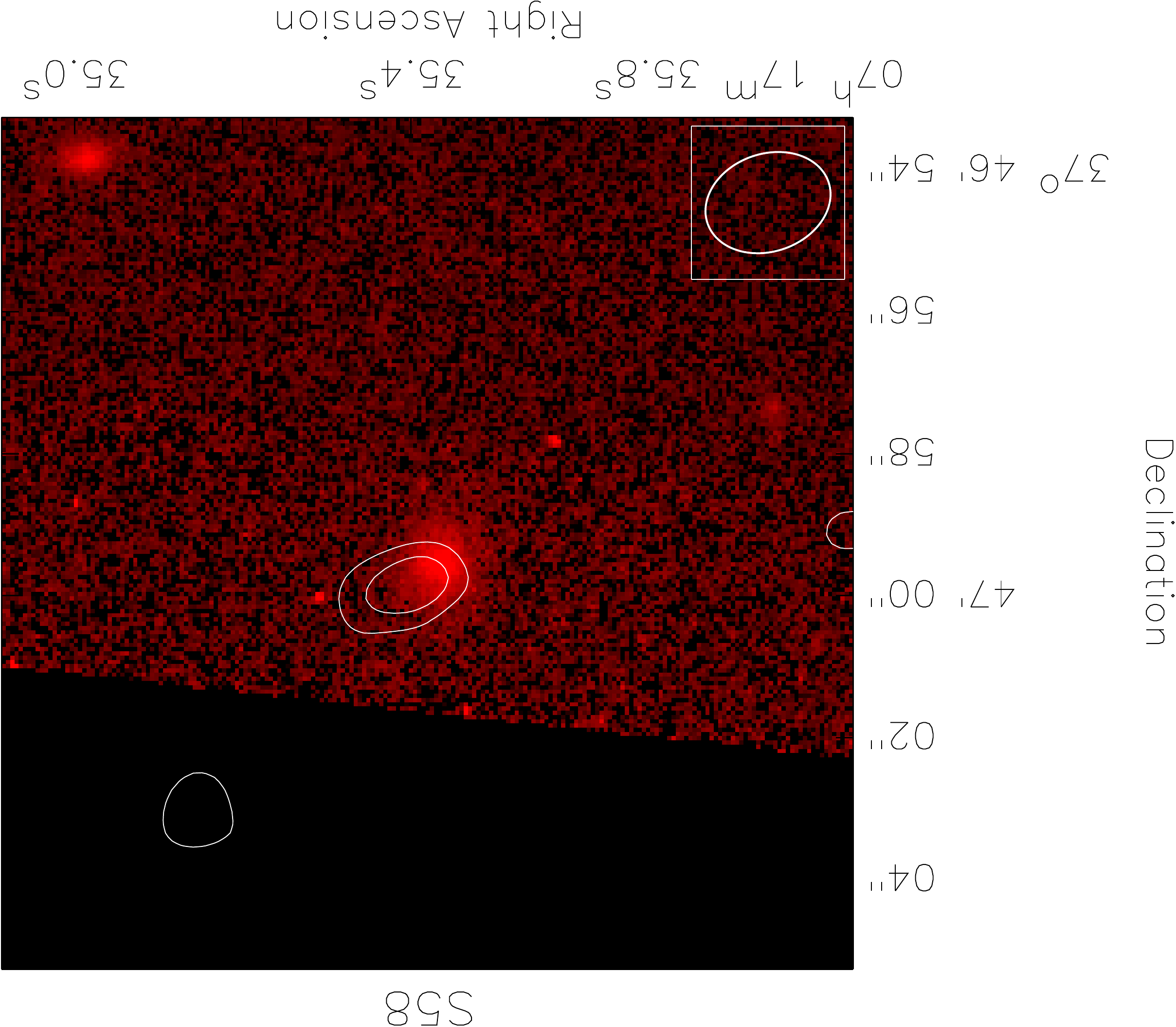} % 33
\includegraphics[angle =180, trim =0cm 0cm 0cm 0cm,width=0.24\textwidth]{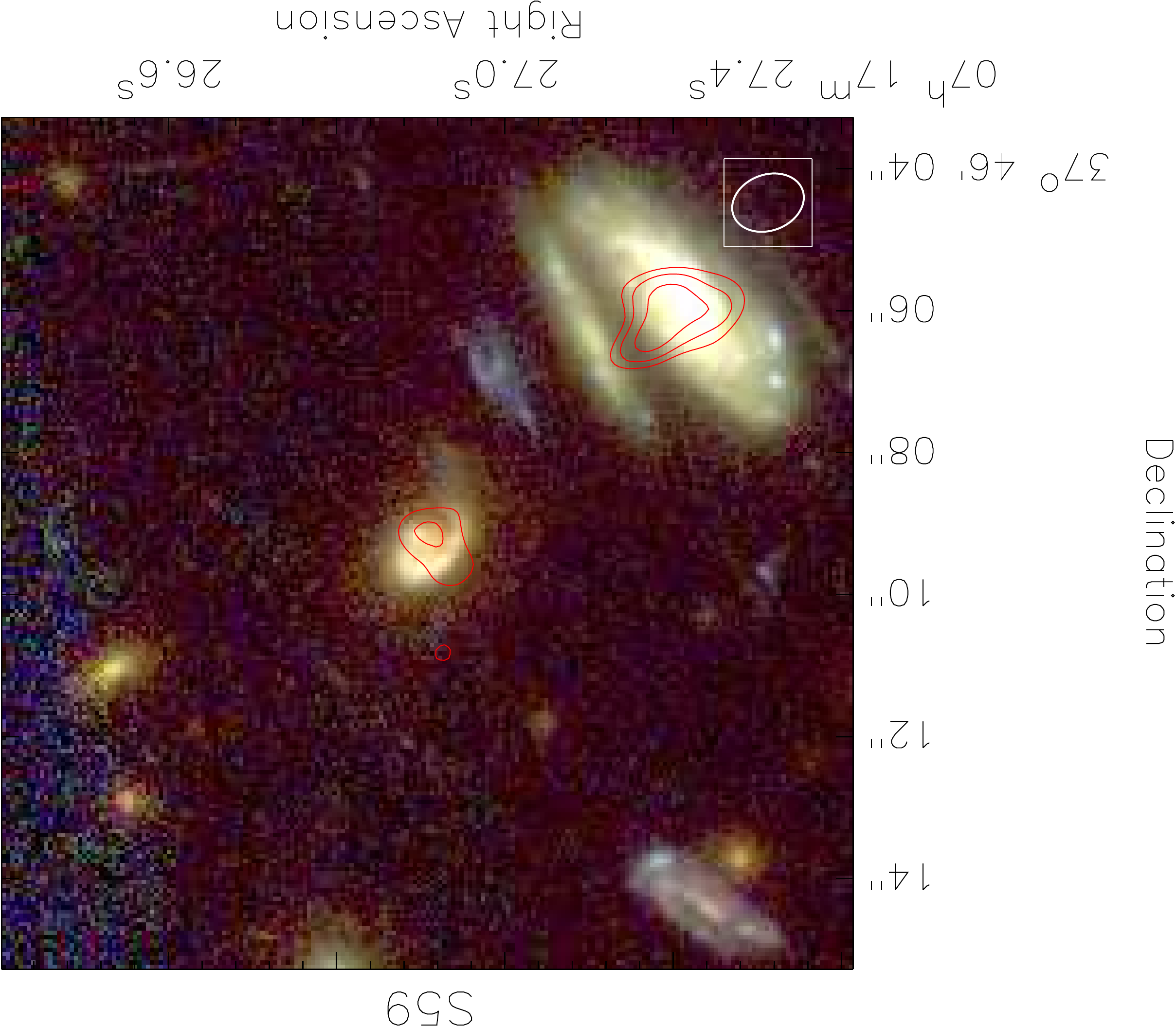}
\includegraphics[angle =180, trim =0cm 0cm 0cm 0cm,width=0.24\textwidth]{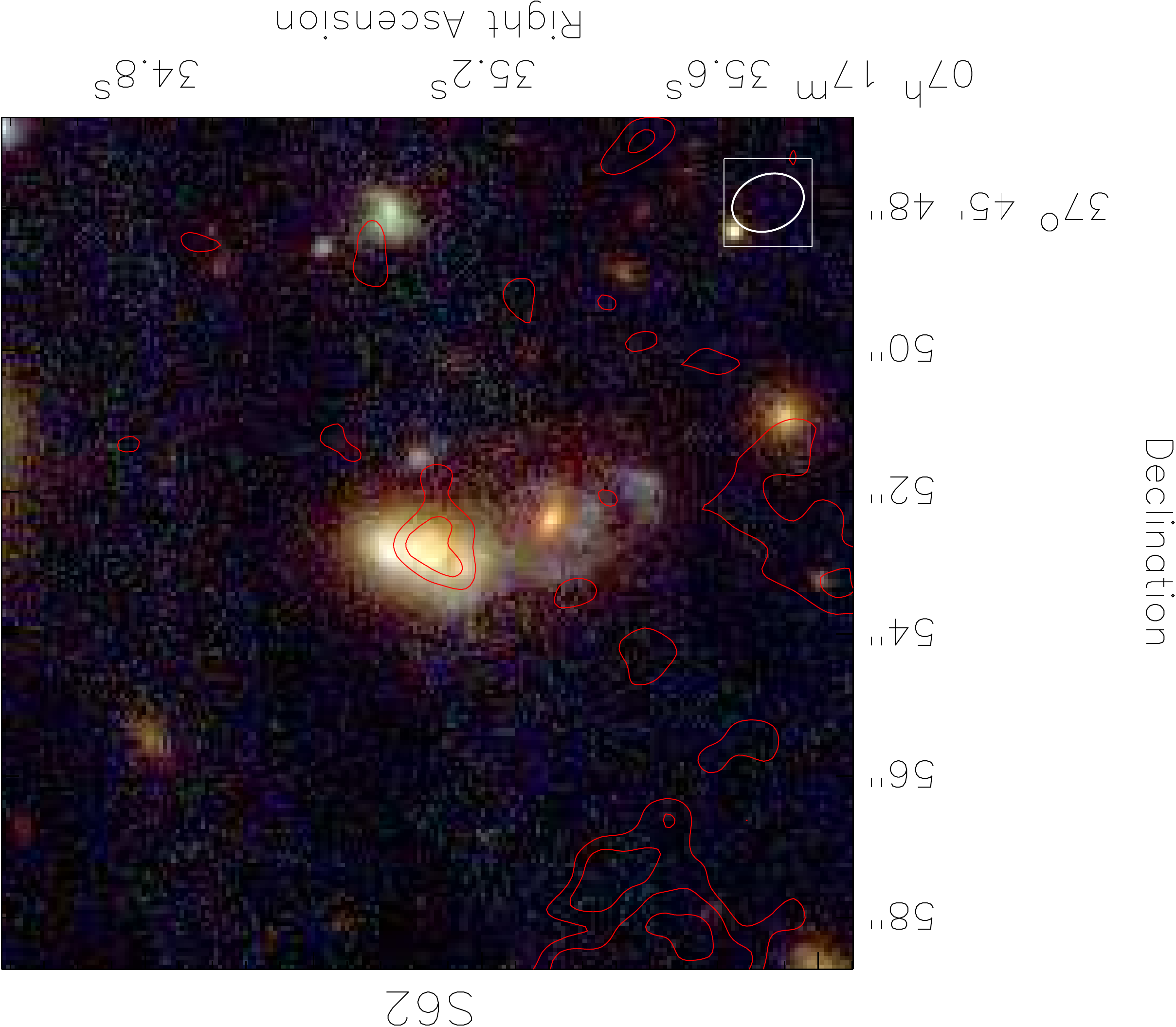} % 35
\caption{Continued.}
\end{figure*}

\bibliography{ref_filaments.bib}

\end{document}